%% file: dijet_arXiv_v3.tex
\documentclass[aps,preprint,preprintnumbers,superscriptaddress,nofootinbib,
floatfix] {revtex4-1}
\pdfoutput=1 

\usepackage[T1]{fontenc}

\usepackage{color}
\usepackage[usenames,dvipsnames,svgnames,table]{xcolor}
\usepackage{placeins}
\usepackage{hyperref}

\usepackage{amsmath}
\usepackage{amssymb}
\RequirePackage{mathrsfs}
\usepackage{dsfont}
\usepackage{graphicx}
\usepackage{slashed}
\usepackage{soul}
\usepackage{multirow}
\usepackage{caption,subcaption}
\usepackage{setspace}
\usepackage[normalem]{ulem}

\usepackage{array}
\newcolumntype{L}[1]{>{\raggedright\let\newline\\\arraybackslash\hspace{0pt}}m{#1}}
\newcolumntype{C}[1]{>{\centering\let\newline\\\arraybackslash\hspace{0pt}}m{#1}}
\newcolumntype{R}[1]{>{\raggedleft\let\newline\\\arraybackslash\hspace{0pt}}m{#1}}

\graphicspath{{./img/}}
\newcommand{\unit}[1]{\,\mathrm{#1}}
\newcommand{\gev}{\unit{GeV}}
\newcommand{\tev}{\unit{TeV}}
\newcommand{\fb}{\unit{fb}}
\newcommand{\invfb}{\fb^{-1}}

\newcommand{\ee}{\mathrm{e}}
\newcommand{\dd}{\mathrm{d}}

\renewcommand{\epsilon}{\varepsilon}

\newcommand{\reffig}[1]{fig.~\ref{#1}}
\newcommand{\refeq}[1]{eq.~\eqref{#1}}
\newcommand{\reftab}[1]{tab.~\ref{#1}}
\newcommand{\sect}{section}
\newcommand{\refsec}[1]{\sect~\ref{#1}}
\newcommand{\Ref}{ref.}
\newcommand{\Refs}{refs.}
\newcommand{\app}{appendix}

\renewcommand{\vec}[1]{\boldsymbol{#1}}
\renewcommand{\bar}{\overline}

\newcommand{\tabhead}[1]{\multicolumn{1}{c}{#1}}

\begin{document}

\title{Constraints on New Scalar and Vector Mediators from LHC Dijet Searches}

\author{Sebastian Baum}
\email{sbaum@fysik.su.se}
\affiliation{Oskar Klein Centre, Department of Physics, Stockholm University, 
AlbaNova, Stockholm SE-10691, Sweden}
\affiliation{Nordita, KTH Royal Institute of Technology and Stockholm 
University, Roslagstullsbacken 23, 10691 Stockholm, Sweden}

\author{Riccardo Catena}
\email{catena@chalmers.se}
\affiliation{Chalmers University of Technology, Department of Physics, SE-412 
96 G\"oteborg, Sweden}

\author{and Martin B. Krauss}
\email{martin.krauss@chalmers.se}
\affiliation{Chalmers University of Technology, Department of Physics, SE-412 
96 G\"oteborg, Sweden}
\affiliation{Dipartimento di Matematica e Fisica, Università di Roma Tre, Via 
della Vasca Navale 84, 00146 Rome, Italy}

\begin{abstract}
We present a reanalysis of the latest results from CMS dijet searches for an integrated luminosity of $36\invfb$ together with preliminary results for $78\invfb$ in the framework of simplified models for dark matter interacting with quarks through the exchange of a scalar, pseudoscalar, vector or pseudovector mediator particle.~Within the same framework, we also project the sensitivity of dijet searches in future LHC runs and study how well the parameters of a simplified model could be reconstructed in case of a future discovery at the high luminosity (HL) LHC.~Finally, we explore the possibility of discriminating different mediator scenarios by extending the sensitivity of dijet searches for simplified models through the use of angular information.~It is the first time that these studies are performed systematically for the case of spin 0 mediators.~Among other results we find:~1) no evidence for a dijet signal in the simplified model framework;~2) improvements due to an increased luminosity at the HL-LHC are significant, but mostly for heavy mediators, where dijet searches are limited by statistical, rather than systematical uncertainties;~3)~Information on the angular separation of dijets could be used at the HL-LHC to discriminate different mediator scenarios.
\end{abstract}

\maketitle
\flushbottom

\section{Introduction}
In the quest for physics Beyond the Standard Model (BSM) at the Large Hadron Collider (LHC), {\it dijet} searches are a powerful probe of new particles coupling to quarks or gluons. Such searches generically focus on the resonant production of a new heavy particle in proton--proton collisions and the subsequent decay of the new particle into pairs of quarks, gluons, or a quark and a gluon.\footnote{Of course, BSM physics can give rise to many different final states. Here, we focus on dijet final states as a probe of BSM physics.}  Due to hadronization, these quarks or gluons will then be seen as a pair of hadronic jets in the detector.

Neither the ATLAS nor the CMS collaboration has reported a significant excess to date, instead, they established upper limits on resonant dijet production cross sections. Currently, ATLAS has published results for generic (high-mass) dijet searches from $37\invfb$ of data taken during 2015 and 2016~\cite{Aaboud:2017yvp}, whereas CMS has presented results for $36\invfb$ of data from the 2016 dataset~\cite{CMS-PAS-EXO-16-056,Sirunyan:2018xlo}, as well as preliminary results for $78\invfb$ from the combined 2016 and 2017 datasets~\cite{CMS-PAS-EXO-17-026}. More targeted analyses have been carried out beyond the generic (high-mass) dijet searches, for example focusing on low-mass dijet resonances~\cite{Aaboud:2018fzt}, or final states including $b$-{\it tagged} jets~\cite{Sirunyan:2018pas}, additional charged leptons~\cite{ATLAS-CONF-2018-015}, or additional photons~\cite{Aaboud:2018zba}. While such searches can make new regions of the parameter space available, we will focus on traditional dijet searches aimed at the multi-TeV mediator mass range in this work.

An interesting application of dijet searches is in the context of dark matter (DM). If the interaction between the visible and the dark sector is mediated by a new (heavy) particle, this \textit{mediator} can potentially be produced at the LHC and decay into a pair of quarks. While the DM particle is not directly involved in such processes, one can obtain valuable information about the underlying DM model. For example, the width and branching ratios (BRs) of the mediator depend on the properties of the DM candidate, which thus may be inferred in such an analysis. Focusing on the role of DM in dijet events, we study the interplay of dijet searches at the LHC and DM direct detection experiments in the simplified model framework in a companion paper~\cite{Baum:2018lua}. 

Scenarios where the interaction between DM and the Standard Model (SM) is communicated by as yet undiscovered mediator particles are realized in many models of new physics, the most well-known being the various realizations of supersymmetry. For phenomenological purposes it often suffices to study {\it simplified models} which extend the SM's particle content only by the DM candidate and the DM--SM mediator~\cite{Abdallah:2014hon,Buchmueller:2013dya}. Compared to the effective field theory (EFT) approach which has been widely used in the past to study classes of ultra-violet completions of the SM, the simplified model approach has the advantage that it remains valid when the momentum transfer in the process becomes of order of the mass of the mediator. This is particularly relevant for resonant dijet searches where one searches for the on-shell production of the mediator.

A list of simplified models for dark matter can be found in \Refs~\cite{Abdallah:2014hon,Buchmueller:2013dya,Dent:2015zpa}. Here, we will focus on simplified models with spin 0 or 1 mediators. Models with fermionic mediators do not give rise to resonant dijet signatures and are thus not considered in this work. The ATLAS and CMS collaboration usually report limits in the simplified models framework for vector and axial-vector, i.e. spin 1, mediators only. Moreover, results are often reported for particular values of the coupling of mediators to DM particles, or correspondingly the width of the mediator, only. Reanalyses of the experimental data often allow for a range of values of the coupling to DM particles/the mediator's width, but nonetheless focus mostly on spin 1 mediators, see e.g.~refs.~\cite{Frandsen:2012rk,An:2012va,An:2012ue,Dobrescu:2013coa,Alves:2013tqa,Arcadi:2013qia,Chala:2015ama,Fairbairn:2016iuf}. In this work, we obtain exclusion limits and future discovery projections for a wider class of mediators. In particular, we study the scalar, pseudoscalar, vector, and axial-vector mediator cases. 

The remainder of this work is structure as follows: In \refsec{sec:framework} we introduce the simplified model framework used in this analysis. In \refsec{sec:dijet_LHC} we describe dijet searches at the LHC in more detail. In \refsec{sec:background} we discuss the SM background for dijet searches, and in \refsec{sec:signal} we discuss how we produce signal samples for the simplified models for our reanalysis of the LHC dijet searches. The statistical framework we use to set exclusion limits and sensitivity projections is described in \refsec{sec:statistic}. In \refsec{sec:limits} we present exclusion limits on the simplified models from the current dijet results reported by the LHC collaborations, before showing discovery prospects for dijet searches in future LHC runs in \refsec{sec:projections}. In \refsec{sec:param} we discuss how well the parameters of a simplified model could be reconstructed in case of a future discovery and point out how angular information of the dijet events could both be used to further improve the sensitivity of dijet searches and to distinguish between different dijet models in case of a discovery. We conclude in \refsec{sec:conclusions}.

\section{Theoretical Framework}\label{sec:framework}
In our reanalysis of the latest LHC dijet searches, we focus on a framework where new scalar or vector mediators couple to quarks and invisible particles which could form the DM component of the Universe.~The models comprising this framework are often referred to as simplified models for DM.~A list of simplified models for DM and their full Lagrangians can be found in \Refs~\cite{Abdallah:2014hon,Buchmueller:2013dya,Dent:2015zpa} and appendix~\ref{app:simp}.~Such models are characterised by the mediator mass and coupling to quarks, and by the mass and couplings of the additional invisible particles, which, as already anticipated, could, but not necessarily, play the role of DM.

The resonant dijet searches discussed here are sensitive to pairs of quarks produced via an $s$-channel resonance. Thus, we can restrict ourselves to simplified models containing mediators with spin $s_{\rm med} = \{0,1\}$. Fermionic mediators do not give rise to dijet final states via resonant processes and are thus not relevant for the searches discussed here. Further, dijet searches probe only the quark-quark-mediator vertex of simplified models. While couplings of the mediator to other particles can be crucial, e.g. when studying the DM phenomenology of the model or mono-jet signatures at the LHC, for dijet searches, their only effect is that they alter the decay width of the mediator and the mediator's branching ratios. In this work, we consider two benchmark cases. In the first, we assess the sensitivity of dijet searches on simplified models in the limit where all couplings between the mediator and other particles, e.g. DM, are set to zero. Then, the width of the mediator and its branching ratios are determined solely by the couplings between the SM and the mediator and we can set limits on these couplings. In the second case, we consider the width of the mediator as a free parameter, thus allowing for additional couplings of the mediator, e.g. to DM particles. Then, we can set limits on the dijet production cross section as a function of the mass and width of the mediator.

The relevant part of the Lagrangian of simplified models with $s_{\rm med} = 0$ mediators is
\begin{equation}
\label{eq:spin0}
  {\cal L}_\phi \supset \frac{1}{2}\partial_\mu \phi \partial^\mu \phi - \frac{m_\phi^2}{2}\phi^2 - \frac{\mu_1 m_\phi}{3} \phi^3 - \frac{\mu_2}{4} \phi^4 - h_1 \bar{q} q \phi - i h_2 \bar{q} \gamma_5 q \phi\,.
\end{equation}
Here, $m_\phi$ is the mass of the real scalar mediator $\phi$ and the $\mu_i$ are dimensionless self-couplings. Since the self-couplings of the mediators have no effect on the dijet signatures of the model, we set the $\mu_i = 0$ in the rest of this work. The couplings of $\phi$ to quarks $h_i$ should in general be understood as $(6\times6)$ matrices and the quark fields $q$ as vectors in flavor space, $q = (u,d,c,s,t,b)$. In this work, we will assume flavor-diagonal and universal quark couplings such that the $h_i$ can be understood as numbers instead of matrices. We will study the cases where only one of the (universal) $h_i$ is different from zero at a time: in the following we refer to the $h_1 \neq 0$ case as a {\it scalar} mediator and the $h_2 \neq 0$ case as a {\it pseudoscalar} mediator.

Similarly, the relevant part of the Lagrangian for simplified models with $s_{\rm med} = 1$ mediators is 
\begin{equation}
\label{eq:spin1}
  {\cal L}_G \supset - \frac{1}{4} {\cal G}_{\mu\nu} {\cal G}^{\mu\nu} + \frac{m_G^2}{2} G_\mu G^\mu - h_3 \bar{q} \gamma_\mu q G^\mu - h_4 \bar{q} \gamma_\mu \gamma_5 q G^\mu \,,
\end{equation}
where ${\cal G}^{\mu\nu}$ is the field-strength tensor of the mediator $G^\mu$, $m_G$ its mass, and the $h_i$ are the dimensionless couplings of $G^\mu$ to quarks. As in the $s_{\rm med} = 0$ case the $h_i$ in general should be understood as $(6\times6)$ matrices, but can be understood as numbers for the purposes of this work because we consider universal and flavor-diagonal quark couplings. We will only consider one of the $h_i$ different from zero at a time; the $h_3 \neq 0$ case is referred to as the {\it vector} mediator case, and the $h_4 \neq 0$ case as {\it axial-vector} in the following.

When not talking about a specific model, we will use $m_{\rm med}$ ($\Gamma_{\rm med}$) to refer to the mass (width) of the respective mediator, and $g_q$ to refer to the quark-mediator coupling different from zero. Depending on our assumptions on the width of the mediator, we thus obtain models specified by two or three parameters. In the case where we assume the mediator to couple only to quarks, the model is specified by the parameters $\{m_{\rm med}, g_q\}$, and the width $\Gamma_{\rm med}$ is a function of $m_{\rm med}$ and $g_q$ only. In the case where we treat the width as a free parameter to include the possibility that the mediator decays into particles other than quarks, the model is specified by the parameters $\{m_{\rm med}, g_q, \Gamma_{\rm med}\}$. 

The latter case includes in particular the region of parameter space where the mediator has a sizeable branching into pairs of DM candidates. Note though, that the larger the branching ratio into pairs of DM particles is, the less sensitive dijet searches are. This is because under the narrow width approximation, for fixed values of $m_{\rm med}$ and $g_q$, the dijet production cross section scales as ($\sigma = \sigma_{\rm prod} \times {\rm BR}_{qq} \propto \Gamma_{qq}^2 / \Gamma_{\rm med}$), where $\sigma_{\rm prod}$ is the production cross section of the mediator, $\Gamma_{qq}$ its partial decay width into quarks, and ($\Gamma_{\rm med} = \Gamma_{qq} + \Gamma_{\rm DM} + \ldots$) with $\Gamma_{\rm DM}$ the partial width into pairs of DM particles and the ``$\ldots$'' indicating the partial width corresponding to any additional decay channels of the mediator. The broadening of the resonant dijet spectrum with increasing decay width further depreciates the sensitivity of dijet searches towards such models: the wider the resonance is, the more challenging it is to differentiate the dijet spectrum from resonant production of the mediator from the smoothly falling SM background discussed in the following sections.

\section{Dijet Searches at the LHC}\label{sec:dijet_LHC}
Dijet searches are performed by both the ATLAS and the CMS collaborations. As of yet, the collaborations have not reported observations of statistically significant excesses over the SM expectation and instead set upper limits on resonant dijet processes. Typically, such results are presented in a model-independent fashion as upper limits on the dijet production cross section as a function of mediator mass. In addition, the experimental collaborations often reported results in the simplified model framework, setting constraints on the mediator-quark-quark coupling $g_q$ as a function of mediator mass. However, such results are presented for spin 1 (vector and axial-vector) mediators only, and usually only for particular values of the mediators width. In this work, we present exclusion limits from current LHC data and projections for future LHC runs for spin 1 as well as for spin 0 (scalar and pseudoscalar) mediators and for a range of mediator widths. 

CMS defines a low mass region from 0.6 TeV to 1.6 TeV and a separate region for resonances above 1.6\,TeV~\cite{CMS-PAS-EXO-16-056,Sirunyan:2018xlo} for dijet searches. We will focus on the high-mass region in this study and use the latest published results corresponding to $36\invfb$ of data at $\sqrt{s} = 13\tev$~\cite{CMS-PAS-EXO-16-056,Sirunyan:2018xlo} as well as preliminary results for $78\invfb$ of data~\cite{CMS-PAS-EXO-17-026}. For our analysis, we simulate signal events as well as SM background samples as described further below. We use cuts similar to those employed in the CMS searches, cf. \Refs~\cite{CMS-PAS-EXO-16-056,Sirunyan:2018xlo}: We use anti-$k_T$ jets~\cite{Cacciari:2008gp,Cacciari:2005hq} with a distance parameter of $\Delta R = 0.4$. Only jets with transverse momentum $p_T > 30\,$GeV and pseudo-rapidity $|\eta| < 2.5$ are considered in the analysis. Further, the two jets with largest $p_T$ in the event (the {\it leading jets}) are used as seeds for the construction of so-called {\it widejets} by merging all jets within $\Delta R = \sqrt{(\Delta \eta)^2 + (\Delta \phi)^2} < 1.1$ of the leading jets with the seed jets. In order to suppress $t$-channel background events, the resulting widejets must be separated in pseudorapidity by less than $|\Delta\eta_{jj}| < 1.3$. The CMS analyses further make a number of quality cuts on the reconstructed jets. In our analysis, we only require that less than $90\,\%$ of each of the two leading jets' respective energies are deposited in the electromagnetic calorimeter. 

Note that the ATLAS collaboration uses somewhat different selection cuts in their analysis, e.g. they require $p_T > 400\,$GeV for the leading jet~\cite{Aaboud:2017yvp}. However, we expect the differences in the selection cuts between the ATLAS and CMS analyses to only have minor impact on the reported limits and projected sensitivity. Thus, results found in this work are expected to apply to both the ATLAS and CMS collaborations' results despite our use of CMS-like cuts.

\subsection{Background}\label{sec:background}
The dominant background source for dijet searches at the LHC are ($pp \to qq/qg/gg + X$) SM processes, where $p$ is a proton, $q$ a quark, $g$ a gluon, and $X$ indicates additional SM particles. In their analysis, both ATLAS and CMS use data-driven methods to estimate the SM background. The collaborations fit the measured background spectra by~\cite{Aaboud:2017yvp,CMS-PAS-EXO-16-056,Sirunyan:2018xlo,CMS-PAS-EXO-17-026}
\begin{equation}\label{eq:bg_fit}
   \frac{\dd\sigma}{\dd m_{jj}} = p_0 \frac{(1-x)^{p_1}}{x^{p_2+p_3\log(x)}}\;,
\end{equation}
where $x \equiv m_{jj}/\sqrt{s}$ with $m_{jj}$ the invariant mass of the (wide)jets and $\sqrt{s}$ the proton-proton center of mass energy, and the $p_i$ are the fit parameters. We use the same fit-function as given in \refeq{eq:bg_fit} which we fit to data from the CMS collaboration to obtain the $p_i$. We show the CMS dijet data for $36\invfb$~\cite{CMS-PAS-EXO-16-056,Sirunyan:2018xlo} together with the fitted background parameterization in the left panel of \reffig{fig:sig_bg}. A similar plot for $78\invfb$ can be found in \Ref~\cite{CMS-PAS-EXO-17-026}.~Note that this parameterization of the background allows us to predict the background for larger luminosities without the need to carry out dedicated collider simulations of the SM dijet background.

In order to perform an independent test of the background parameterization and in particular to obtain double-differential distributions $\dd^2 \sigma/\dd m_{jj} \dd\Delta\eta_{jj}$ we will make use of in \refsec{sec:projections}, we have performed our own SM background simulations using \texttt{MadGraph5}~\cite{Alwall:2014hca} for the simulation of the ($pp \to qq/qg/gg + X$) events, \texttt{pythia8}~\cite{Sjostrand:2006za,Sjostrand:2014zea} for showering, and \texttt{Delphes3}~\cite{deFavereau:2013fsa} for fast detector simulation using the cuts described above.

\subsection{Signal}\label{sec:signal}
\begin{figure}
   \includegraphics[width=0.47\linewidth]{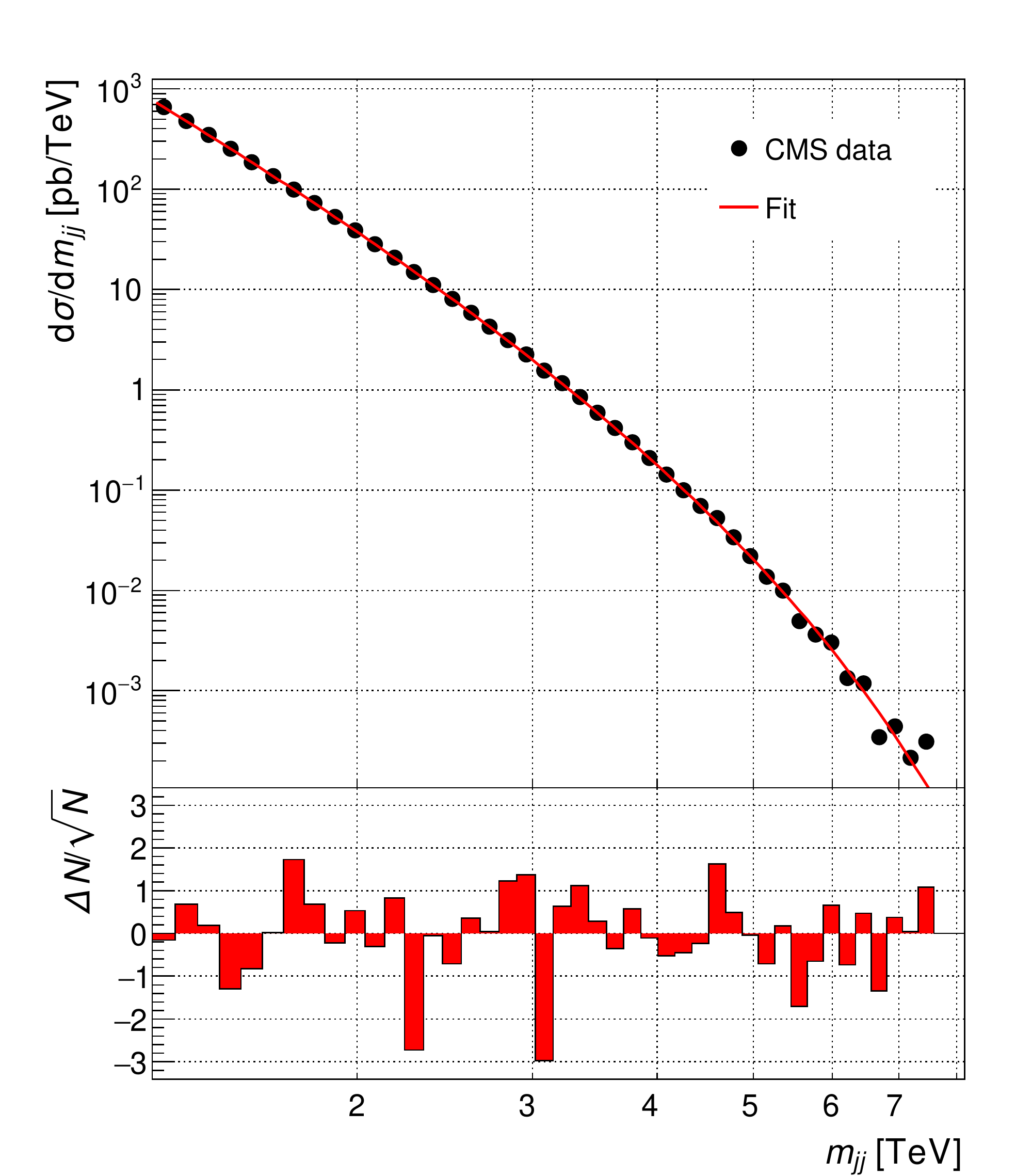}
   \includegraphics[width=0.47\linewidth]{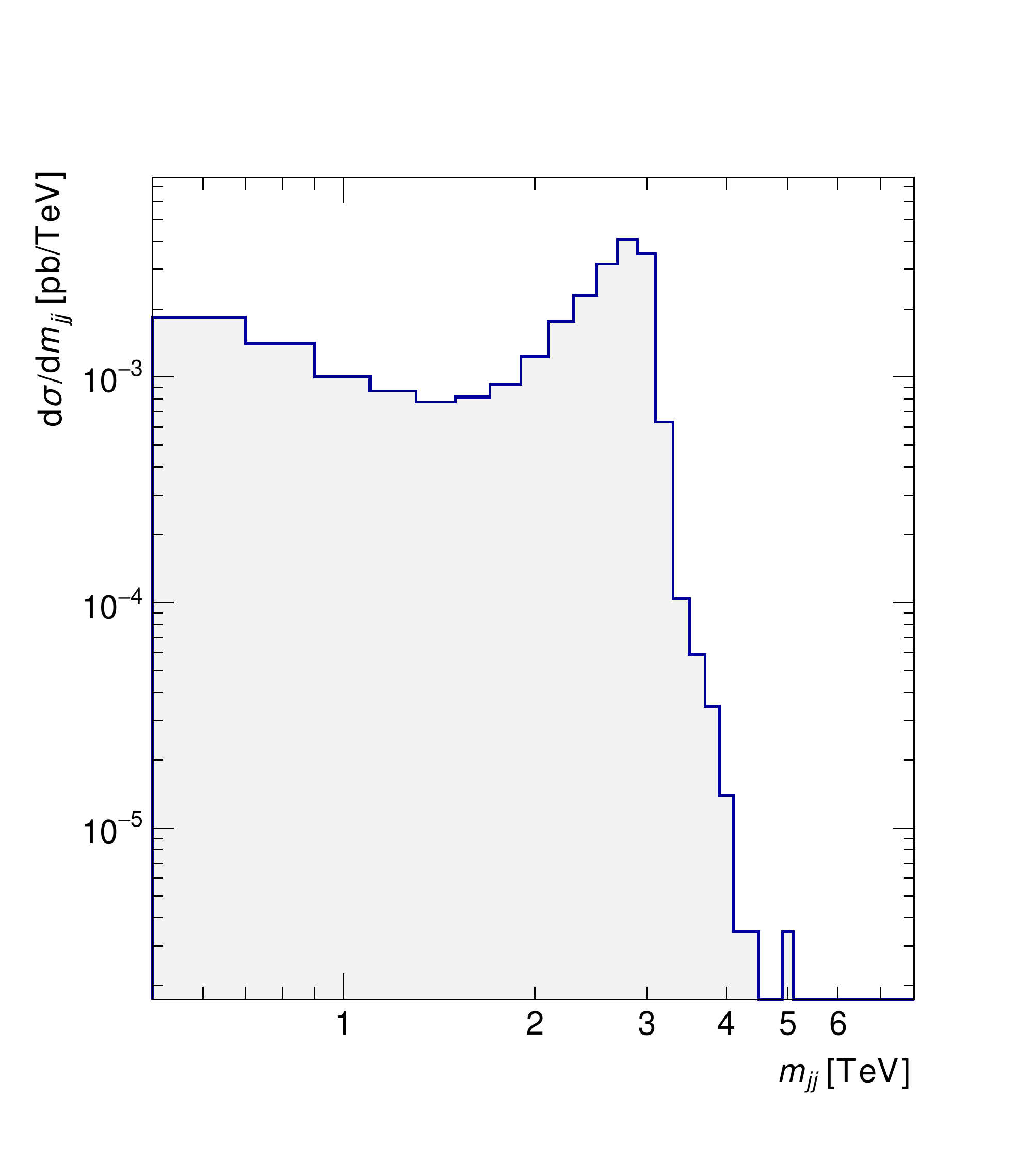}
   \caption{{\it Left:} Invariant dijet mass spectrum measured by CMS for the $36\invfb$~\cite{CMS-PAS-EXO-16-056,Sirunyan:2018xlo} dataset obtained from the \texttt{hepdata} server~\cite{1676214}. In the upper panel, the black dots show the CMS data and the red line shows the background parameterization in \refeq{eq:bg_fit} after fitting. The lower panel shows the residuals of the data with respect to the fitted background parameterization divided by the statistical error of the data.
   {\it Right:} Dijet mass spectrum of the simulated signal for a vector mediator with $\{m_{\rm med}=3.0\tev,g_q=0.1\}$ under the assumption that $\Gamma_{\rm med}$ is determined by decays into quarks only. The bin width is set to $25\gev$. Note that the scale of the axes differs between the panels.}
   \label{fig:sig_bg}
\end{figure}

In order to obtain limits from current dijet searches and to make projections for future searches we need both the SM background spectra and the signal spectra in the respective simplified model cases we consider. For our main analysis, we first simulate all possible signal processes 
\begin{equation}\label{eq:sig_proccess}
   pp \to {\rm Mediator} \to \bar{q}q + X\;,
\end{equation}
for each of the simplified model cases considered here \{(pseudo)scalar, (axial-)vector\} over a grid of model parameters $m_{\rm med}$, $g_{q}$ (and $\Gamma_{\rm med}$ when treating the width as an independent parameter) with \texttt{WHIZARD}~\cite{Kilian:2007gr,Moretti:2001zz}.\footnote{We use implementations of the simplified models for \texttt{WHIZARD} developed in \Ref~\cite{Baum:2017kfa}.} In \refeq{eq:sig_proccess}, $X$ stands for additional SM particles. For example, when simulating $pp \to {\rm Mediator} \to \bar{t}t$ events we also simulate the ($t \to W b$) decays and the subsequent decays of the $W$-bosons into pairs of SM particles in \texttt{WHIZARD}. Note, that no $b$-jet veto is employed in the dijet analyses considered here.  Parton distribution functions are implemented via the {\it CT14lo} set as obtained from \texttt{LHAPDF6}~\cite{Buckley:2014ana}. For showering and hadronization we use \texttt{pythia8}. For the (fast) detector simulation we use \texttt{Delphes3} with \texttt{FASTJET}~\cite{Cacciari:2011ma} for jet reconstruction and the default CMS configuration file, except that we modified the jet-reconstruction parameters to match those employed in the CMS analyses. We use our own \texttt{C++} code together with several \texttt{ROOT}~\cite{Brun:1997pa} libraries to analyze the signal, including the application of selection cuts, widejet reconstruction, and signal fitting.
 
We have validated our tool-chain for event generation and analysis with an independent simulation chain using \texttt{MadGraph5} for the generation of the hard event, \texttt{pythia8} for showering and \texttt{Delphes3} for fast detector simulation.

Note that there are a number of sources of uncertainty in our simulations, e.g. parton distribution functions (pdfs) or the jet-energy scale used in the reconstruction. The detailed assessment of the effects of such uncertainties is beyond the scope of this work, but we have checked that pdf uncertainties would not have large effects on the exclusion limits/sensitivity projections presented in this work.

We show an example of a signal spectrum for the vector mediator model for $\{ m_{\rm med} = 3.0\tev, g_q = 0.1\}$ in the right panel of \reffig{fig:sig_bg}. Here, we assume that the mediator couples only to quarks, thus, its width is determined by $m_{\rm med}$ and $g_q$. The tail of the signal at low invariant masses of the widejet system $m_{jj} \ll m_{\rm med}$ arises from showering and hadronization. 

Input parameters in the calculation of signal spectra are $m_{\rm med}$, $g_q$ and $\Gamma_{\rm med}$, when treating the mediator width as an independent parameter.~At the same time, we find that signal spectra can conveniently be fit by
\begin{equation}\label{eq:signal_fit}
   \frac{\dd\sigma}{\dd m_{jj}} = p_0 V(m_{jj}-p_2,p_3,p_1)\ee^{p_4\frac{m_{jj}}{p_2}} \left( \frac{m_{jj}}{p_2} \right)^{p_5}\,,
\end{equation}
where the $p_i$ are fit parameters. The parameter $p_0$ has dimensions of mass$^{-3}$, $\{p_1, p_2, p_3\}$ have dimension of mass, and $\{p_4,p_5\}$ are dimensionless.
\begin{equation} \label{eq:logLsig}
   V(x,\sigma,\gamma) = \int_{-\infty}^{\infty}\frac{1}{\sqrt{2\pi}\sigma} \ee^{-\frac{x'^2}{2\sigma^2}} \frac{\gamma}{[(x-x')^2+\gamma^2]}\dd x'\,,
\end{equation}
is a Voigt profile, the convolution of a Lorentzian profile with a Gaussian. Here, the Lorentzian profile accounts for the shape of the resonance, while the Gaussian profile accounts for the broadening due to the finite detector resolution. Out of the parameters, $p_1$ is related to the mediator's width, $p_2$ to its mass, and $p_3$ reflects the invariant mass resolution of the detector. $p_0$ controls the overall normalization of the differential cross section. The parameters $p_4$ and $p_5$ control the exponential and power-law tails of the differential cross section. They are purely phenomenologically motivated and account for the low energy tail of the spectrum.

We use the profile likelihood to fit this parameterization to our simulated spectra. The goodness-of-fit is measured via the ratio of the likelihood of the best fit point to the likelihood of fitting the simulated spectra to themselves,
\begin{equation}
   \lambda \equiv \frac{\mathscr{L}}{\mathscr{L}_S} = \left\{ \prod_{i=1}^N \frac{M_i^{S_i}(p)}{S_i!}\ee^{-M_i(p)} \right\} / \left\{ \prod_{i=1}^N \frac{S_i^{S_i}}{S_i!}\ee^{-S_i} \right\}\;,
\end{equation}
where $M_i$ and $S_i$ are respectively the values in our signal fit (with parameters $p$) and our simulated signal spectra of $d\sigma/dm_{jj}$ averaged over the $i$-th bin in the invariant mass of the dijet system $m_{jj}$. 

In \reffig{fig:sig_tab}, located in the appendix, we show simulated signal spectra with the respective fits for several parameter points. The parameterization of the signal spectra in \refeq{eq:signal_fit} suffices to fit the signal well for the wide range of parameters $\{m_{\rm med}, g_q\}$ shown in \reffig{fig:sig_tab}.

In \app.~\ref{app:signal_fit} we list the fit parameters for several models and parameter points, which can be used for further analyses without the need for a full simulation. In the following analyses, we use the simulated signal spectra, scaled to the appropriate luminosity, and not the parametric fits to these signal spectra.

\section{Statistical Method}\label{sec:statistic}
We compute exclusion limits and sensitivity projections using the profile likelihood method. Our analysis is similar to the one carried out by the CMS collaboration, and follows the procedure outlined in \Ref~\cite{Cowan:2010js}.~Computing exclusion limits, we test the background plus signal hypothesis, $H_1$, against the background only hypothesis, $H_0$, and compute the significance with which a point in parameter space can be excluded.~When projecting the sensitivity of dijet searches in future LHC runs, we test the null hypothesis $H_0$ against the alternative $H_1$ and compute the significance with which a point in parameter space can be observed.

The significance,  
\begin{equation}\label{eq:significance}
   Z = \Phi^{-1}(1-p)
\end{equation}
is related to the $p$-value via the quantile $\Phi^{-1}$ of a Gaussian distribution with mean 0 and variance 1. For example, the usual 95\,\% confidence level (C.L.) with $p$-value 0.05 corresponds to $Z=1.64$. In general, the significance is computed from the likelihood function $\mathscr{L}$ for finding a given signal $\boldsymbol{s}=\{s_1,\dots,s_N\}$ over a background $\boldsymbol{b}=\{b_1,\dots,b_N\}$ for a dataset $\vec{n}=(n_1,\dots,n_N)$, where $N$ is the number of considered bins in the dijet invariant mass,
\begin{equation}
  \mathscr{L}\left(\boldsymbol{s},\boldsymbol{\theta}\right) = \prod_{i=1}^N \frac{(s_i + 
b_i(\vec{\theta}))^{n_i}}{n_i!}\ee^{-[s_i + b_i(\vec{\theta})]}\,,
\end{equation}
and $\vec{\theta}$ is an array of nuisance parameters, which in our case correspond to the background fit parameters from \refeq{eq:bg_fit}. In order to calculate the significance from the likelihood, we construct a profile likelihood ratio, $\lambda$.~The exact form of $\lambda$ depends on whether we calculate the significance for exclusion limits or sensitivity projections. For exclusion limits, we use the profile likelihood ratio
\begin{equation} \label{eq:profL}
   \lambda  =  \frac{\mathscr{L}(\boldsymbol{s},\widehat{\boldsymbol{\theta}})}{\mathscr{L}(\boldsymbol{0},\widehat{\widehat{\boldsymbol{\theta}}})} \,,
\end{equation}
where $\widehat{\boldsymbol{\theta}}$ and $\widehat{\widehat{\boldsymbol{\theta}}}$ are the nuisance parameters that maximize the likelihood function $\mathscr{L}$ for the signals $\boldsymbol{s}$ and $\boldsymbol{0}$, respectively. Similarly, for sensitivity projections we use the profile likelihood ratio 
\begin{equation} \label{eq:profL2}
\lambda  =  \frac{\mathscr{L}(\boldsymbol{0},\widehat{\widehat{\boldsymbol{\theta}}})}{\mathscr{L}(\boldsymbol{s},\widehat{\boldsymbol{\theta}})} \,.
\end{equation}
In both cases, from the profile likelihood ratio we construct the test statistic $q = -2 \ln \lambda$, from which we obtain~\cite{Cowan:2010js}
\begin{equation}
Z\simeq\sqrt{q} \,.
\end{equation}
In our analysis, we are interested in {\it observed} and {\it expected} exclusion limits.~In the former case, we compute the significance using $n_i=n_i^{\rm CMS}$, where $n_i^{\rm CMS}$ is the number of observed dijet events at CMS in the $i$-th dijet mass bin.\footnote{When setting limits based on the CMS result from $36\invfb$ of data, $n_i^{\rm CMS}$ is the measured dijet spectrum obtained from~\cite{1676214}. For limits based on $78\invfb$, $n_i^{\rm CMS}$ is taken from~\cite{CMS-PAS-EXO-17-026}.}~.In the latter case, we assume $n_i=b_i(\boldsymbol{\theta}_{\rm bf})$, where $\boldsymbol{\theta}_{\rm bf}$ is the value of $\boldsymbol{\theta}$ that maximises $\mathscr{L}\left(\boldsymbol{0},\boldsymbol{\theta}\right)$ for $n_i=n_i^{\rm CMS}$.~For the expected limit, we are also interested in the expected deviation of the limit. For e.g. the 95\,\%\;C.L. limit, which is the $Z = 1.64$ contour, the $\pm 1\,\sigma$ deviation is then given by the $Z = 1.64 \pm 1$ contours.~As already anticipated, we are also interested in computing the significance for sensitivity projections.~In this case, we use the dataset $n_i=b_i(\boldsymbol{\theta}_{\rm bf})+s_i$.~In all cases discussed above, when maximizing the likelihood with respect to $\boldsymbol{\theta}$ to obtain $\widehat{\boldsymbol{\theta}}$ or $\widehat{\widehat{\boldsymbol{\theta}}}$ at a given point in parameter space, we exclude a window around the mediator mass in the dijet invariant mass spectrum\footnote{The masked region is defined by excluding the region  bounded by the bins with largest and smallest invariant mass in which the profile likelihood ratio $\lambda$, defined in \refeq{eq:profL}, takes values $\lambda > 1$.}.

\section{Exclusion Limits}\label{sec:limits}

In this section we present exclusion limits for the (pseudo)scalar and (axial-)vector mediator cases from the $36\,\invfb$~\cite{CMS-PAS-EXO-16-056,Sirunyan:2018xlo} and $78\invfb$~\cite{CMS-PAS-EXO-17-026} CMS data. We obtain 95\,\%\,C.L. exclusion limits by simulating signals for each parameter point for a grid in $\{m_{\rm med}, g_q\}$ ($\{m_{\rm med}, g_q, \Gamma_{\rm med}\}$) space as described in \refsec{sec:signal} and computing the profile likelihood for the null hypothesis according to the procedure outlined in \refsec{sec:statistic}.

\begin{figure}
   \begin{subfigure}{\linewidth}
      \includegraphics[width=0.49\linewidth]{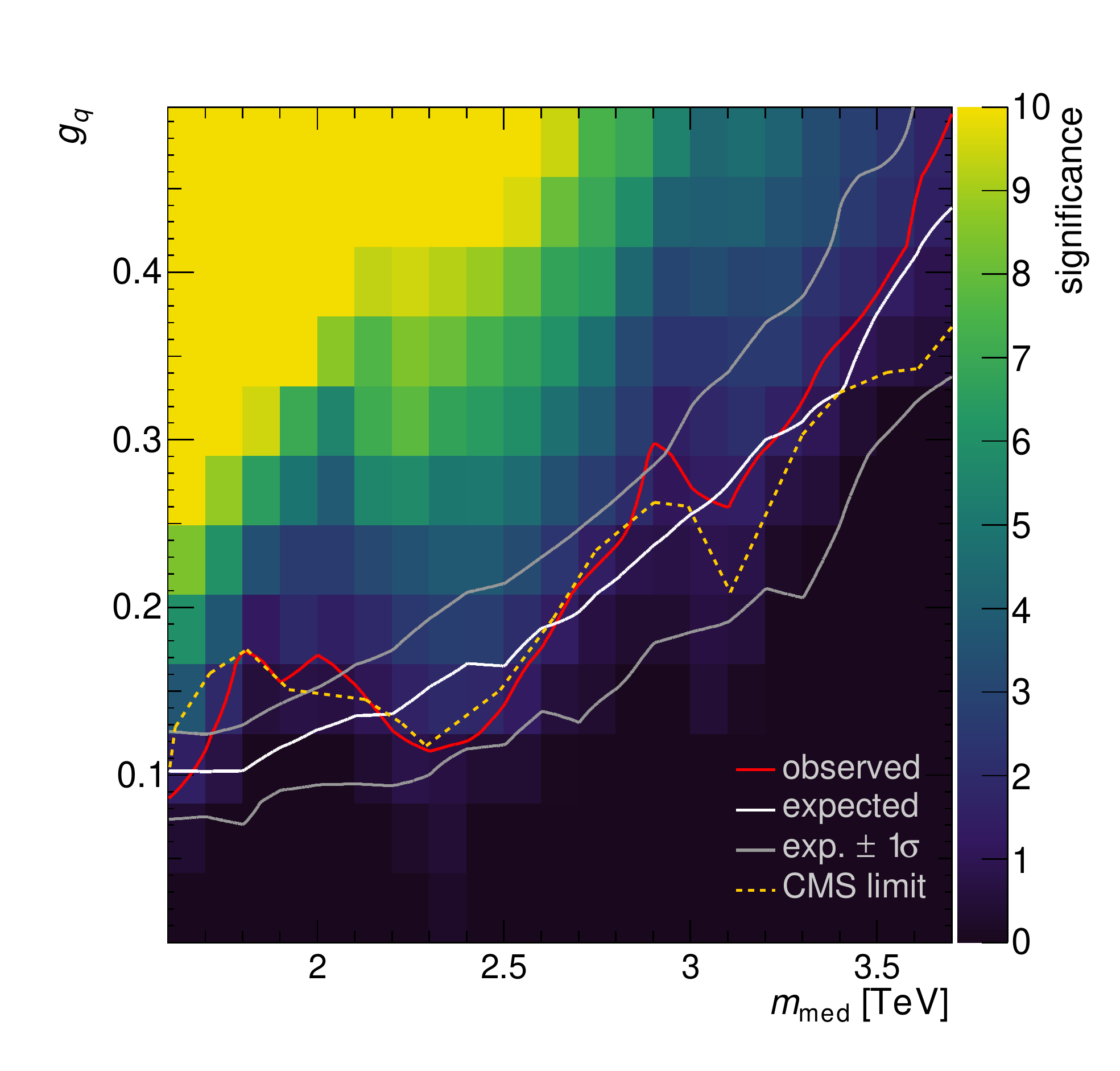}
      \includegraphics[width=0.49\linewidth]{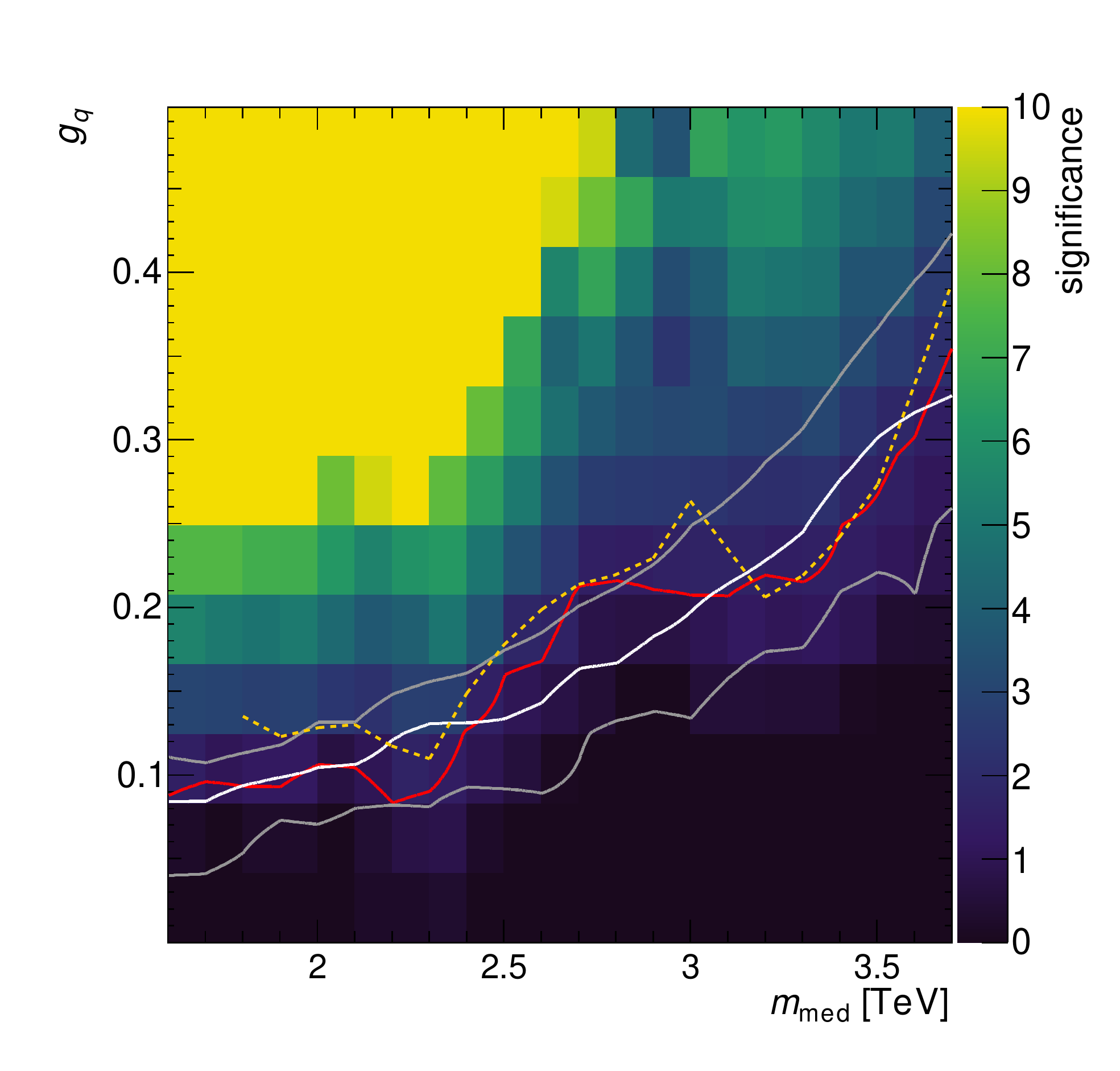}
      \caption{Vector Mediator ($h_3 \neq 0$)}
   \end{subfigure}

   \begin{subfigure}{\linewidth}
      \includegraphics[width=0.49\linewidth]{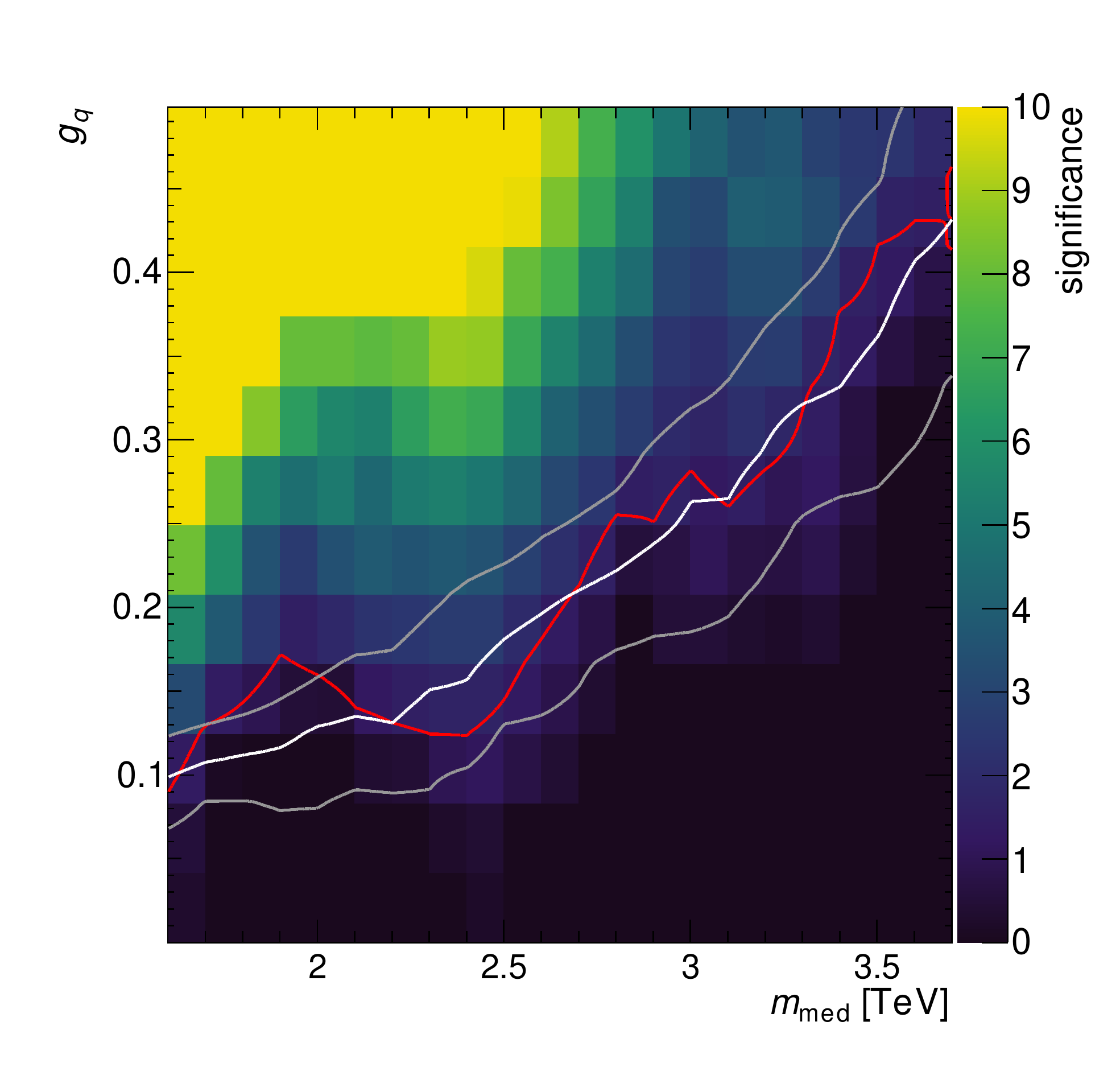}
      \includegraphics[width=0.49\linewidth]{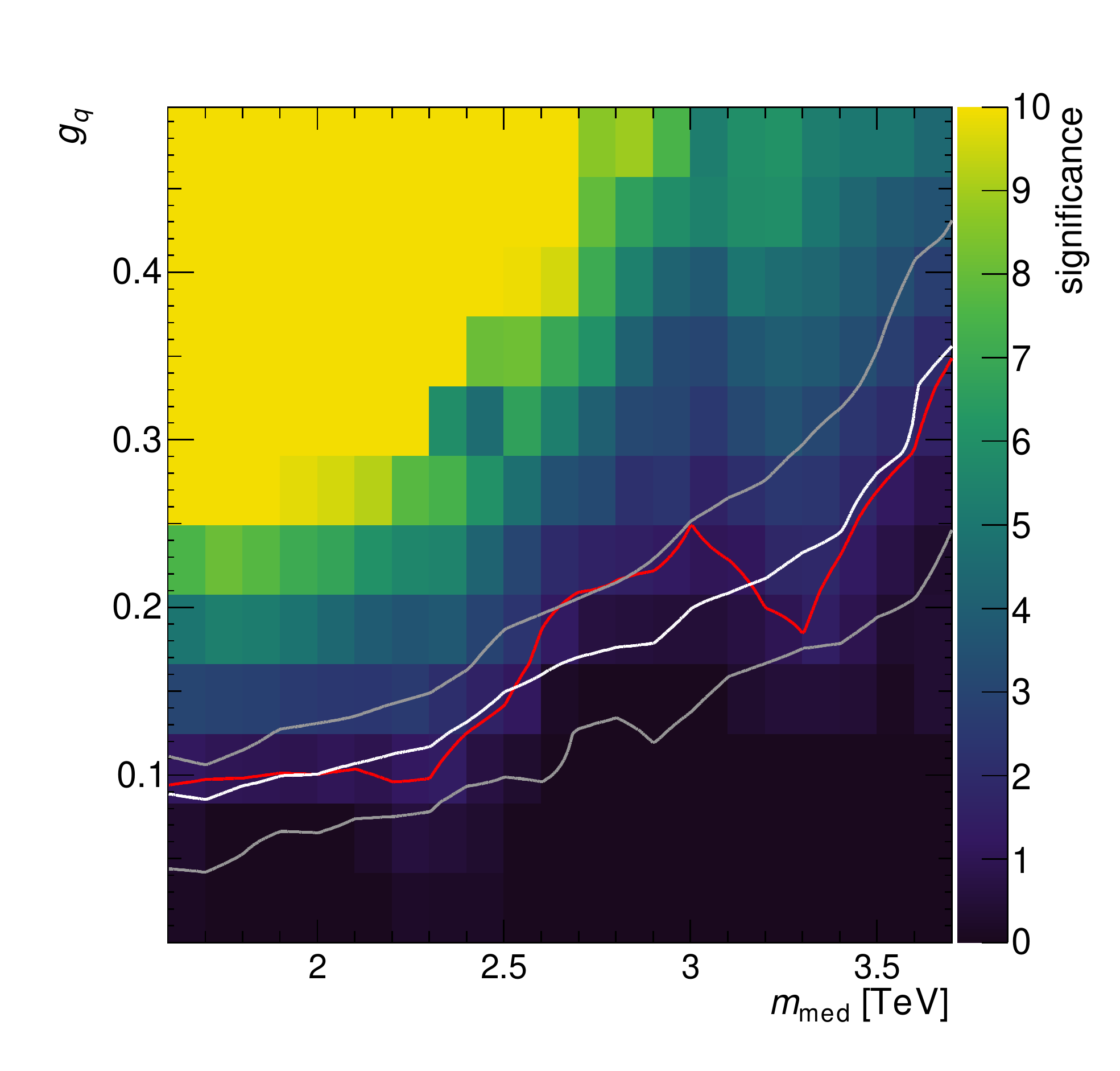}
      \caption{Axial-Vector Mediator ($h_4 \neq 0$)}
   \end{subfigure}
   \caption{Exclusion limits for the vector mediator (upper panels) and axial-vector mediator (lower panels) under the assumptions that the mediator couples to quarks only. The left panels are for the $36\,\invfb$ CMS data~\cite{CMS-PAS-EXO-16-056,Sirunyan:2018xlo} while the right panels are for the $78\invfb$ data~\cite{CMS-PAS-EXO-17-026}. The color scales show the significance [defined in \refeq{eq:significance}] for which a parameter point in the $m_{\rm med} - g_q$ plane could be excluded. The white line shows the expected 95\,\%\,C.L. exclusion limit with the gray lines indicating the $1\,\sigma$ error bands around the expected limit. The red line shows the 95\,\%\,C.L. exclusion limit obtained from the observed data. For comparison, in the upper panels we show the 95\,\%\,C.L. exclusion limit reported by the CMS collaboration for the vector mediator case with the dashed yellow line.}
   \label{fig:EL_vector_gqonly}
\end{figure}

\begin{figure}
   \begin{subfigure}{\linewidth}
      \includegraphics[width=0.49\linewidth]{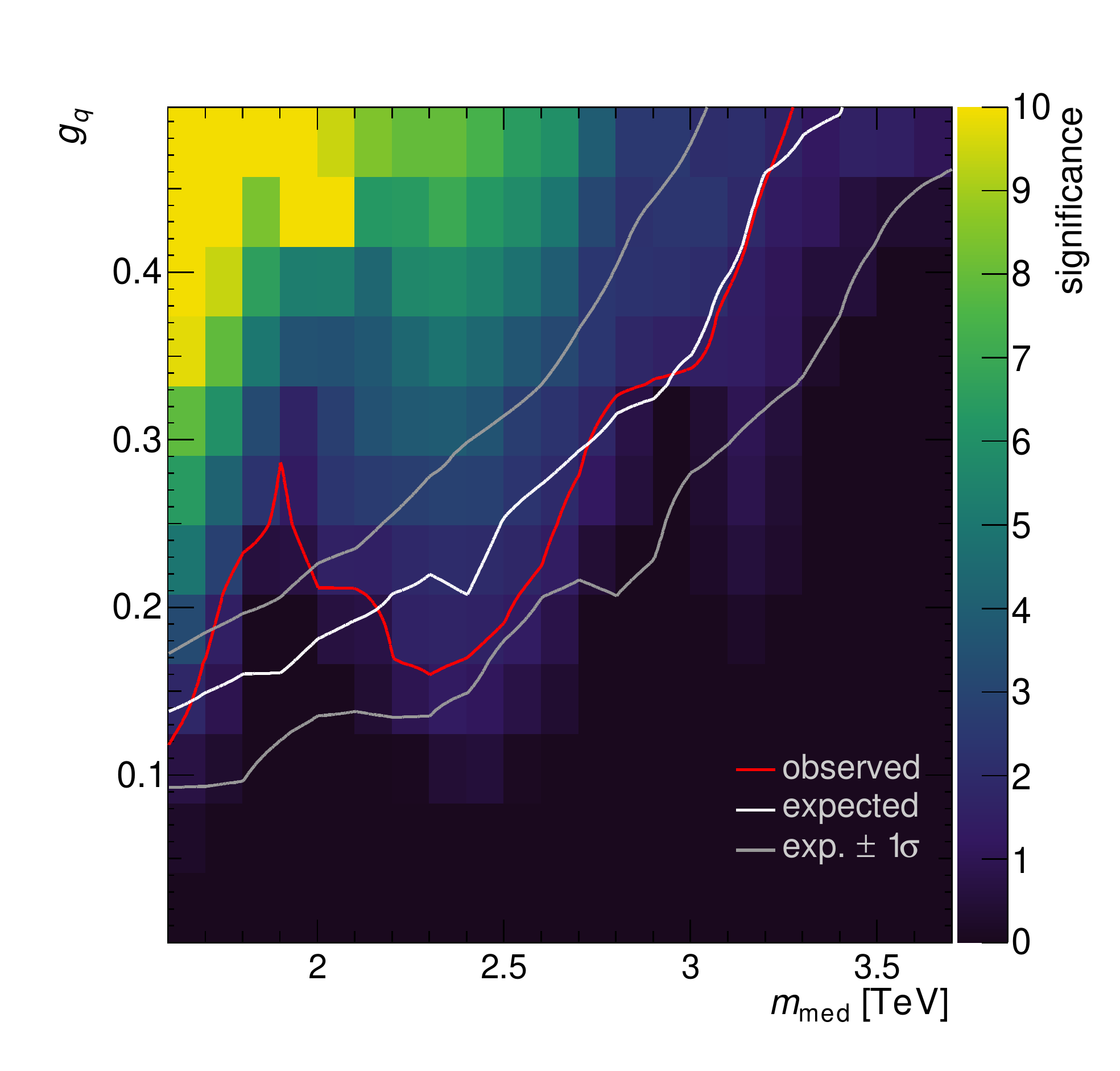}
      \includegraphics[width=0.49\linewidth]{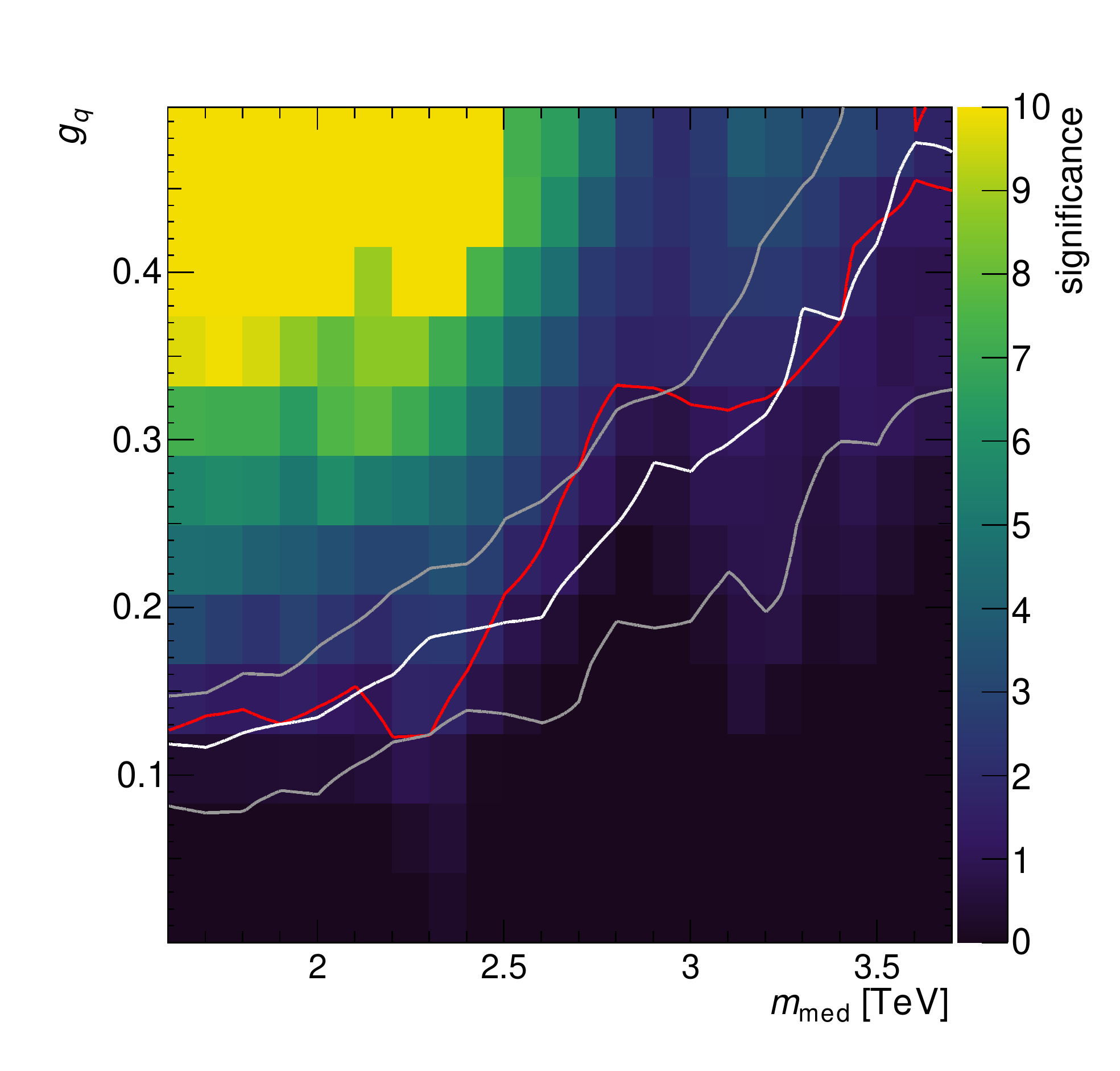}
      \caption{Scalar Mediator ($h_1 \neq 0$)}
   \end{subfigure}

   \begin{subfigure}{\linewidth}
      \includegraphics[width=0.49\linewidth]{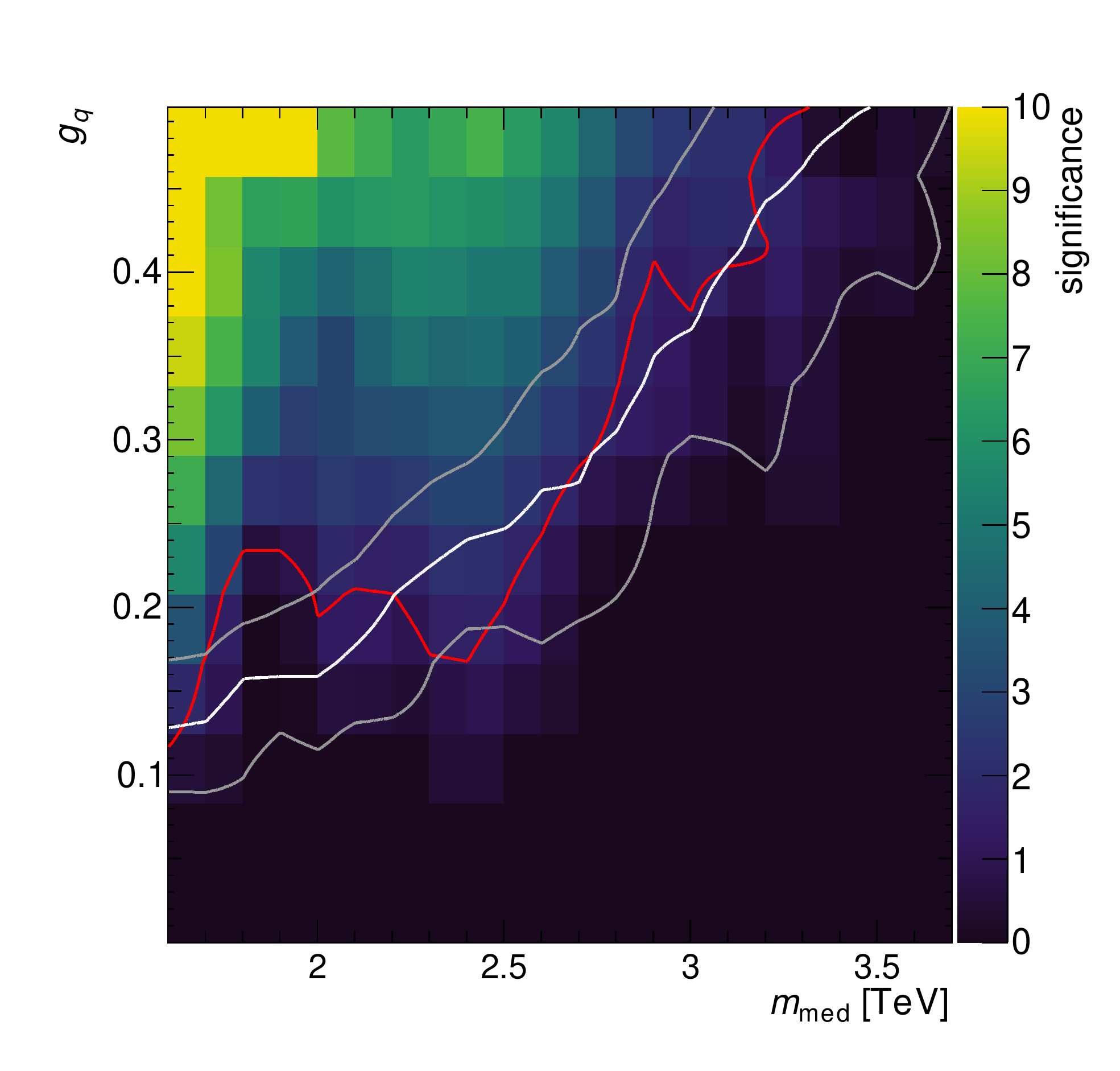}
      \includegraphics[width=0.49\linewidth]{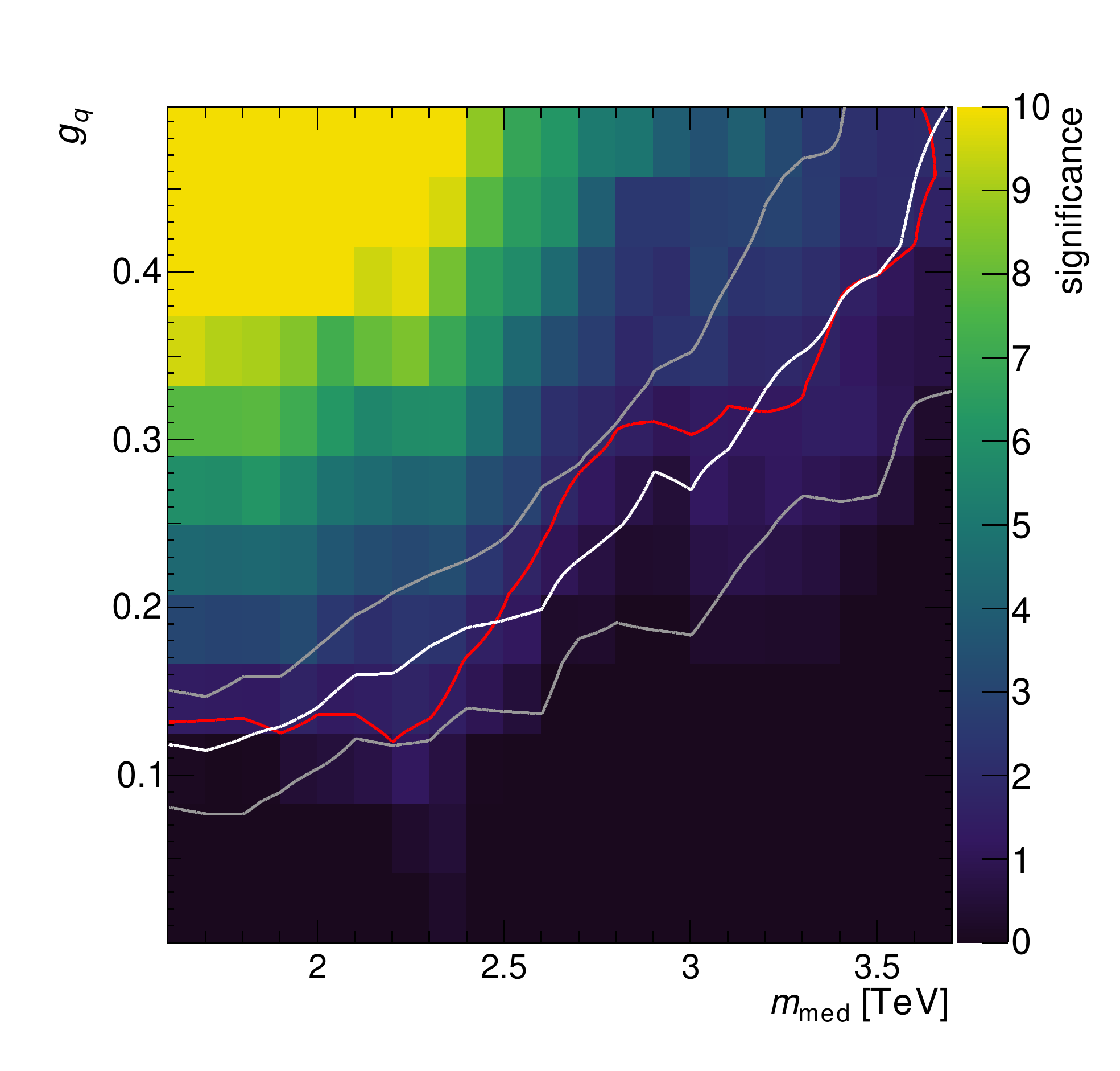}
      \caption{Pseudoscalar Mediator ($h_2 \neq 0$)}
   \end{subfigure}
   \caption{Same as \reffig{fig:EL_vector_gqonly} but for the scalar mediator (upper panels) and pseudoscalar mediator (lower panels) cases.}
   \label{fig:EL_scalar_gqonly}
\end{figure}

We show our results for the vector and axial-vector mediator cases in \reffig{fig:EL_vector_gqonly} and for the scalar and pseudoscalar cases in \reffig{fig:EL_scalar_gqonly} under the assumption that the mediator couples to quarks only, thus its width $\Gamma_{\rm med}$ is a function of $m_{\rm med}$ and $g_q$ only. In these figures, we show both the expected and observed limits as well as the significance [as defined in \refeq{eq:significance}] for which a parameter point in the $m_{\rm med} - g_q$ plane is excluded by the CMS data. For comparison, we also show the limit obtained by the CMS collaboration for the vector mediator case.

We begin by discussing our results for the vector mediator case, shown in the upper panels of \reffig{fig:EL_vector_gqonly}. For both the $36\invfb$ and the $78\invfb$ data, the limits we obtain are within $\sim 1\,\sigma$ of the respective expected limits, indicating that both CMS datasets are compatible with the background-only hypothesis and show no evidence for a dijet signal from a simplified model with a vector mediator. For the $36\invfb$, the largest deviations from the expected limits are for mediator masses of $\sim 2\tev$ and $\sim 3\tev$ where the observed limits are $\sim 1.5\,\sigma$ and $\sim 1\,\sigma$ weaker, respectively, than the expected limit. Although of small statistical significance, theses excesses can be understood from the correlated residuals in the CMS data with respect to the background fit at the corresponding invariant masses, cf. the left panel of \reffig{fig:sig_bg}. Our limits are in good agreement with those obtained by the CMS collaboration for vector mediator from the same dataset. Note that while both the CMS collaboration and we employ similar statistical techniques, our analysis is independent from those carried out by the CMS collaboration. The only common input are the measured dijet spectra.

Considering the results from the $36\,\invfb$ data, we find the strongest limits for the smallest mediator masses considered, excluding (universal) couplings of the mediator to quarks larger than $g_q \sim 0.1$ for mediator masses $m_{\rm med} \sim 1.6\tev$ at 95\,\%\,C.L. With increasing mediator mass, the limits weaken. This is because the background spectrum (signal cross section) is falling steeply with $m_{jj}$ ($m_{\rm med}$). The limits are driven mostly by the data in the bins of the invariant mass spectrum comparable to $m_{\rm med}$. Thus, with increasing mediator mass, the relative statistical uncertainty in the relevant part of the dijet spectrum is increasing, leading to weaker exclusion limits. For mediator masses $m_{\rm med} \sim 3.5\tev$ we exclude couplings to quarks $g_q \gtrsim 0.4$. 

Comparing the results from the $36\invfb$ data to those from the $78\invfb$ data, we find that the larger luminosity allows for more stringent exclusion limits, in particular for larger mediator masses. Recall that the exclusion limits are driven by the portion of the observed data where the invariant mass of the widejet system, $m_{jj}$, approximately corresponds to the mass of the mediator. The SM background spectrum is monotonically falling with $m_{jj}$. For the smallest mediator masses considered here of $m_{\rm med} \sim 1.6\tev$, the relevant portion of the dijet spectrum is already dominated by systematic errors, thus, without changing the analysis techniques no improvements in the limits are expected by increasing the luminosity. For heavier mediators on the other hand, the obtained exclusion limits are limited by the statistical error of the relevant portion of observed dijet spectra, thus, we find sizable differences in the observed limits between the $36\invfb$ and the $78\invfb$ datasets. For mediator masses of $m_{\rm med} \sim 2\tev$, the 95\,\%\,C.L. exclusion limits strengthen from $g_q \lesssim 0.15$ from the $36\invfb$ dataset to $g_q \lesssim 0.10$ from the $78\invfb$ dataset. For heavier vector mediators $m_{\rm med} \sim 3.5\tev$ we find that the limits strengthen from $g_q \lesssim 0.4$ to $g_q \lesssim 0.25$.

For the axial-vector case, cf. the lower panels of \reffig{fig:EL_vector_gqonly}, we find results very similar to the vector mediator case. This is because the limits are driven by decays of the mediator into the lighter quark flavors $q = \{u,d,c,s,b\}$. In the limit of vanishing quark masses, the dijet cross sections for vector and axial-vector mediator become equal.

Considering (pseudo)scalar mediators, cf. \reffig{fig:EL_scalar_gqonly}, we again find almost identical limits when comparing the scalar and the pseudoscalar cases. As in the case of vector and axial-vector mediators, this is because the dijet production cross sections for scalar and pseudoscalar mediators become equal in the limit of massless quarks. Comparing the limits for (pseudo)scalar mediators to the (axial-)vector case, we find that while the limits on the coupling to quarks are similar for mediator masses $m_{\rm med} \sim 1.6\tev$, the limits for the (pseudo)scalar cases are considerably weaker for heavier mediators. For mediator masses of $m_{\rm med} \sim 3.5\tev$ couplings of the mediator to quarks larger than $g_q \sim 0.55$ are excluded for (pseudo)scalar mediators from the $36\invfb$ dataset, while for (axial-)vector mediators the data constrained $g_q \lesssim 0.4$ for such mediator masses. The $78\invfb$ dataset excludes couplings larger than $g_q \sim 0.4$ for (pseudo)scalar mediators with $m_{\rm med} \sim 3.5\tev$ while for (axial-)vector mediators the same data constrains the couplings to $g_q \lesssim 0.25$.

\begin{figure}
   \centering
   \begin{subfigure}{.49\linewidth}
      \includegraphics[width=\linewidth]{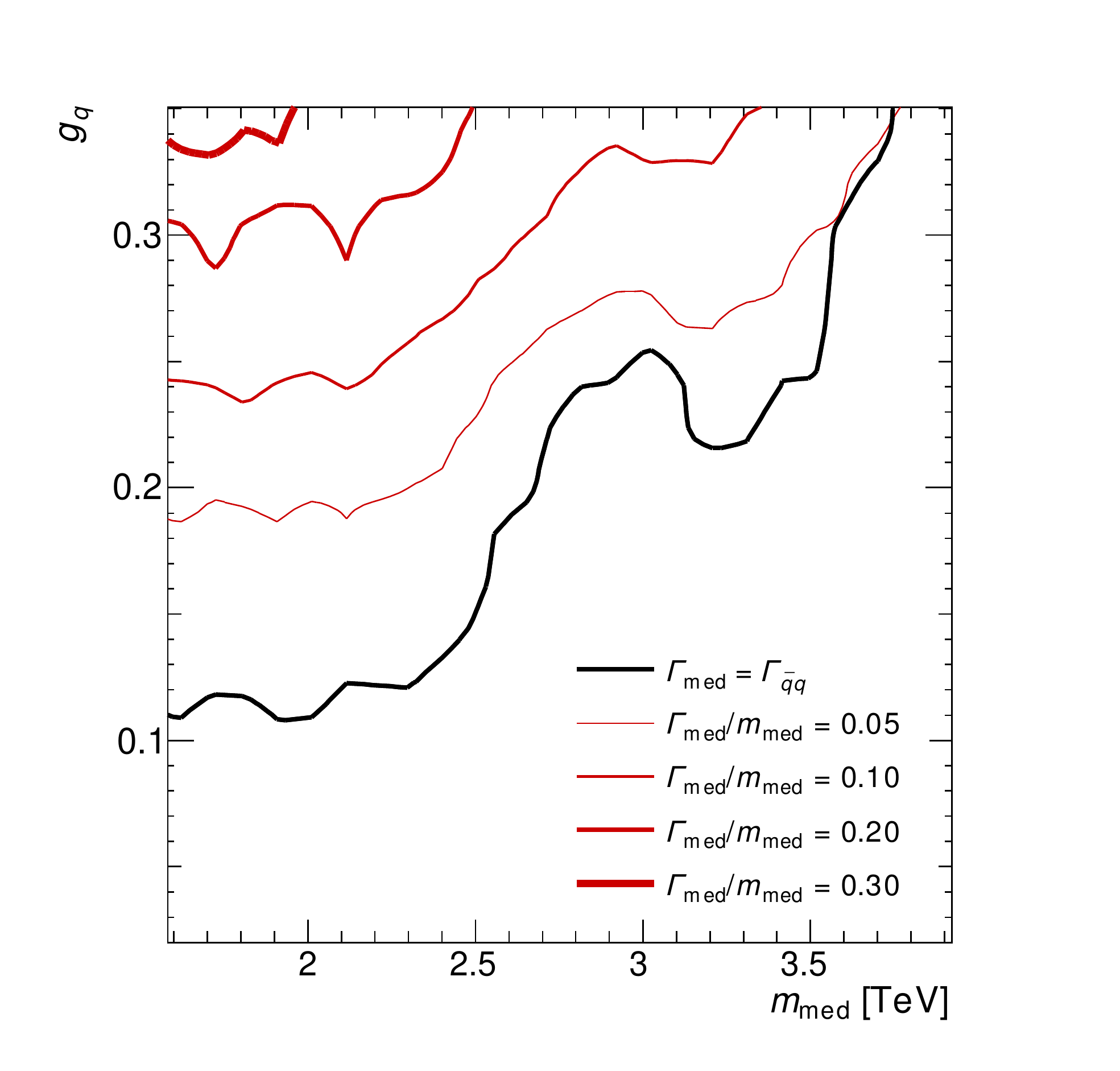}
      \caption{vector mediator}
   \end{subfigure}
   \begin{subfigure}{.49\linewidth}
      \includegraphics[width=\linewidth]{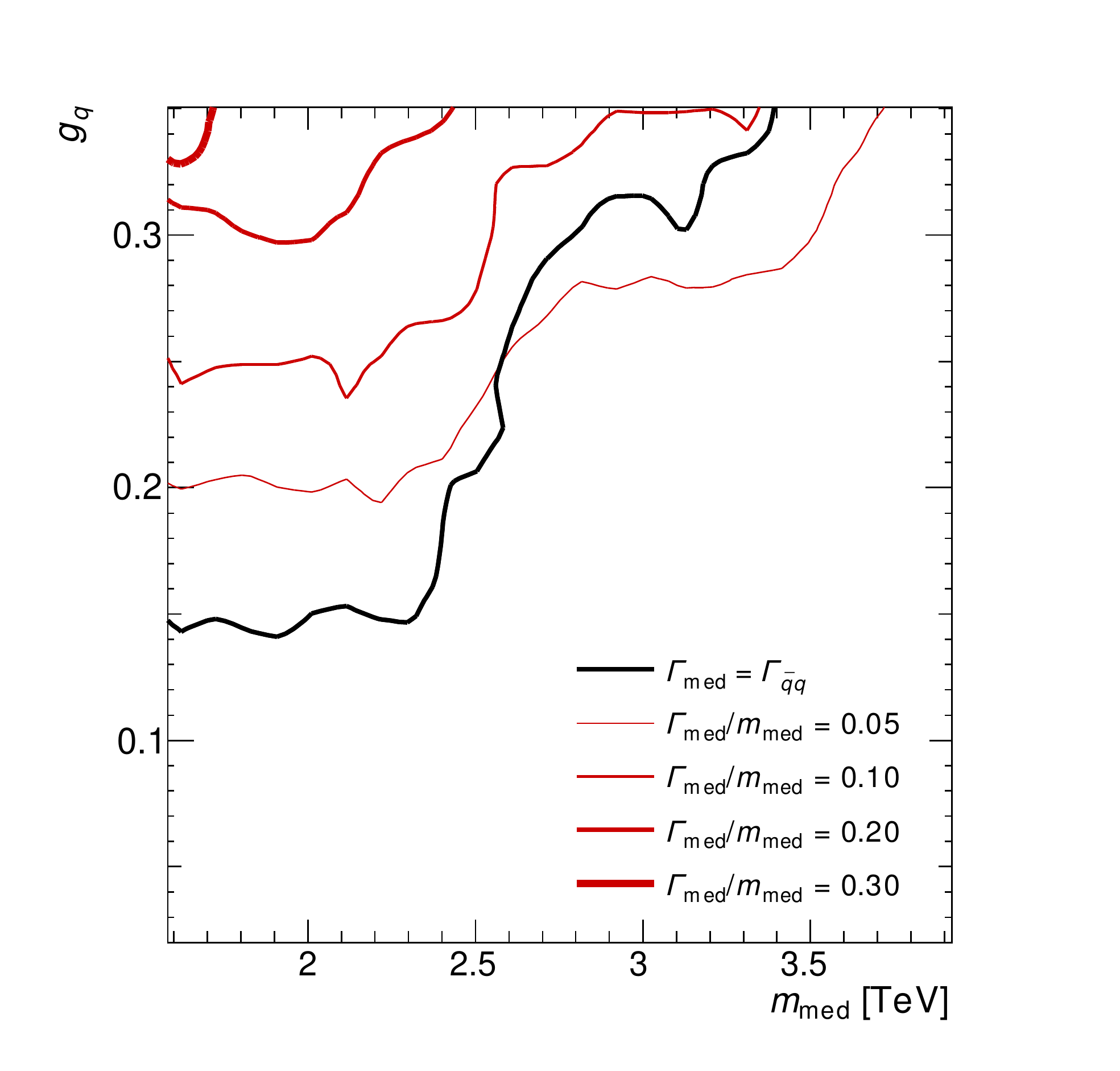}
      \caption{scalar mediator}
   \end{subfigure}
   \caption{95\,\%\,C.L. exclusion limits for the vector mediator (left panel) and scalar mediator (right panel) cases from the $78\invfb$ data. In \reffig{fig:EL_vector_gqonly} and \reffig{fig:EL_scalar_gqonly} we showed exclusion limits under the assumption that the mediator couples to quarks only and thus the mediator's width $\Gamma_{\rm med}$ is a function of its mass $m_{\rm med}$ and the coupling to quarks $g_q$ only. Here, we treat $\Gamma_{\rm med}$ as a free parameter accounting for additional decay channels of the mediator, for example into pairs of DM candidate particles. The different lines are for different values of $\Gamma_{\rm med}$ as indicated in the legend. The axial-vector and pseudoscalar cases not shown here would look very similar to the vector and scalar cases, respectively.}
   \label{fig:EL_width}
\end{figure}

The results shown in \reffig{fig:EL_vector_gqonly} and \reffig{fig:EL_scalar_gqonly} assume that the mediator decays only into quarks, thus, its width $\Gamma_{\rm med}$ is a function of its mass $m_{\rm med}$ and the coupling to quarks $g_q$ only. In \reffig{fig:EL_width} we show 95\,\%\,C.L. exclusion limits for the vector and scalar mediator case treating $\Gamma_{\rm med}$ as a free parameter to account for additional decay channels of the mediator. In particular, such decay channels may include decays of the mediator into pairs of DM candidate particles. We do not show results for the axial-vector and pseudoscalar cases, since the limits would be very similar to the vector and scalar cases, respectively, cf. the discussion above. 

Increasing the width of the mediator with respect to the width given by the decay into quarks only weakens the limits via two effects: 1) Increasing the width results in a broader signal spectrum which is more difficult to distinguish from the smooth SM background. 2) Increasing the width reduces the overall dijet production cross section for the signal, since ($\sigma \propto \Gamma_{qq}^2/\Gamma_{\rm med}$) in the narrow width approximation. Here, $\Gamma_{qq}$ is the partial width into pairs of quarks fixed by the mediator's mass and the coupling to quarks, and ($\Gamma_{\rm med} = \Gamma_{qq} + \ldots$), with ``$\ldots$'' indicating the partial width into additional decay channels, is the width of the mediator.

In the limit of massless quarks and for a scalar mediator we can use the approximation $\Gamma_{qq} / m_\text{med} \sim 18 h_1^2/8\pi$, where the factor of 18 accounts for the summation over all possible flavor and color states. For a vector mediator we obtain $\Gamma_{qq} / m_\text{med}\sim (2/3) \times18 h_3^2/8\pi$, where the additional factor of $2/3$ comes from averaging over the polarization states of the vector mediator. The same approximations would apply for a pseudoscalar and an axial-vector, respectively, since their decay width into quarks differs only for non-zero quark masses. The exact formula for $\Gamma_{qq}$ for all mediators discussed here can be found in \Ref~\cite{Catena:2017xqq}. For a vector mediator we obtain then, for instance, [$\Gamma_{qq} / m_\text{med} (g_q = h_3 = 0.1) = 0.005$] and [$\Gamma_{qq} / m_\text{med} (0.3) = 0.04$], while for a scalar mediator we have [$\Gamma_{qq} / m_\text{med} (g_q = h_1 = 0.1) = 0.007$] and [$\Gamma_{qq} / m_\text{med} (0.3) = 0.06$]. Thus, we can understand the results shown in in \reffig{fig:EL_width}: In the case of a vector mediator, the limit assuming decay into quarks only is stronger than the limit for $\Gamma_{qq} / m_\text{med} = 0.05$ for the entire range of mediator masses shown. For a scalar mediator, instead, the limits in the $\Gamma_{\rm med} = \Gamma_{qq}$ case exclude quark couplings $g_q \gtrsim 0.25$ for mediator masses $m_{\rm med} \sim 2.7\tev$ and even larger $g_q$ for larger $m_{\rm med}$. The partial width into quarks $\Gamma_{qq}$ is larger than $\Gamma_{qq}/m_{\rm med} = 0.05$ for $g_q \gtrsim 0.25$. Thus, for the $\Gamma_{\rm med}/m_{\rm med} = 0.05$ case we exclude couplings $g_q$ smaller than for the $\Gamma_{\rm med} = \Gamma_{qq}$ case for $m_{\rm med} \gtrsim 2.7\tev$, although it should be noted that limits for $\Gamma_\text{med} < \Gamma_{qq}$ are of course purely hypothetical and do not correspond to a physical scenario.

\section{Sensitivity Projections} \label{sec:projections}

\begin{figure}
   \begin{subfigure}{\linewidth}
      \includegraphics[width=0.49\linewidth]{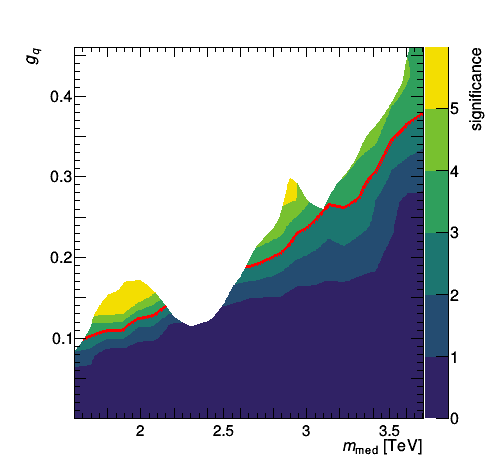}
      \includegraphics[width=0.49\linewidth]{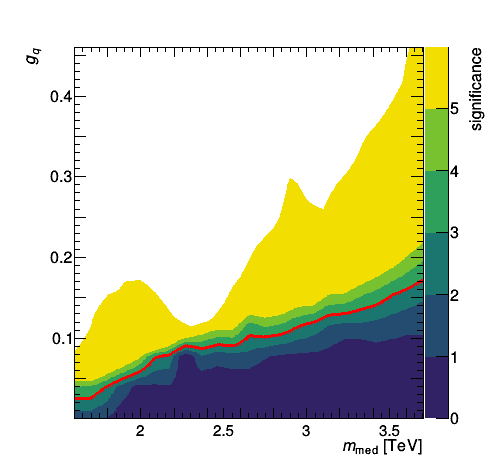}
      \caption{Vector Mediator ($h_3 \neq 0$)}
   \end{subfigure}

   \begin{subfigure}{\linewidth}
      \includegraphics[width=0.49\linewidth]{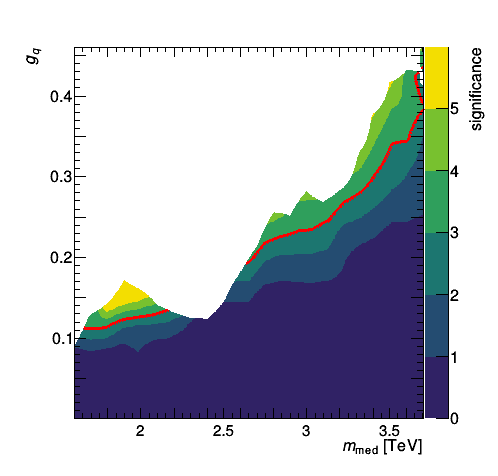}
      \includegraphics[width=0.49\linewidth]{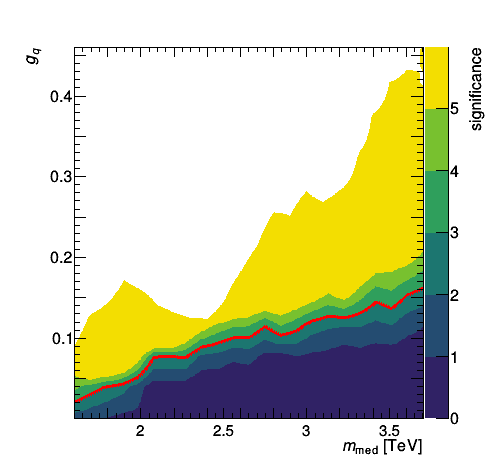}
      \caption{Axial-Vector Mediator ($h_4 \neq 0$)}
   \end{subfigure}
   \caption{Sensitivity projections for the vector mediator (upper panels) and axial-vector mediator (lower panels) in future LHC runs under the assumption that the width of the mediator is determined by decays into quarks only. The left panels are for $300\invfb$ of data expected to be collected after Run 3 of the LHC. The right panels are for $3000\invfb$ of data as expected to be collected by the High Luminosity (HL) upgrade of the LHC. The color scale shows the significance [defined in \refeq{eq:significance}] with which a parameter point in the $m_{\rm med} - g_q$ parameter plane could be observed. The red lines denotes the $3\,\sigma$ evidence contour for a dijet signal. The white region is the parameter space excluded by the $36\invfb$ data, cf. \reffig{fig:EL_vector_gqonly}.}
   \label{fig:DISCO_vector_gqonly}
\end{figure}

\begin{figure}
   \begin{subfigure}{\linewidth}
      \includegraphics[width=0.49\linewidth]{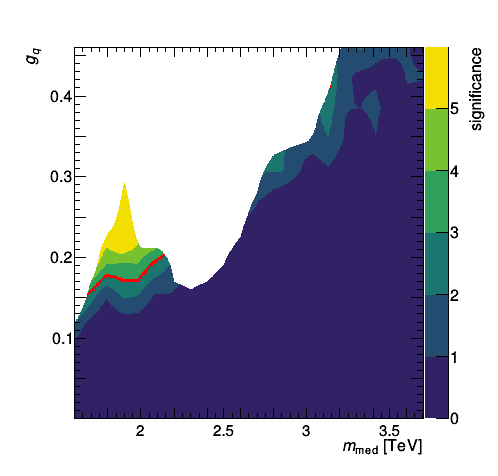}
      \includegraphics[width=0.49\linewidth]{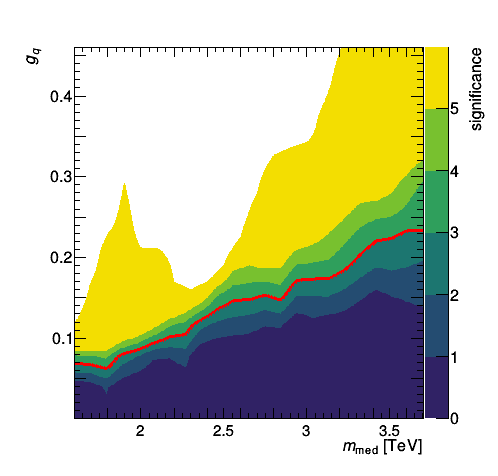}
      \caption{Scalar Mediator ($h_1 \neq 0$)}
   \end{subfigure}

   \begin{subfigure}{\linewidth}
      \includegraphics[width=0.49\linewidth]{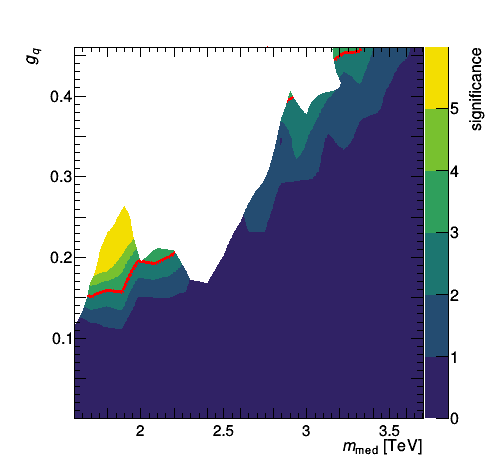}
      \includegraphics[width=0.49\linewidth]{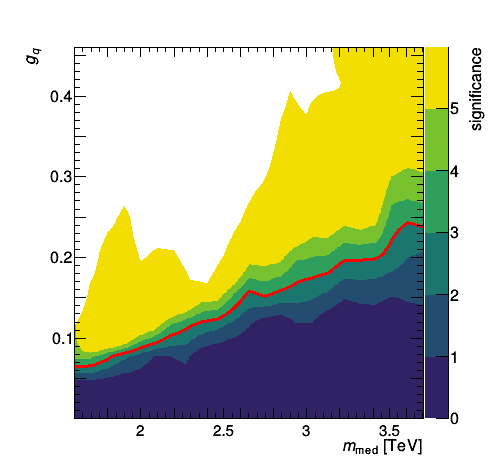}
      \caption{Pseudoscalar Mediator ($h_2 \neq 0$)}
   \end{subfigure}
   \caption{Same as \reffig{fig:DISCO_vector_gqonly} but for the scalar (upper panel) and the pseudoscalar (lower panel) mediator case.}
   \label{fig:DISCO_scalar_gqonly}
\end{figure}
   
Assuming that the selection cuts will remain unchanged with respect to those employed in the current dijet analyses, we can project the sensitivity of dijet searches in future LHC runs without the need to run dedicated SM background simulations. To this end, we use the background fit to the $36\invfb$ CMS dataset described in \refsec{sec:background} and rescale the background to the appropriate luminosity. Here, we consider two benchmark cases: an integrated luminosity of $300\invfb$ as expected to be collected at the end of Run 3 of the LHC in 2023, and an integrated luminosity of $3000\invfb$ as expected to be collected by $\sim 2037$ after the high luminosity (HL) upgrade of the LHC. Using the fit of the background spectrum to the existing $36\invfb$ dataset entails that our projections do not account for improvements in sensitivity from a possible increase of the LHC's center of mass energy from the current $\sqrt{s} = 13\tev$ to $\sqrt{s} = 14\tev$. However, the impact of such a modest increase in $\sqrt{s}$ on the projected sensitivity would be much smaller than from the increase in luminosity by a factor of $10-100$ with respect to $36\invfb$ of data. Together with our simulated signal samples, we can project the future sensitivities using the statistical procedure outlined in \refsec{sec:statistic}.

We show sensitivity projections for the vector and axial-vector cases in \reffig{fig:DISCO_vector_gqonly} and for the scalar and pseudoscalar cases in \reffig{fig:DISCO_scalar_gqonly} under the assumption that the mediator couples to quarks only, thus, its width is a function of $m_{\rm med}$ and $g_q$ only. The improvements with respect to the currently excluded region of parameter space is rather similar for all cases considered here. As discussed in \refsec{sec:limits}, for the smallest mediator masses $m_{\rm med} \sim 1.6\tev$ considered here, the dijet search is already limited by the systematic uncertainty on the background spectra. Thus, increasing the luminosity to $300\invfb$ or $3000\invfb$ does not allow to probe smaller couplings to quarks than those already excluded. For heavier mediators, increasing the luminosity does allow to probe sizeable portions of currently unconstrained parameter space. 

Considering (axial-)vector mediators, cf. \reffig{fig:DISCO_vector_gqonly}, we find that at mediator masses of $m_{\rm med} \sim 3.5\tev$ a signal could be discovered with a significance of at least $5\,\sigma$ if the coupling to quarks is larger than $g_q \sim 0.2$. Recall that the $36\invfb$ data excludes couplings larger than $g_q \sim 0.4$ for such mediator masses,  while the $78\invfb$ data excludes couplings larger than $g_q \sim 0.25$. In the absence of a signal (and assuming that ``discovery'' and exclusion limits are comparable), $300\invfb$ ($3000\invfb$) of data would allow to exclude couplings larger than $g_q \sim 0.25$ ($g_q \sim 0.1$) at 95\,\%\,C.L. for (axial-)vector mediators.\footnote{Recall that a 95\,\%\,C.L. exclusion limit corresponds to a significance of $Z = 1.64$.}

For (pseudo)scalar mediators, cf. \reffig{fig:DISCO_scalar_gqonly}, we find similar increases in sensitivity with respect to current limits. For mediators with mass $m_{\rm med} \sim 3.5\tev$ the HL-LHC with $3000\invfb$ of data would allow for a detection with a significance of at least $5\,\sigma$ if the coupling to quarks is larger than $g_q \sim 0.3$. In the absence of a signal, the HL-LHC could exclude (pseudo)scalar mediators with masses of $m_{\rm med} \sim 3.5\tev$ at 95\,\%\,C.L. if the coupling to quarks is larger than $g_q \sim 0.2$ (again assuming that ``discovery'' and exclusion limits are comparable).

\section{Parameter Reconstruction} \label{sec:param}
\begin{figure}
   \includegraphics[width=0.49\linewidth]{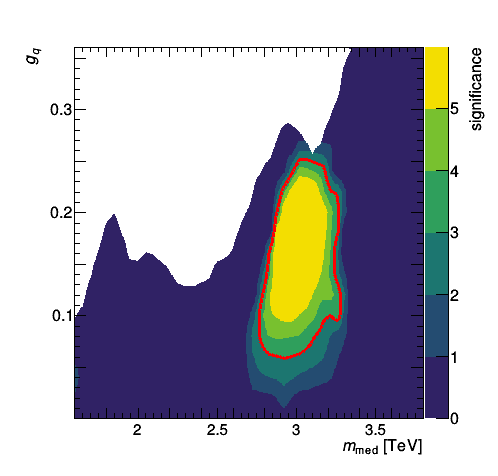}
   \includegraphics[width=.49\linewidth]{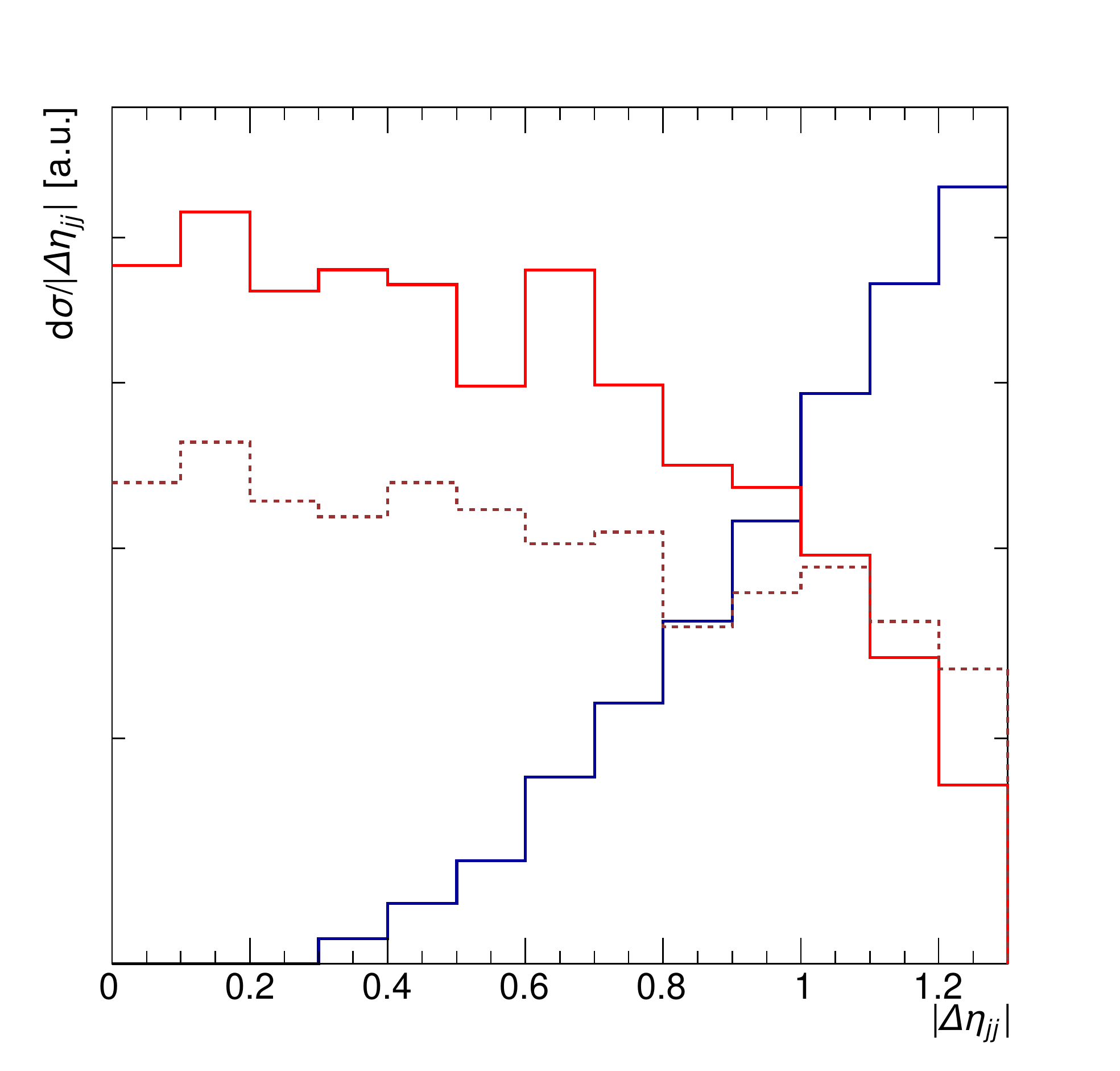} 
   \caption{{\it Left:} Potential for parameter reconstruction if a signal from a vector mediator with a true mass of $m_{\rm med} = 3.0\tev$ coupling to quarks only with $g_q = 0.2$ is observed at the HL-LHC. The color scale shows the expected significance $Z$ with which the background only hypothesis can be rejected in favor of a hypothetical signal at a given point in parameter space based on data generated for the true signal and assuming a luminosity of $3000\invfb$. The red line shows the $3\,\sigma$ error ellipse for parameter reconstruction. 
   {\it Right:} Angular distribution $\dd\sigma/\dd|\Delta\eta_{jj}|$ of the dijet spectra for the SM background (solid blue), a scalar mediator (solid red), and a vector mediator (dashed burgundy). Note that the spectra are not shown to scale in order to emphasize the differences in shape. Both signal spectra are for $m_{\rm med} = 2\tev$ and a decay width of $\Gamma_{\rm med}/m_{\rm med} = 0.05$. We consider only events near the resonance, $1.5\tev < m_{\rm jj} < 2.5\tev$. Note that both background and signal events have been generated with our \texttt{MadGraph} instead of our default \texttt{WHIZARD} tool chain.}
   \label{fig:Param_angular}
\end{figure}

In this section, we study how well the parameters of a simplified model could be reconstructed at the HL-LHC in case of a significant future discovery. Thus, we shift from projecting discovery/exclusion limits to parameter reconstruction. Confidence regions for parameter reconstruction are computed as in the case of sensitivity projections, cf. \refsec{sec:statistic}, except that the profile likelihood ratio in \refeq{eq:profL2} is now calculated from the dataset $n_i=b_i(\boldsymbol{\theta}_{\rm bf}) + \bar{s}_i$,where $\bar{s}_i$ is the number of dijet events in the $i$-th mass bin for the true signal parameters $m_{\rm med}$, $g_q$, and $\Gamma_{\rm med}$. 

As an illustrative example, in the left panel of \reffig{fig:Param_angular} we show the prospects for parameter reconstruction for the benchmark case of a vector mediator with a true mass of $m_{\rm med} = 3.0\tev$ coupling only to quarks with strength $g_q = 0.20$ at the HL-LHC. These results indicate that for this parameter point, the mass could be reconstructed with a relative error of $\lesssim 10\,\%$ while the coupling strength could be reconstructed to $g_q \sim 0.18^{+0.05}_{-0.10}$. The best fit mass is $m_{\rm med} = 3.05\tev$. Not surprisingly, the relative error on the mediator mass, fixed by the location of the excess over the background in the $m_{jj}$ direction, is smaller than the relative error on $g_q$ which is determined by measuring the cross-section of the signal.

Differentiating different simplified models is difficult based on the differential dijet spectra with respect to the invariant mass, $\dd \sigma/\dd m_{jj}$, discussed until here. This is because hadronization effects as well as the finite energy resolution of a real detector like CMS make it challenging to observe the minute differences in the spectra from different models. However, the dijet production cross sections have different dependence on the scattering angle for the different mediator cases considered here. Such dependence can be exploited by considering double differential dijet spectra $\dd^2 \sigma / \dd m_{jj} \dd \Delta\eta_{jj}$, where $\Delta\eta_{jj}$ is the pseudorapidity difference between the two widejets. In the right panel of \reffig{fig:Param_angular} we show the $\dd\sigma/\dd|\Delta\eta_{jj}|$ distribution of the SM background and those for a scalar and vector mediator in comparison. The signal and background spectra shown in that panel are simulated with our \texttt{MadGraph} tool chain described in \refsec{sec:dijet_LHC} instead of our default \texttt{WHIZARD} tool chain. This figure shows that the angular distributions can in principle be used for distinguishing different simplified models. Further, considering the differences in the angular distributions between the signal and background spectra, the sensitivity of dijet searches may be further enhanced with respect to the results shown in \refsec{sec:limits} and \refsec{sec:projections} by using angular information.

Although seemingly promising, an analysis extending the sensitivity of dijet searches for simplified models with angular information is beyond the scope of this work. Note that both the ATLAS and the CMS collaboration have used angular information in dijet searches for BSM physics~\cite{Sirunyan:2018wcm,Sirunyan:2017ygf,Aad:2015eha,Khachatryan:2014cja}. Regarding differentiating different simplified models in case of a discovery, calculations based on a naive profile likelihood indicate that the number of events in the region of parameter space not already excluded by current limits would not suffice to differentiate models. However, a more sophisticated analysis with optimized selection cuts may yield more promising results. We leave such investigations for future work.

\section{Conclusions}\label{sec:conclusions}

We presented a reanalysis of the latest results from CMS dijet searches for an integrated luminosity of $36\invfb$ together with preliminary results for $78\invfb$ in the framework of simplified models for (scalar, fermionic and vector) DM interacting with quarks through the exchange of a scalar, pseudoscalar, vector or pseudovector mediator particle~\cite{Baum:2017kfa,Dent:2015zpa}.~Within the same framework, we also projected the sensitivity of dijet searches in future LHC runs, assuming that the selection cuts will remain unchanged with respect to those employed in the current dijet analyses, and rescaling our background fit to the $36\invfb$ CMS dataset described in \refsec{sec:background} accordingly.~Finally, we also studied how well the parameters of a simplified model could be reconstructed in case of a significant future discovery at the HL-LHC.

From our reanalysis of the latest results from CMS dijet searches, we obtained 95\,\%\,C.L. exclusion limits on the coupling of the mediator to quarks as a function of the mediator mass for spin 0 and spin 1 mediators.~Exclusion limits for scalar and pseudoscalar mediators were derived here for the first time.~In this study, we considered two cases separately:~1) The mediator couples to quarks only, and thus its width, $\Gamma_{\rm med}$, is determined by $m_{\rm med}$ and $g_q$.~2)~$\Gamma_{\rm med}$ is a free parameter to account for additional decay channels of the mediator, e.g. into pairs of DM particles.~For both the $36\invfb$ and the $78\invfb$ data, the limits that we obtained are within $1\,\sigma$ of the respective expected limits, indicating that both CMS datasets show no evidence for a dijet signal in the simplified model framework.~Comparing the results from the $36\invfb$ data to those from the $78\invfb$ data, we found that the larger luminosity allows for more stringent exclusion limits, but only for large mediator masses where exclusion limits are limited by statistical rather than systematical errors.

When projecting the sensitivity of dijet searches in future LHC runs, we focused on two benchmark cases:~1) An integrated luminosity of $300\invfb$ as expected to be collected at the end of Run 3 of the LHC in 2023.~2)~An integrated luminosity of $3000\invfb$ as expected to be collected by $\sim 2037$ after the high luminosity upgrade of the LHC.~We obtained sensitivity projections under the assumption that the mediator couples to quarks only, and thus its width is determined by $m_{\rm med}$ and $g_q$ only.~The improvements found with respect to the currently excluded regions in the parameter space of simplified models is rather small for mediator masses $m_{\rm med} \sim 1.6\tev$ -- where the dijet searches are already limited by the systematic uncertainty on the background spectra --  but they become significant for heavier mediators, where dijet searches are limited by statistics.~We found that in this latter case sizable portions of currently unconstrained parameter space will be within reach at the HL-LHC. 

Assessing how well the parameters of a given simplified model can be reconstructed in case of a significant future discovery at the HL-LHC, we focused on a single benchmark case where a vector mediator with a true mass of $m_{\rm med} = 3.0\tev$ couples only to quarks with strength $g_q = 0.20$.~In this specific case, we found that the mediator mass could be reconstructed with a relative error of $\lesssim 10\,\%$, while the reconstructed coupling strength was $g_q \sim 0.2^{+0.05}_{-0.10}$.

Finally, we also explored the possibility of discriminating different mediator scenarios by extending the sensitivity of dijet searches for simplified models through the use of angular information.~The feasibility of this approach relies on the fact that dijet production cross sections at the LHC have a different dependence on the scattering angle for the different mediator cases considered here.~Although seemingly promising, we left such investigation for 
future work.

\acknowledgments
We thank Jan Conrad and Katherine Freese for their contribution to the early stages of this project.~We are grateful to Caterina Doglioni and Felix Kahlhoefer for useful insights into how to model dijet signal and background events at the LHC.~Finally, it is a pleasure to thank K\r{a}re Fridell and Vanessa Zema for helpful discussions on non-relativistic effective theories and simplified models for dark matter.~This work is performed within the Swedish Consortium for Dark Matter Direct Detection (SweDCube), and was supported by the Knut and Alice Wallenberg Foundation (PI, Jan Conrad), by the Vetenskapsr\r{a}det (Swedish Research Council) through contract No. 638-2013-8993, and the Oskar Klein Centre for Cosmoparticle Physics. 

\appendix

\section{Examples of simplified models}\label{app:simp}
For concreteness, below we list selected examples of interaction Lagrangians where a real scalar mediator, $\phi$, or a real vector mediator, $G_\mu$, simultaneously couples to quark $q$ [as in eqs.~\eqref{eq:spin0} and \eqref{eq:spin1}] and to scalar, fermionic or vector particles which could in principle be DM candidates~\cite{Baum:2018lua}:
\begin{align}
   \mathcal{L}_1 &=-h_1\bar{q} q\phi -g_1m_SS^\dagger S\phi \,;& \quad &(S\otimes S)_0 \nonumber \\
   \mathcal{L}_2 &=-h_3(\bar{q}\gamma_{\mu}q)G^{\mu}-ig_4(S^{\dagger}\partial_{\mu}S-\partial_{\mu}S^{\dagger}S)G^{\mu}\,;& \quad &(V\otimes i\partial)_0 \nonumber \\
   \mathcal{L}_3 &=-h_1\phi\bar{q} q-\lambda_1\phi\bar{\chi}\chi \,;& \quad &(S\otimes S)_{1/2} \nonumber \\
   \mathcal{L}_4 &=-h_1\phi\bar{q} q-i\lambda_2\phi\bar{\chi}\gamma^{5}\chi\,;& \quad &(S\otimes PS)_{1/2} \nonumber \\
   \mathcal{L}_5 &=-h_3\bar{q}\gamma_{\mu}qG^{\mu}-\lambda_{3}\bar\chi\gamma^\mu\chi G_{\mu} \,;& \quad &(V\otimes V)_{1/2} \nonumber \\
   \mathcal{L}_6 &=-h_3\bar{q}\gamma_{\mu}qG^{\mu}-\lambda_{4}\bar\chi\gamma^\mu\gamma^5\chi G_{\mu}  \,;& \quad &(V\otimes A)_{1/2} \nonumber \\
   \mathcal{L}_7 &=-h_4\bar{q}\gamma_{\mu}\gamma^{5}qG^{\mu}-\lambda_{4}\bar\chi\gamma^\mu\gamma^5\chi G_{\mu}  \,; &\quad &(A\otimes A)_{1/2} \nonumber \\
   \mathcal{L}_8 &=-h_1\phi\bar{q}q-b_1m_X\phi X_{\mu}^{\dagger}X^{\mu} \,;& \quad &(S\otimes S)_1 \nonumber \\
   \mathcal{L}_9 &=-h_3G_\mu\bar{q}\gamma^\mu q-ib_{5}(X_{\nu}^{\dagger}\partial_{\mu}X^{\nu}-X^\nu \partial_\mu X^\dagger_\nu)G^\mu\,;& \quad &(V\otimes i\partial)_1
   \label{eq:simpL}
\end{align}
where $g_q=h_1, h_2, h_3$, $h_4$ is the mediator coupling to quarks, and $g_{\rm DM}= \lambda_1, \lambda_2, \lambda_3, \lambda_4, g_1, g_4, b_1$, $b_5$ is the mediator coupling to the additional invisible particle, e.g. DM.~In \refeq{eq:simpL}, scalar, fermionic and vector DM are described by the complex scalar field $S$, the spinor field $\chi$ or the complex vector field $X_\nu$, respectively.~In the quark bilinears a summation over quark flavours is understood.~The label next to each interaction Lagrangian characterises the corresponding simplified model.~For example, $(S\otimes S)_0$, is the label for a model where a spin 0 invisible particle $S$ couples to quarks via a scalar $S$-$S$-mediator vertex and a scalar quark-quark-mediator vertex.~Analogously, $(V\otimes i\partial)_1$ refers to a model where a spin 1 invisible particle $X$ couples to quarks via a vector quark-quark-mediator vertex and a derivative $X$-$X$-mediator vertex.~In the remaining cases, $PS$ refers to pseudo-scalar coupling and $A$ to axial coupling.

For the particle $S$ or $G$ to be a DM candidate, stability over cosmological time scales is a necessary condition.~The stability of $S$ and $G$ is not guarantied by \refeq{eq:simpL} or eqs.~\eqref{eq:spin0} and \eqref{eq:spin1}, and must be imposed by hand, e.g., by postulating the existence of an additional symmetry in Nature under which DM and the SM particles have opposite charges, such as R-party in Sypersymmetric models, or a $Z_2$ symmetry in simplified models for scalar DM, like in \refeq{eq:simpL}.

The DM relic density for the models in \refeq{eq:simpL} has been computed in \Ref~\cite{Catena:2017xqq} focusing on regions in the parameter space associated with a detectable signal at XENONnT.~For example, it has been found that model ($S\otimes S)_0$ is compatible with the detection of about 100 signal events at XENONnT (as expected if a 50 GeV DM candidate couples to nucleons with a strength just below current exclusion limits) only for $m_{\rm med}$ in a narrow window around 100 GeV.~Similar results have been found for model $(A\otimes A)_{1/2}$ and model $(V\otimes A)_{1/2}$~\cite{Catena:2017xqq}.~We refer to \Ref~\cite{Catena:2017xqq} for a comprehensive study of the DM relic density in simplified models, including the ones in \refeq{eq:simpL}.

A reanalysis of LHC monojet searches within the framework of the simplified models in \refeq{eq:simpL} has been performed in \Ref~\cite{Baum:2017kfa}.~Also in this case, the focus has been on finding regions in the parameter space of the simplified models where DM signals can simultaneously be identified in LHC monojet searches and future direct detection experiments, such as XENONnT.~In general, monojet and dijet searches at the LHC are found to be complementary within the simplified model framework:~the former search is especially sensitive to large $g_{DM}$ values, a region in parameter space where $\Gamma_{\rm med}$ is expected to be large, which in turn decreases the LHC sensitivity to narrow resonances in dijet final states.~We refer to \Ref~\cite{Baum:2017kfa} for further details.

\section{Signal fit and spectra}\label{app:signal_fit}
\begin{figure}[h!]
   \hspace{-.7cm}
   \begin{center}
   \includegraphics[width=.96\linewidth]{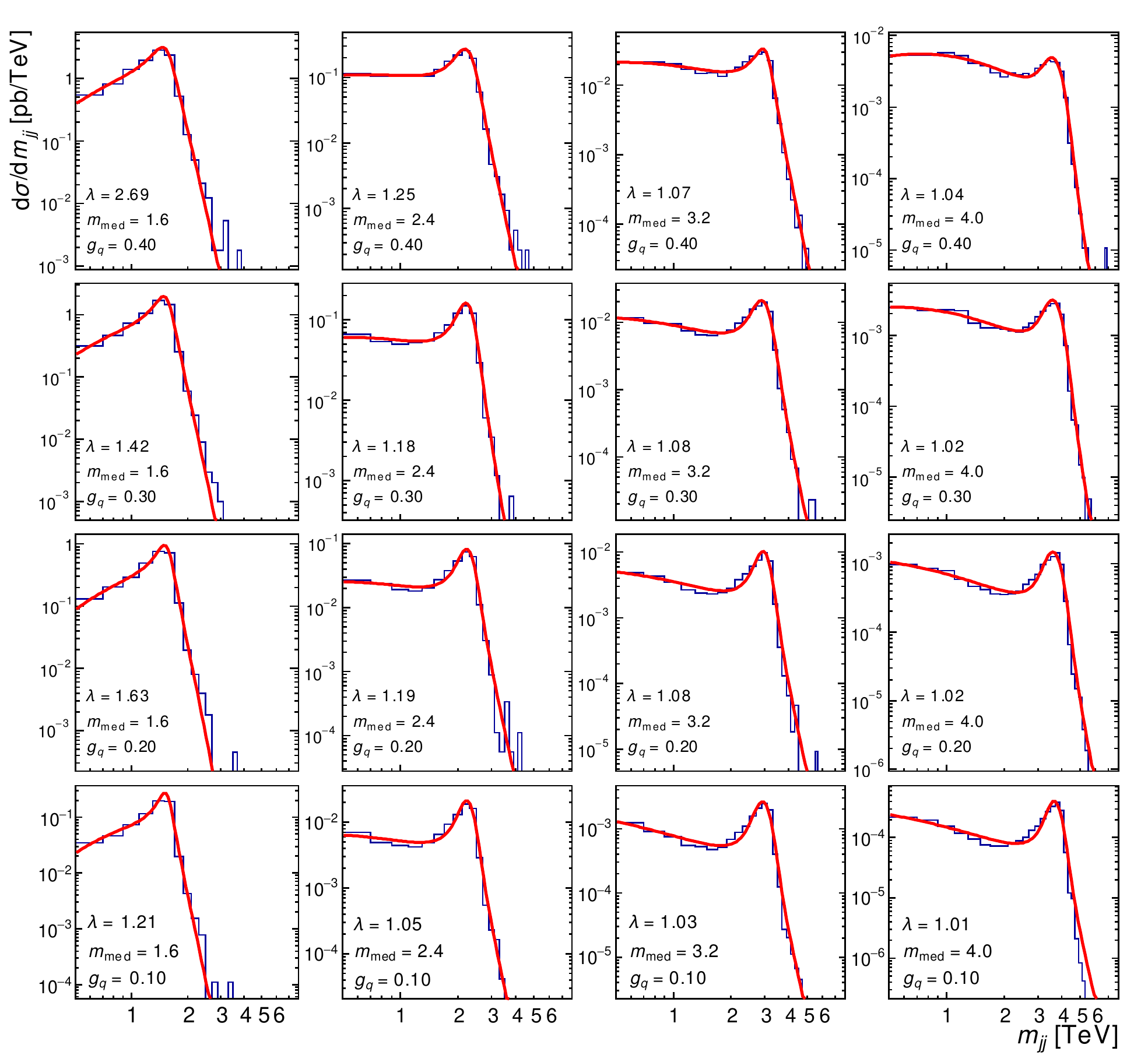}
   \end{center}
   \caption{\small Simulated signal spectra (blue histogram) and corresponding fit [\refeq{eq:signal_fit}, red line] for several parameter points in the vector mediator case assuming that the mediator couples to quarks only. The value of the likelihood ratio $\lambda$ at the best fit point, cf. \refeq{eq:logLsig}, the mediator mass $m_{\rm med}$ in TeV, and the coupling to quarks $g_q$ are listed in the respective panels. Note that the scale of the $y$-axis differs between the respective panels. Panels in the same column share the value for the mediator mass, while panels in the same row share the value for the coupling of the mediator to quarks. The signal spectra for the scalar, pseudoscalar, and axial-vector case are similar to the vector mediator case shown here.}
   \label{fig:sig_tab}
\end{figure}

In this appendix we list the fit parameters corresponding to \refeq{eq:signal_fit} for selected parameter points for a vector, axial-vector, scalar and pseudoscalar mediator in \reftab{tab:signal_fit1}-\ref{tab:signal_fit4}, which can be used for further analysis without the need to simulate the signal processes again. We assume universal couplings to the quarks and no additional decay channels for the mediator. In \reffig{fig:sig_tab} we show signal spectra and the corresponding fits assuming a vector mediator for illustration for a subset of parameter points.

Note that in tabs.~\ref{tab:signal_fit1}-\ref{tab:signal_fit4} we report the parameters for the signal distributions with $6-7$ significant digits. This is because the propagation of errors from the parameters to the shape of the signal spectra is non-trivial. The number of significant digits reported in these tables should not be understood to imply a relative precision of $\mathcal{O}(10^{-6})$ of the signal spectra; when using the spectra to set/project limits, a statistical procedure properly accounting for systematic errors of the signal spectra of at least a few percent as well as for statistical errors must be employed.

\newpage

\begin{table}[h!]
\begin{center}
\vspace*{-3em}
\def\arraystretch{.5}
\begin{footnotesize}
   \begin{tabular}{C{1cm} C{2.3cm} L{2.3cm} R{1.5cm} C{2.5cm} L{2.5cm} R{1.5cm} R{2cm}}
    \hline\hline \\[-.4em]
 $\dfrac{m_\text{med}}{\text{GeV}}$ & $g_q$ & 
\tabhead{$\dfrac{p_0}{\text{fb/GeV}}$} 
& \tabhead{$\dfrac{p_1}{\text{GeV}}$} & \tabhead{$\dfrac{p_2}{\text{GeV}}$} & 
\tabhead{$\dfrac{p_3}{\text{GeV}}$} & \tabhead{$p_4$ }& \tabhead{$p_5$} \\[.6em]
      \hline\\[-.4em]
      \input{tab_h3lam0.dat}
      \hline\hline
   \end{tabular}
    \vspace*{-1em}
   \end{footnotesize}
   \end{center}
   \caption{Fit values for the signal for a vector mediator. 
   \label{tab:signal_fit1}}
\end{table}

\newpage
\begin{table}[h!]
\begin{center}
\vspace*{-3em}
\def\arraystretch{.5}
\begin{footnotesize}
   \begin{tabular}{C{1cm} C{2.3cm} L{2.3cm} R{1.5cm} C{2.5cm} L{2.5cm} R{1.5cm} R{2cm}}
    \hline\hline \\[-.4em]
      $\dfrac{m_\text{med}}{\text{GeV}}$ & $g_q$ & 
\tabhead{$\dfrac{p_0}{\text{fb/GeV}}$} 
& \tabhead{$\dfrac{p_1}{\text{GeV}}$} & \tabhead{$\dfrac{p_2}{\text{GeV}}$} & 
\tabhead{$\dfrac{p_3}{\text{GeV}}$} & \tabhead{$p_4$ }& \tabhead{$p_5$} \\[.6em]
      \hline\\[-.4em]
      \input{tab_h4lam0.dat}
      \hline\hline
   \end{tabular}
    \vspace*{-1em}
\end{footnotesize}
   \end{center}
   \caption{Fit values for the signal for an axial-vector mediator.}
\end{table}

\newpage
\begin{table}[h!]
\begin{center}
\vspace*{-3em}
\def\arraystretch{.5}
\begin{footnotesize}
   \begin{tabular}{C{1cm} C{2.3cm} L{2.3cm} R{1.5cm} C{2.5cm} L{2.5cm} R{1.5cm} R{2cm}}
    \hline\hline \\[-.4em]
 $\dfrac{m_\text{med}}{\text{GeV}}$ & $g_q$ & 
\tabhead{$\dfrac{p_0}{\text{fb/GeV}}$} 
& \tabhead{$\dfrac{p_1}{\text{GeV}}$} & \tabhead{$\dfrac{p_2}{\text{GeV}}$} & 
\tabhead{$\dfrac{p_3}{\text{GeV}}$} & \tabhead{$p_4$ }& \tabhead{$p_5$} \\[.6em]
      \hline\\[-.4em]
      \input{tab_h1g0.dat}
      \hline\hline
   \end{tabular}
   \vspace*{-1em}
\end{footnotesize}
   \end{center}
   \caption{Fit values for the signal for a scalar mediator.}
\end{table}

\newpage
\begin{table}[h!]
\begin{center}
\vspace*{-3em}
\def\arraystretch{.5}
\begin{footnotesize}
   \begin{tabular}{C{1cm} C{2.3cm} L{2.3cm} R{1.5cm} C{2.5cm} L{2.5cm} R{1.5cm} R{2cm}}
    \hline\hline \\[-.4em]
 $\dfrac{m_\text{med}}{\text{GeV}}$ & $g_q$ & 
\tabhead{$\dfrac{p_0}{\text{fb/GeV}}$} 
& \tabhead{$\dfrac{p_1}{\text{GeV}}$} & \tabhead{$\dfrac{p_2}{\text{GeV}}$} & 
\tabhead{$\dfrac{p_3}{\text{GeV}}$} & \tabhead{$p_4$ }& \tabhead{$p_5$} \\[.6em]
      \hline\\[-.4em]
      \input{tab_h2g0.dat}
      \hline\hline
   \end{tabular}
    \vspace*{-1em}
   \end{footnotesize}
   \end{center}
   \caption{Fit values for the signal for a pseudo-scalar 
mediator.\label{tab:signal_fit4}}
\end{table}

\FloatBarrier

\bibliography{TheBib}

\end{document}

%% file: tab_h3lam0.dat
1600 & 0.05 & 5.43357$\times 10^{5}$ & 282.612 & 1546.450 & 6.05008$\times 10^{-1}$ & -9.86023 & 4.53496\\ 
1600 & 0.10 & 3.68206$\times 10^{6}$ & 270.945 & 1551.107 & 4.54318 & -10.41522 & 4.79370\\ 
1600 & 0.15 & 5.97351$\times 10^{5}$ & 301.361 & 1532.236 & 2.73148$\times 10^{-1}$ & -7.71246 & 3.22862\\ 
1600 & 0.20 & 3.12253$\times 10^{6}$ & 318.096 & 1542.951 & 5.62014$\times 10^{-1}$ & -8.79416 & 3.96967\\ 
1600 & 0.25 & 2.80549$\times 10^{6}$ & 334.138 & 1538.902 & 1.51016 & -8.24228 & 3.62487\\ 
1600 & 0.30 & 2.72932$\times 10^{6}$ & 363.043 & 1535.325 & 4.41062$\times 10^{-1}$ & -7.83232 & 3.45908\\ 
1600 & 0.35 & 7.31285$\times 10^{6}$ & 383.076 & 1541.762 & 1.57555$\times 10^{-1}$ & -8.53564 & 3.94213\\ 
1600 & 0.40 & 2.15259$\times 10^{6}$ & 429.462 & 1523.177 & 6.49771$\times 10^{-1}$ & -6.95947 & 3.16951\\ 
1600 & 0.45 & 1.60324$\times 10^{6}$ & 466.606 & 1516.691 & 6.91752$\times 10^{-1}$ & -6.44059 & 2.90848\\ 
2000 & 0.05 & 3.93500$\times 10^{3}$ & 296.440 & 1914.366 & 6.69305$\times 10^{1}$ & -6.05549 & 1.28131\\ 
2000 & 0.10 & 1.29671$\times 10^{4}$ & 325.795 & 1914.354 & 5.45899 & -5.87183 & 1.16092\\ 
2000 & 0.15 & 4.22177$\times 10^{4}$ & 330.627 & 1913.267 & 9.58694$\times 10^{-1}$ & -6.23354 & 1.44063\\ 
2000 & 0.20 & 4.42496$\times 10^{4}$ & 285.340 & 1916.568 & 8.51642$\times 10^{1}$ & -5.71231 & 0.92549\\ 
2000 & 0.25 & 1.52420$\times 10^{5}$ & 354.664 & 1920.033 & 1.30534$\times 10^{-1}$ & -6.53100 & 1.60575\\ 
2000 & 0.30 & 2.33439$\times 10^{5}$ & 396.727 & 1928.272 & 5.59112$\times 10^{-1}$ & -6.56657 & 1.75425\\ 
2000 & 0.35 & 1.41725$\times 10^{5}$ & 389.932 & 1899.717 & 7.21201$\times 10^{1}$ & -5.73011 & 1.31740\\ 
2000 & 0.40 & 1.15572$\times 10^{5}$ & 464.323 & 1899.313 & 7.60358$\times 10^{1}$ & -5.22676 & 1.17490\\ 
2000 & 0.45 & 3.06705$\times 10^{5}$ & 483.640 & 1913.765 & 9.87537$\times 10^{1}$ & -6.00754 & 1.63213\\ 
2400 & 0.05 & 5.59414$\times 10^{2}$ & 346.703 & 2286.943 & 5.36397$\times 10^{-1}$ & -5.13274 & 0.45203\\ 
2400 & 0.10 & 2.65611$\times 10^{3}$ & 303.767 & 2293.762 & 9.87371$\times 10^{1}$ & -5.32930 & 0.44016\\ 
2400 & 0.15 & 1.05415$\times 10^{5}$ & 346.163 & 2305.575 & 2.75506$\times 10^{-1}$ & -8.24659 & 2.31603\\ 
2400 & 0.20 & 1.26468$\times 10^{4}$ & 317.490 & 2297.326 & 1.03349$\times 10^{2}$ & -5.51095 & 0.54670\\ 
2400 & 0.25 & 1.48876$\times 10^{4}$ & 331.346 & 2287.527 & 1.16979$\times 10^{2}$ & -5.22197 & 0.38458\\ 
2400 & 0.30 & 3.39259$\times 10^{4}$ & 370.681 & 2293.786 & 1.16980$\times 10^{2}$ & -5.71855 & 0.70399\\ 
2400 & 0.35 & 2.84299$\times 10^{4}$ & 470.683 & 2287.043 & 8.57590$\times 10^{1}$ & -5.17932 & 0.62286\\ 
2400 & 0.40 & 5.18226$\times 10^{4}$ & 465.051 & 2292.506 & 1.13611$\times 10^{2}$ & -5.49941 & 0.78452\\ 
2400 & 0.45 & 9.09806$\times 10^{4}$ & 540.594 & 2278.128 & 6.61869$\times 10^{1}$ & -5.82709 & 1.13088\\ 
2800 & 0.05 & 9.68462$\times 10^{1}$ & 403.791 & 2661.798 & 3.55910$\times 10^{-1}$ & -4.30635 & -0.10629\\ 
2800 & 0.10 & 3.52737$\times 10^{1}$ & 441.458 & 2632.252 & 1.80878$\times 10^{-2}$ & -1.84460 & -1.43052\\ 
2800 & 0.15 & 2.13433$\times 10^{3}$ & 325.072 & 2667.846 & 1.34877$\times 10^{2}$ & -5.27060 & 0.20649\\ 
2800 & 0.20 & 3.08903$\times 10^{3}$ & 353.656 & 2662.574 & 1.41426$\times 10^{2}$ & -5.06439 & 0.12572\\ 
2800 & 0.25 & 2.94567$\times 10^{3}$ & 493.844 & 2649.974 & 3.02310$\times 10^{1}$ & -4.50275 & 0.07269\\ 
2800 & 0.30 & 8.64732$\times 10^{3}$ & 422.868 & 2663.485 & 1.40325$\times 10^{2}$ & -5.25596 & 0.33216\\ 
2800 & 0.35 & 1.22637$\times 10^{4}$ & 463.833 & 2655.475 & 1.45209$\times 10^{2}$ & -5.26814 & 0.43607\\ 
2800 & 0.40 & 3.60051$\times 10^{4}$ & 460.848 & 2671.361 & 1.64151$\times 10^{2}$ & -6.15834 & 0.80327\\ 
2800 & 0.45 & 3.57938$\times 10^{4}$ & 539.634 & 2657.619 & 1.58678$\times 10^{2}$ & -5.85375 & 0.82222\\ 
3200 & 0.05 & 5.39058$\times 10^{1}$ & 327.513 & 3036.885 & 1.61773$\times 10^{2}$ & -4.70282 & -0.27394\\ 
3200 & 0.10 & 1.88780$\times 10^{2}$ & 338.731 & 3027.014 & 1.71367$\times 10^{2}$ & -4.52350 & -0.28728\\ 
3200 & 0.15 & 4.49073$\times 10^{2}$ & 347.328 & 3026.539 & 1.57853$\times 10^{2}$ & -4.58585 & -0.29001\\ 
3200 & 0.20 & 1.74776$\times 10^{3}$ & 374.101 & 3039.745 & 1.48830$\times 10^{2}$ & -5.40473 & 0.16547\\ 
3200 & 0.25 & 1.51618$\times 10^{3}$ & 541.023 & 3038.987 & 1.14012$\times 10^{1}$ & -4.74749 & 0.12095\\ 
3200 & 0.30 & 4.56523$\times 10^{3}$ & 427.180 & 3025.041 & 1.88443$\times 10^{2}$ & -5.53016 & 0.29508\\ 
3200 & 0.35 & 1.33440$\times 10^{4}$ & 449.081 & 3061.446 & 1.97443$\times 10^{2}$ & -6.35182 & 0.64895\\ 
3200 & 0.40 & 1.38316$\times 10^{3}$ & 731.751 & 2980.596 & 2.82204$\times 10^{-1}$ & -3.64044 & -0.37884\\ 
3200 & 0.45 & 2.75520$\times 10^{4}$ & 549.746 & 3076.220 & 1.85274$\times 10^{2}$ & -6.56618 & 0.82067\\ 
3600 & 0.05 & 2.34396$\times 10^{1}$ & 314.122 & 3394.798 & 2.22928$\times 10^{2}$ & -4.71483 & -0.33471\\ 
3600 & 0.10 & 5.34590$\times 10^{1}$ & 502.538 & 3411.750 & 1.13208$\times 10^{1}$ & -4.09866 & -0.33191\\ 
3600 & 0.15 & 2.32948$\times 10^{2}$ & 398.755 & 3400.779 & 1.86235$\times 10^{2}$ & -4.83444 & -0.20208\\ 
3600 & 0.20 & 3.29207$\times 10^{2}$ & 559.950 & 3413.971 & 1.12794$\times 10^{1}$ & -4.57082 & -0.15471\\ 
3600 & 0.25 & 1.90275$\times 10^{3}$ & 410.062 & 3408.040 & 2.13609$\times 10^{2}$ & -5.94925 & 0.27389\\ 
3600 & 0.30 & 2.72615$\times 10^{3}$ & 525.893 & 3414.426 & 1.67678$\times 10^{2}$ & -5.88904 & 0.42731\\ 
3600 & 0.35 & 3.57320$\times 10^{3}$ & 544.856 & 3403.284 & 1.84992$\times 10^{2}$ & -5.86356 & 0.34646\\ 
3600 & 0.40 & 7.04552$\times 10^{3}$ & 709.095 & 3421.111 & 2.01746 & -6.25530 & 0.67730\\ 
3600 & 0.45 & 7.93546$\times 10^{3}$ & 751.939 & 3408.811 & 1.95324 & -6.14830 & 0.56989\\ 
4000 & 0.05 & 9.44562 & 345.979 & 3773.699 & 2.37458$\times 10^{2}$ & -4.67954 & -0.41206\\ 
4000 & 0.10 & 2.41455$\times 10^{1}$ & 545.594 & 3775.325 & 1.12972$\times 10^{1}$ & -4.17045 & -0.37513\\ 
4000 & 0.15 & 6.16856$\times 10^{1}$ & 615.578 & 3767.127 & 1.51870$\times 10^{1}$ & -4.31146 & -0.27883\\ 
4000 & 0.20 & 3.60401$\times 10^{2}$ & 392.794 & 3771.092 & 2.45381$\times 10^{2}$ & -5.55022 & -0.01319\\ 
4000 & 0.25 & 4.91869$\times 10^{2}$ & 623.840 & 3790.407 & 1.06424$\times 10^{1}$ & -5.39636 & 0.12325\\ 
4000 & 0.30 & 3.49562$\times 10^{3}$ & 425.593 & 3800.649 & 2.52977$\times 10^{2}$ & -7.06489 & 0.61386\\ 
4000 & 0.35 & 4.87073$\times 10^{3}$ & 561.559 & 3803.419 & 2.05524$\times 10^{2}$ & -7.06022 & 0.74989\\ 
4000 & 0.40 & 2.13261$\times 10^{4}$ & 493.077 & 3843.718 & 2.66624$\times 10^{2}$ & -8.42267 & 1.11784\\ 
4000 & 0.45 & 2.21280$\times 10^{1}$ & 366.246 & 3812.175 & 1.01225$\times 10^{3}$ & -0.81612 & -2.05223\\ 

%% file: tab_h4lam0.dat
1600 & 0.05 & 7.22471$\times 10^{5}$ & 261.524 & 1548.861 & 1.74381$\times 10^{-1}$ & -10.16781 & 4.67876\\ 
1600 & 0.10 & 2.07119$\times 10^{6}$ & 268.846 & 1546.786 & 3.15229$\times 10^{1}$ & -9.83374 & 4.40196\\ 
1600 & 0.15 & 2.11243$\times 10^{6}$ & 291.078 & 1540.869 & 7.29639$\times 10^{-2}$ & -9.00341 & 3.99077\\ 
1600 & 0.20 & 8.60864$\times 10^{4}$ & 320.473 & 1518.118 & 3.48798$\times 10^{-2}$ & -5.16218 & 1.55340\\ 
1600 & 0.25 & 4.40812$\times 10^{6}$ & 329.067 & 1543.481 & 1.04830$\times 10^{-1}$ & -8.72852 & 3.93547\\ 
1600 & 0.30 & 4.12705$\times 10^{5}$ & 355.997 & 1527.036 & 3.22511$\times 10^{-1}$ & -5.93029 & 2.10857\\ 
1600 & 0.35 & 1.72044$\times 10^{6}$ & 384.237 & 1524.828 & 6.08628 & -7.02556 & 3.10563\\ 
1600 & 0.40 & 3.70595$\times 10^{6}$ & 405.785 & 1528.178 & 5.11102$\times 10^{-2}$ & -7.55400 & 3.47877\\ 
1600 & 0.45 & 4.99712$\times 10^{6}$ & 443.776 & 1520.235 & 4.32108$\times 10^{1}$ & -7.56885 & 3.71086\\ 
2000 & 0.05 & 4.35505$\times 10^{3}$ & 277.945 & 1919.574 & 7.47707$\times 10^{1}$ & -6.20024 & 1.21594\\ 
2000 & 0.10 & 1.09272$\times 10^{4}$ & 273.862 & 1913.306 & 7.75026$\times 10^{1}$ & -5.70715 & 0.93570\\ 
2000 & 0.15 & 3.98160$\times 10^{4}$ & 319.899 & 1915.552 & 4.38707$\times 10^{-2}$ & -6.20363 & 1.30892\\ 
2000 & 0.20 & 3.26680$\times 10^{4}$ & 318.109 & 1908.711 & 6.48833$\times 10^{1}$ & -5.38423 & 0.87602\\ 
2000 & 0.25 & 7.34075$\times 10^{4}$ & 345.473 & 1906.653 & 7.02021$\times 10^{1}$ & -5.75907 & 1.18835\\ 
2000 & 0.30 & 1.01296$\times 10^{5}$ & 344.519 & 1906.232 & 8.46143$\times 10^{1}$ & -5.71224 & 1.20405\\ 
2000 & 0.35 & 1.08614$\times 10^{5}$ & 381.387 & 1898.578 & 7.89286$\times 10^{1}$ & -5.47725 & 1.11100\\ 
2000 & 0.40 & 1.43755$\times 10^{5}$ & 463.510 & 1899.285 & 5.29813 & -5.45658 & 1.31249\\ 
2000 & 0.45 & 3.63814$\times 10^{5}$ & 515.704 & 1917.195 & 5.90913 & -6.14050 & 1.82965\\ 
2400 & 0.05 & 7.98471$\times 10^{2}$ & 311.333 & 2299.673 & 7.74125$\times 10^{1}$ & -5.49934 & 0.58689\\ 
2400 & 0.10 & 2.71015$\times 10^{3}$ & 268.995 & 2284.975 & 1.15770$\times 10^{2}$ & -5.32861 & 0.39071\\ 
2400 & 0.15 & 7.57449$\times 10^{3}$ & 309.078 & 2296.197 & 9.52726$\times 10^{1}$ & -5.57612 & 0.54633\\ 
2400 & 0.20 & 8.26778$\times 10^{3}$ & 294.482 & 2285.577 & 1.37785$\times 10^{2}$ & -5.08423 & 0.23697\\ 
2400 & 0.25 & 1.18606$\times 10^{4}$ & 361.258 & 2280.840 & 1.03076$\times 10^{2}$ & -4.99099 & 0.34222\\ 
2400 & 0.30 & 1.21103$\times 10^{4}$ & 406.246 & 2280.300 & 1.00911$\times 10^{2}$ & -4.61060 & 0.21267\\ 
2400 & 0.35 & 4.03849$\times 10^{4}$ & 414.390 & 2286.582 & 1.15030$\times 10^{2}$ & -5.54181 & 0.77573\\ 
2400 & 0.40 & 2.62471$\times 10^{4}$ & 478.025 & 2272.751 & 1.06082$\times 10^{2}$ & -4.81027 & 0.40772\\ 
2400 & 0.45 & 4.54467$\times 10^{4}$ & 534.710 & 2264.979 & 8.72913$\times 10^{1}$ & -5.11589 & 0.70109\\ 
2800 & 0.05 & 7.99366$\times 10^{1}$ & 414.681 & 2659.754 & 6.63948 & -4.09824 & -0.16491\\ 
2800 & 0.10 & 6.89306$\times 10^{2}$ & 329.295 & 2660.279 & 1.25516$\times 10^{2}$ & -4.92936 & 0.03736\\ 
2800 & 0.15 & 1.44819$\times 10^{3}$ & 334.034 & 2655.127 & 1.35885$\times 10^{2}$ & -4.84239 & 0.00717\\ 
2800 & 0.20 & 3.05854$\times 10^{3}$ & 330.974 & 2669.421 & 1.61384$\times 10^{2}$ & -5.02812 & 0.07033\\ 
2800 & 0.25 & 3.06514$\times 10^{3}$ & 458.479 & 2656.949 & 3.60328$\times 10^{1}$ & -4.55270 & 0.04771\\ 
2800 & 0.30 & 1.25037$\times 10^{4}$ & 404.238 & 2665.745 & 1.56778$\times 10^{2}$ & -5.64260 & 0.52216\\ 
2800 & 0.35 & 1.43054$\times 10^{4}$ & 418.868 & 2659.954 & 1.71818$\times 10^{2}$ & -5.48547 & 0.40376\\ 
2800 & 0.40 & 2.09166$\times 10^{4}$ & 460.904 & 2665.642 & 1.62621$\times 10^{2}$ & -5.61546 & 0.48066\\ 
2800 & 0.45 & 4.56375$\times 10^{4}$ & 524.035 & 2676.133 & 1.62444$\times 10^{2}$ & -6.16940 & 0.86434\\ 
3200 & 0.05 & 5.19532$\times 10^{1}$ & 327.093 & 3032.071 & 1.62118$\times 10^{2}$ & -4.64117 & -0.25025\\ 
3200 & 0.10 & 1.65045$\times 10^{2}$ & 372.236 & 3029.917 & 1.33993$\times 10^{2}$ & -4.36593 & -0.28234\\ 
3200 & 0.15 & 7.38357$\times 10^{2}$ & 319.697 & 3039.065 & 1.90814$\times 10^{2}$ & -5.12650 & -0.08700\\ 
3200 & 0.20 & 1.41773$\times 10^{3}$ & 348.891 & 3028.817 & 1.94602$\times 10^{2}$ & -5.19704 & -0.00471\\ 
3200 & 0.25 & 2.00614$\times 10^{3}$ & 459.088 & 3041.581 & 1.30894$\times 10^{2}$ & -5.04587 & 0.15280\\ 
3200 & 0.30 & 2.87860$\times 10^{3}$ & 460.310 & 3028.983 & 1.74669$\times 10^{2}$ & -5.04724 & 0.05934\\ 
3200 & 0.35 & 1.17676$\times 10^{4}$ & 472.940 & 3055.053 & 1.61058$\times 10^{2}$ & -6.24580 & 0.59337\\ 
3200 & 0.40 & 1.33870$\times 10^{4}$ & 451.069 & 3041.077 & 2.36791$\times 10^{2}$ & -6.07216 & 0.48080\\ 
3200 & 0.45 & 4.25546$\times 10^{4}$ & 528.994 & 3072.086 & 2.06203$\times 10^{2}$ & -7.06165 & 1.04056\\ 
3600 & 0.05 & 9.58131 & 514.402 & 3403.385 & 1.50979$\times 10^{1}$ & -3.74820 & -0.48711\\ 
3600 & 0.10 & 1.09692$\times 10^{2}$ & 381.489 & 3413.096 & 1.77672$\times 10^{2}$ & -4.88150 & -0.14278\\ 
3600 & 0.15 & 2.79017$\times 10^{2}$ & 373.126 & 3421.325 & 1.99754$\times 10^{2}$ & -4.99646 & -0.13504\\ 
3600 & 0.20 & 7.45370$\times 10^{2}$ & 438.439 & 3403.592 & 1.75254$\times 10^{2}$ & -5.41304 & 0.15901\\ 
3600 & 0.25 & 7.33213$\times 10^{2}$ & 589.579 & 3402.359 & 1.43299$\times 10^{1}$ & -4.89556 & 0.04284\\ 
3600 & 0.30 & 3.71762$\times 10^{3}$ & 495.444 & 3413.360 & 1.68682$\times 10^{2}$ & -6.23221 & 0.53117\\ 
3600 & 0.35 & 4.17688$\times 10^{3}$ & 561.193 & 3406.906 & 1.58027$\times 10^{2}$ & -5.98192 & 0.45757\\ 
3600 & 0.40 & 1.54159$\times 10^{4}$ & 689.369 & 3426.519 & 1.09322$\times 10^{1}$ & -7.08610 & 1.07841\\ 
3600 & 0.45 & 3.87852$\times 10^{4}$ & 519.889 & 3458.267 & 2.51578$\times 10^{2}$ & -7.92126 & 1.13583\\ 
4000 & 0.05 & 8.64212 & 362.063 & 3779.525 & 2.29793$\times 10^{2}$ & -4.60167 & -0.43791\\ 
4000 & 0.10 & 2.04269$\times 10^{1}$ & 552.916 & 3781.710 & 1.12972$\times 10^{1}$ & -3.99881 & -0.45205\\ 
4000 & 0.15 & 1.09514$\times 10^{2}$ & 426.865 & 3789.181 & 2.12923$\times 10^{2}$ & -4.92267 & -0.21230\\ 
4000 & 0.20 & 1.63621$\times 10^{2}$ & 618.757 & 3787.468 & 1.11946$\times 10^{1}$ & -4.70702 & -0.11656\\ 
4000 & 0.25 & 1.23778$\times 10^{3}$ & 442.761 & 3799.985 & 2.30898$\times 10^{2}$ & -6.38462 & 0.36140\\ 
4000 & 0.30 & 4.78918$\times 10^{3}$ & 423.961 & 3822.639 & 2.53584$\times 10^{2}$ & -7.43131 & 0.74585\\ 
4000 & 0.35 & 2.22022$\times 10^{3}$ & 708.023 & 3793.527 & 1.68849$\times 10^{1}$ & -6.23818 & 0.47919\\ 
4000 & 0.40 & 1.21003$\times 10^{4}$ & 562.075 & 3832.288 & 2.70009$\times 10^{2}$ & -7.80832 & 0.95964\\ 
4000 & 0.45 & 4.12952$\times 10^{4}$ & 481.870 & 3843.512 & 3.24035$\times 10^{2}$ & -8.87296 & 1.28449\\ 

%% file: tab_h1g0.dat
1600 & 0.05 & 3.38630$\times 10^{5}$ & 273.491 & 1549.757 & 1.09411$\times 10^{-2}$ & -9.95714 & 4.41330\\ 
1600 & 0.10 & 1.12530$\times 10^{6}$ & 284.372 & 1547.002 & 1.73230$\times 10^{-1}$ & -9.73847 & 4.45723\\ 
1600 & 0.15 & 2.98621$\times 10^{6}$ & 297.111 & 1546.167 & 2.60991 & -9.90257 & 4.57945\\ 
1600 & 0.20 & 2.32618$\times 10^{6}$ & 311.388 & 1543.277 & 6.38288$\times 10^{-1}$ & -9.05902 & 4.07785\\ 
1600 & 0.25 & 5.28996$\times 10^{5}$ & 356.664 & 1529.356 & 9.28856$\times 10^{-1}$ & -7.09259 & 2.91046\\ 
1600 & 0.30 & 2.02690$\times 10^{6}$ & 414.445 & 1532.818 & 1.01555 & -8.03318 & 3.83000\\ 
1600 & 0.35 & 8.63365$\times 10^{5}$ & 439.023 & 1518.524 & 1.52358 & -6.83339 & 3.21348\\ 
1600 & 0.40 & 1.55652$\times 10^{6}$ & 507.714 & 1519.528 & 4.36909$\times 10^{-1}$ & -7.13384 & 3.54007\\ 
1600 & 0.45 & 1.78774$\times 10^{6}$ & 558.898 & 1507.142 & 3.07215$\times 10^{-1}$ & -7.03292 & 3.59298\\ 
2000 & 0.05 & 2.21953$\times 10^{3}$ & 283.601 & 1918.759 & 5.51110$\times 10^{1}$ & -6.03719 & 1.11727\\ 
2000 & 0.10 & 1.35137$\times 10^{4}$ & 273.508 & 1919.628 & 7.47989$\times 10^{1}$ & -6.49663 & 1.35314\\ 
2000 & 0.15 & 2.06325$\times 10^{4}$ & 336.546 & 1920.426 & 4.69462 & -6.07970 & 1.29375\\ 
2000 & 0.20 & 1.49975$\times 10^{4}$ & 298.465 & 1902.048 & 1.03528$\times 10^{2}$ & -5.15059 & 0.70149\\ 
2000 & 0.25 & 3.59638$\times 10^{4}$ & 325.094 & 1912.701 & 1.06304$\times 10^{2}$ & -5.60558 & 1.03251\\ 
2000 & 0.30 & 6.97361$\times 10^{4}$ & 408.102 & 1916.164 & 4.34773 & -5.89053 & 1.33345\\ 
2000 & 0.35 & 8.12983$\times 10^{4}$ & 466.877 & 1900.340 & 1.15069$\times 10^{1}$ & -5.70185 & 1.38174\\ 
2000 & 0.40 & 9.72522$\times 10^{4}$ & 573.844 & 1892.438 & 1.57794 & -5.56130 & 1.57844\\ 
2000 & 0.45 & 2.59981$\times 10^{5}$ & 574.148 & 1928.985 & 6.70369$\times 10^{1}$ & -6.38070 & 1.91570\\ 
2400 & 0.05 & 2.76182$\times 10^{1}$ & 372.769 & 2268.313 & 7.73057$\times 10^{-2}$ & -2.57987 & -0.91755\\ 
2400 & 0.10 & 6.84684$\times 10^{2}$ & 338.406 & 2280.030 & 6.00796$\times 10^{1}$ & -4.42790 & 0.02632\\ 
2400 & 0.15 & 3.38227$\times 10^{3}$ & 297.440 & 2295.105 & 1.05922$\times 10^{2}$ & -5.30579 & 0.32251\\ 
2400 & 0.20 & 6.48120$\times 10^{3}$ & 355.433 & 2295.967 & 9.69624$\times 10^{1}$ & -5.37400 & 0.55380\\ 
2400 & 0.25 & 2.02793$\times 10^{4}$ & 380.192 & 2306.561 & 9.34515$\times 10^{1}$ & -6.08384 & 0.98010\\ 
2400 & 0.30 & 2.29152$\times 10^{4}$ & 401.847 & 2291.080 & 1.29997$\times 10^{2}$ & -5.84090 & 0.86003\\ 
2400 & 0.35 & 1.71312$\times 10^{4}$ & 557.853 & 2282.679 & 5.49647 & -5.16012 & 0.76220\\ 
2400 & 0.40 & 4.25894$\times 10^{4}$ & 538.422 & 2293.188 & 1.08403$\times 10^{2}$ & -5.86726 & 1.04146\\ 
2400 & 0.45 & 6.97560$\times 10^{4}$ & 665.235 & 2290.991 & 2.05692 & -6.09851 & 1.37050\\ 
2800 & 0.05 & 6.59744$\times 10^{1}$ & 297.030 & 2658.200 & 1.39159$\times 10^{2}$ & -4.49309 & -0.31930\\ 
2800 & 0.10 & 4.02940$\times 10^{2}$ & 417.669 & 2670.717 & 5.17397 & -4.88655 & 0.23576\\ 
2800 & 0.15 & 5.56311$\times 10^{2}$ & 444.785 & 2653.833 & 5.64693 & -4.38572 & -0.00415\\ 
2800 & 0.20 & 2.43438$\times 10^{3}$ & 329.094 & 2665.414 & 1.60913$\times 10^{2}$ & -5.37478 & 0.22541\\ 
2800 & 0.25 & 5.38415$\times 10^{3}$ & 369.633 & 2665.801 & 1.71251$\times 10^{2}$ & -5.73452 & 0.46508\\ 
2800 & 0.30 & 1.79229$\times 10^{2}$ & 659.302 & 2556.159 & 3.73320$\times 10^{-1}$ & -1.71914 & -1.23592\\ 
2800 & 0.35 & 1.05995$\times 10^{4}$ & 524.597 & 2667.605 & 1.57665$\times 10^{2}$ & -5.66616 & 0.68458\\ 
2800 & 0.40 & 6.34529$\times 10^{2}$ & 809.514 & 2553.878 & 5.92711$\times 10^{-2}$ & -2.37558 & -0.75458\\ 
2800 & 0.45 & 5.85787$\times 10^{4}$ & 548.858 & 2709.360 & 2.27062$\times 10^{2}$ & -7.03406 & 1.24782\\ 
3200 & 0.05 & 1.93562$\times 10^{1}$ & 453.534 & 3034.438 & 2.20013 & -4.13108 & -0.24153\\ 
3200 & 0.10 & 1.65120$\times 10^{2}$ & 348.334 & 3035.386 & 1.68005$\times 10^{2}$ & -4.96464 & -0.09652\\ 
3200 & 0.15 & 2.46862$\times 10^{2}$ & 496.400 & 3038.515 & 2.19229 & -4.48388 & -0.06645\\ 
3200 & 0.20 & 1.08979$\times 10^{3}$ & 401.542 & 3035.259 & 1.62011$\times 10^{2}$ & -5.45358 & 0.21655\\ 
3200 & 0.25 & 5.76739$\times 10^{2}$ & 595.962 & 3014.854 & 1.00286$\times 10^{1}$ & -4.29130 & -0.14619\\ 
3200 & 0.30 & 7.00933$\times 10^{1}$ & 778.269 & 2950.095 & 4.70260$\times 10^{-1}$ & -1.73439 & -1.27694\\ 
3200 & 0.35 & 4.12125$\times 10^{3}$ & 702.643 & 3023.723 & 5.02951 & -5.63244 & 0.56299\\ 
3200 & 0.40 & 9.92050$\times 10^{3}$ & 636.208 & 3049.896 & 2.19994$\times 10^{2}$ & -6.31735 & 0.76767\\ 
3200 & 0.45 & 1.84701$\times 10^{4}$ & 799.165 & 3044.615 & 1.42380$\times 10^{2}$ & -6.69027 & 1.08946\\ 
3600 & 0.05 & 1.50667$\times 10^{1}$ & 367.089 & 3411.342 & 1.78689$\times 10^{2}$ & -4.80561 & -0.19737\\ 
3600 & 0.10 & 4.23942$\times 10^{1}$ & 474.500 & 3409.443 & 7.85452$\times 10^{1}$ & -4.44915 & -0.21250\\ 
3600 & 0.15 & 5.60645$\times 10^{1}$ & 572.187 & 3390.615 & 4.92089 & -3.87421 & -0.48165\\ 
3600 & 0.20 & 5.15207$\times 10^{2}$ & 406.564 & 3413.827 & 2.29368$\times 10^{2}$ & -5.59177 & 0.12257\\ 
3600 & 0.25 & 8.70832$\times 10^{2}$ & 645.798 & 3419.739 & 9.93860 & -5.64058 & 0.39936\\ 
3600 & 0.30 & 1.77505$\times 10^{3}$ & 699.554 & 3423.577 & 2.89978 & -5.99481 & 0.57158\\ 
3600 & 0.35 & 3.30752$\times 10^{3}$ & 738.341 & 3404.719 & 9.98038 & -6.31467 & 0.69214\\ 
3600 & 0.40 & 9.61949$\times 10^{3}$ & 793.028 & 3442.838 & 1.31419 & -7.15104 & 1.04170\\ 
3600 & 0.45 & 1.31058$\times 10^{4}$ & 894.687 & 3437.337 & 1.00726$\times 10^{2}$ & -7.23459 & 1.12404\\ 
4000 & 0.05 & 5.17696 & 382.467 & 3782.988 & 2.19348$\times 10^{2}$ & -4.58906 & -0.38007\\ 
4000 & 0.10 & 1.70025$\times 10^{1}$ & 562.946 & 3794.446 & 1.28497$\times 10^{-1}$ & -4.37029 & -0.29736\\ 
4000 & 0.15 & 3.39362$\times 10^{1}$ & 628.036 & 3766.378 & 5.12344 & -4.20579 & -0.30615\\ 
4000 & 0.20 & 1.41861$\times 10^{1}$ & 788.761 & 3750.000 & 2.04301$\times 10^{-1}$ & -2.73831 & -1.02880\\ 
4000 & 0.25 & 3.47830$\times 10^{2}$ & 702.052 & 3771.040 & 1.00286$\times 10^{1}$ & -5.54813 & 0.21768\\ 
4000 & 0.30 & 9.62600$\times 10^{2}$ & 787.579 & 3784.294 & 1.67365 & -6.21644 & 0.54180\\ 
4000 & 0.35 & 2.70986$\times 10^{3}$ & 802.826 & 3802.423 & 4.43008 & -6.96405 & 0.81097\\ 
4000 & 0.40 & 6.01189$\times 10^{3}$ & 896.294 & 3806.740 & 1.30173 & -7.56984 & 1.03734\\ 
4000 & 0.45 & 7.01371$\times 10^{3}$ & 1086.105 & 3823.621 & 1.07115$\times 10^{1}$ & -7.50088 & 1.01614\\ 

%% file: tab_h2g0.dat
1600 & 0.05 & 1.71887$\times 10^{5}$ & 266.804 & 1552.146 & 3.80328$\times 10^{-1}$ & -9.26805 & 3.87058\\ 
1600 & 0.10 & 5.19261$\times 10^{4}$ & 283.560 & 1532.086 & 9.83644$\times 10^{-3}$ & -6.59533 & 2.38065\\ 
1600 & 0.15 & 1.02629$\times 10^{6}$ & 296.175 & 1549.389 & 2.43648$\times 10^{-3}$ & -8.81629 & 3.82032\\ 
1600 & 0.20 & 2.29493$\times 10^{5}$ & 326.132 & 1525.132 & 3.82526$\times 10^{-1}$ & -6.70108 & 2.54281\\ 
1600 & 0.25 & 1.19460$\times 10^{6}$ & 346.035 & 1529.321 & 3.15404$\times 10^{1}$ & -7.91232 & 3.45775\\ 
1600 & 0.30 & 3.60679$\times 10^{5}$ & 396.829 & 1517.493 & 2.04181 & -6.29540 & 2.63030\\ 
1600 & 0.35 & 1.95418$\times 10^{6}$ & 449.179 & 1533.890 & 1.47218$\times 10^{-2}$ & -7.70149 & 3.63592\\ 
1600 & 0.40 & 1.37037$\times 10^{6}$ & 532.747 & 1513.991 & 4.03759 & -6.98750 & 3.51866\\ 
1600 & 0.45 & 2.26351$\times 10^{6}$ & 541.513 & 1513.512 & 4.86235$\times 10^{-1}$ & -7.25564 & 3.72507\\ 
2000 & 0.05 & 1.27719$\times 10^{3}$ & 276.536 & 1913.254 & 7.82957$\times 10^{1}$ & -5.46148 & 0.82683\\ 
2000 & 0.10 & 6.77471$\times 10^{1}$ & 362.581 & 1868.981 & 4.62106$\times 10^{-2}$ & -1.06832 & -1.62934\\ 
2000 & 0.15 & 1.38323$\times 10^{4}$ & 336.663 & 1906.353 & 1.39983$\times 10^{1}$ & -5.63581 & 1.08759\\ 
2000 & 0.20 & 3.48834$\times 10^{4}$ & 365.236 & 1910.752 & 5.10019 & -5.98936 & 1.39631\\ 
2000 & 0.25 & 5.72775$\times 10^{4}$ & 339.600 & 1911.992 & 8.27745$\times 10^{1}$ & -6.09012 & 1.29559\\ 
2000 & 0.30 & 7.44500$\times 10^{4}$ & 393.038 & 1912.306 & 7.31368$\times 10^{1}$ & -5.96101 & 1.39842\\ 
2000 & 0.35 & 5.84644$\times 10^{4}$ & 487.655 & 1898.756 & 2.85123$\times 10^{1}$ & -5.34587 & 1.20385\\ 
2000 & 0.40 & 1.90144$\times 10^{5}$ & 569.310 & 1913.505 & 8.83831$\times 10^{-1}$ & -6.27267 & 1.92906\\ 
2000 & 0.45 & 5.43384$\times 10^{4}$ & 645.666 & 1890.961 & 8.38206$\times 10^{-1}$ & -4.71444 & 1.16712\\ 
2400 & 0.05 & 5.37168$\times 10^{2}$ & 280.088 & 2294.186 & 1.02472$\times 10^{2}$ & -5.66324 & 0.56621\\ 
2400 & 0.10 & 1.88890$\times 10^{3}$ & 292.382 & 2292.577 & 1.05147$\times 10^{2}$ & -5.51003 & 0.53767\\ 
2400 & 0.15 & 1.86916$\times 10^{3}$ & 392.341 & 2279.426 & 5.49679 & -4.65407 & 0.28131\\ 
2400 & 0.20 & 3.93111$\times 10^{3}$ & 335.391 & 2275.744 & 1.18549$\times 10^{2}$ & -4.85866 & 0.18788\\ 
2400 & 0.25 & 1.21202$\times 10^{4}$ & 371.786 & 2290.595 & 1.24540$\times 10^{2}$ & -5.55269 & 0.64554\\ 
2400 & 0.30 & 1.23655$\times 10^{4}$ & 428.401 & 2289.473 & 1.15821$\times 10^{2}$ & -5.18564 & 0.54813\\ 
2400 & 0.35 & 2.49388$\times 10^{4}$ & 518.589 & 2281.783 & 8.60692$\times 10^{1}$ & -5.56302 & 0.96164\\ 
2400 & 0.40 & 5.36880$\times 10^{4}$ & 594.434 & 2303.149 & 1.81313 & -6.09594 & 1.23979\\ 
2400 & 0.45 & 6.26662$\times 10^{4}$ & 690.873 & 2294.062 & 9.41360$\times 10^{1}$ & -5.97086 & 1.37238\\ 
2800 & 0.05 & 5.78706$\times 10^{1}$ & 408.856 & 2661.673 & 1.62277 & -4.32451 & -0.09269\\ 
2800 & 0.10 & 1.80086$\times 10^{2}$ & 418.365 & 2646.004 & 7.47397 & -4.05888 & -0.22052\\ 
2800 & 0.15 & 8.96101$\times 10^{2}$ & 434.295 & 2669.675 & 9.07279 & -4.89253 & 0.20762\\ 
2800 & 0.20 & 1.46927$\times 10^{3}$ & 472.731 & 2660.955 & 5.94141 & -4.78727 & 0.19150\\ 
2800 & 0.25 & 2.64937$\times 10^{3}$ & 510.572 & 2654.867 & 5.13110 & -4.91616 & 0.35001\\ 
2800 & 0.30 & 6.32701$\times 10^{3}$ & 441.146 & 2670.388 & 1.69019$\times 10^{2}$ & -5.51656 & 0.39329\\ 
2800 & 0.35 & 9.54524$\times 10^{3}$ & 616.327 & 2649.467 & 7.16162$\times 10^{-2}$ & -5.54899 & 0.71668\\ 
2800 & 0.40 & 1.42616$\times 10^{4}$ & 643.059 & 2662.391 & 1.36427$\times 10^{2}$ & -5.69222 & 0.79505\\ 
2800 & 0.45 & 3.12298$\times 10^{4}$ & 779.067 & 2700.072 & 1.68051 & -6.28460 & 1.11991\\ 
3200 & 0.05 & 2.02864$\times 10^{1}$ & 465.864 & 3030.027 & 1.83473$\times 10^{-1}$ & -4.16929 & -0.18236\\ 
3200 & 0.10 & 9.82148$\times 10^{1}$ & 396.659 & 3029.968 & 1.23992$\times 10^{2}$ & -4.38838 & -0.24654\\ 
3200 & 0.15 & 2.02446$\times 10^{2}$ & 509.818 & 3032.444 & 1.00399$\times 10^{1}$ & -4.27369 & -0.13497\\ 
3200 & 0.20 & 4.04878$\times 10^{2}$ & 535.512 & 3027.933 & 7.18709 & -4.39688 & -0.10830\\ 
3200 & 0.25 & 2.78279$\times 10^{3}$ & 463.030 & 3048.710 & 1.50522$\times 10^{2}$ & -5.99517 & 0.49177\\ 
3200 & 0.30 & 3.08515$\times 10^{3}$ & 638.072 & 3033.206 & 1.12689$\times 10^{1}$ & -5.63909 & 0.59223\\ 
3200 & 0.35 & 4.16549$\times 10^{3}$ & 725.028 & 3032.567 & 1.02021$\times 10^{-2}$ & -5.62205 & 0.60431\\ 
3200 & 0.40 & 2.68592$\times 10^{4}$ & 554.152 & 3082.207 & 2.15230$\times 10^{2}$ & -7.41210 & 1.15787\\ 
3200 & 0.45 & 1.02845$\times 10^{5}$ & 460.577 & 3129.373 & 2.98148$\times 10^{2}$ & -8.67939 & 1.52798\\ 
3600 & 0.05 & 7.70261 & 519.734 & 3411.109 & 5.24471 & -4.08693 & -0.31523\\ 
3600 & 0.10 & 7.94554$\times 10^{1}$ & 341.434 & 3405.078 & 2.10733$\times 10^{2}$ & -5.09566 & -0.13925\\ 
3600 & 0.15 & 1.89137$\times 10^{2}$ & 389.034 & 3402.391 & 2.20977$\times 10^{2}$ & -5.17022 & -0.08975\\ 
3600 & 0.20 & 2.24845$\times 10^{2}$ & 605.387 & 3398.702 & 1.17278 & -4.67205 & -0.03972\\ 
3600 & 0.25 & 8.73285$\times 10^{2}$ & 514.405 & 3413.723 & 1.80299$\times 10^{2}$ & -5.66116 & 0.24032\\ 
3600 & 0.30 & 5.33930$\times 10^{3}$ & 486.467 & 3439.010 & 2.47584$\times 10^{2}$ & -7.22246 & 0.87564\\ 
3600 & 0.35 & 1.16349$\times 10^{4}$ & 493.622 & 3440.181 & 2.40164$\times 10^{2}$ & -7.70719 & 1.04787\\ 
3600 & 0.40 & 5.75392$\times 10^{3}$ & 850.145 & 3427.450 & 1.10229$\times 10^{1}$ & -6.62114 & 0.81555\\ 
3600 & 0.45 & 1.65576$\times 10^{4}$ & 783.363 & 3461.780 & 2.02893$\times 10^{2}$ & -7.55571 & 1.10540\\ 
4000 & 0.05 & 9.13344 & 381.622 & 3789.547 & 2.00416$\times 10^{2}$ & -5.17271 & -0.07829\\ 
4000 & 0.10 & 1.64155$\times 10^{1}$ & 568.700 & 3791.270 & 9.68369$\times 10^{-1}$ & -4.31143 & -0.29259\\ 
4000 & 0.15 & 7.92659$\times 10^{1}$ & 431.948 & 3778.164 & 2.25907$\times 10^{2}$ & -5.14046 & -0.16394\\ 
4000 & 0.20 & 3.06745$\times 10^{2}$ & 480.655 & 3781.655 & 2.04211$\times 10^{2}$ & -5.90642 & 0.25400\\ 
4000 & 0.25 & 3.80032$\times 10^{2}$ & 720.398 & 3771.032 & 1.00768$\times 10^{1}$ & -5.62714 & 0.30185\\ 
4000 & 0.30 & 4.46817$\times 10^{3}$ & 491.297 & 3823.530 & 2.43861$\times 10^{2}$ & -7.90173 & 0.98986\\ 
4000 & 0.35 & 4.17238$\times 10^{3}$ & 716.423 & 3834.021 & 1.44441$\times 10^{2}$ & -7.48900 & 0.91814\\ 
4000 & 0.40 & 6.60943$\times 10^{3}$ & 886.825 & 3811.766 & 8.52635$\times 10^{1}$ & -7.69392 & 1.07113\\ 
4000 & 0.45 & 1.40095$\times 10^{4}$ & 901.722 & 3802.722 & 8.10194$\times 10^{-2}$ & -8.24594 & 1.25383\\ 

%% file: dijet_arXiv_v3.bbl
\begin{thebibliography}{39}%
\makeatletter
\providecommand \@ifxundefined [1]{%
 \@ifx{#1\undefined}
}%
\providecommand \@ifnum [1]{%
 \ifnum #1\expandafter \@firstoftwo
 \else \expandafter \@secondoftwo
 \fi
}%
\providecommand \@ifx [1]{%
 \ifx #1\expandafter \@firstoftwo
 \else \expandafter \@secondoftwo
 \fi
}%
\providecommand \natexlab [1]{#1}%
\providecommand \enquote  [1]{``#1''}%
\providecommand \bibnamefont  [1]{#1}%
\providecommand \bibfnamefont [1]{#1}%
\providecommand \citenamefont [1]{#1}%
\providecommand \href@noop [0]{\@secondoftwo}%
\providecommand \href [0]{\begingroup \@sanitize@url \@href}%
\providecommand \@href[1]{\@@startlink{#1}\@@href}%
\providecommand \@@href[1]{\endgroup#1\@@endlink}%
\providecommand \@sanitize@url [0]{\catcode `\\12\catcode `\$12\catcode
  `\&12\catcode `\#12\catcode `\^12\catcode `\_12\catcode `\%12\relax}%
\providecommand \@@startlink[1]{}%
\providecommand \@@endlink[0]{}%
\providecommand \url  [0]{\begingroup\@sanitize@url \@url }%
\providecommand \@url [1]{\endgroup\@href {#1}{\urlprefix }}%
\providecommand \urlprefix  [0]{URL }%
\providecommand \Eprint [0]{\href }%
\providecommand \doibase [0]{http://dx.doi.org/}%
\providecommand \selectlanguage [0]{\@gobble}%
\providecommand \bibinfo  [0]{\@secondoftwo}%
\providecommand \bibfield  [0]{\@secondoftwo}%
\providecommand \translation [1]{[#1]}%
\providecommand \BibitemOpen [0]{}%
\providecommand \bibitemStop [0]{}%
\providecommand \bibitemNoStop [0]{.\EOS\space}%
\providecommand \EOS [0]{\spacefactor3000\relax}%
\providecommand \BibitemShut  [1]{\csname bibitem#1\endcsname}%
\let\auto@bib@innerbib\@empty
\bibitem [{\citenamefont {Aaboud}\ \emph {et~al.}(2017)\citenamefont {Aaboud}
  \emph {et~al.}}]{Aaboud:2017yvp}%
  \BibitemOpen
  \bibfield  {author} {\bibinfo {author} {\bibfnamefont {M.}~\bibnamefont
  {Aaboud}} \emph {et~al.} (\bibinfo {collaboration} {ATLAS}),\ }\href
  {\doibase 10.1103/PhysRevD.96.052004} {\bibfield  {journal} {\bibinfo
  {journal} {Phys. Rev.}\ }\textbf {\bibinfo {volume} {D96}},\ \bibinfo {pages}
  {052004} (\bibinfo {year} {2017})},\ \Eprint
  {http://arxiv.org/abs/1703.09127} {arXiv:1703.09127 [hep-ex]} \BibitemShut
  {NoStop}%
\bibitem [{CMS(2017)}]{CMS-PAS-EXO-16-056}%
  \BibitemOpen
  \href {https://cds.cern.ch/record/2256873} {\emph {\bibinfo {title}
  {{Searches for dijet resonances in pp collisions at
  $\sqrt{s}=13~\mathrm{TeV}$ using data collected in 2016.}}}},\ \bibinfo
  {type} {Tech. Rep.}\ \bibinfo {number} {CMS-PAS-EXO-16-056}\ (\bibinfo
  {institution} {CERN},\ \bibinfo {address} {Geneva},\ \bibinfo {year}
  {2017})\BibitemShut {NoStop}%
\bibitem [{\citenamefont {Sirunyan}\ \emph
  {et~al.}(2018{\natexlab{a}})\citenamefont {Sirunyan} \emph
  {et~al.}}]{Sirunyan:2018xlo}%
  \BibitemOpen
  \bibfield  {author} {\bibinfo {author} {\bibfnamefont {A.~M.}\ \bibnamefont
  {Sirunyan}} \emph {et~al.} (\bibinfo {collaboration} {CMS}),\ }\href
  {\doibase 10.1007/JHEP08(2018)130} {\bibfield  {journal} {\bibinfo  {journal}
  {JHEP}\ }\textbf {\bibinfo {volume} {08}},\ \bibinfo {pages} {130} (\bibinfo
  {year} {2018}{\natexlab{a}})},\ \Eprint {http://arxiv.org/abs/1806.00843}
  {arXiv:1806.00843 [hep-ex]} \BibitemShut {NoStop}%
\bibitem [{CMS(2018)}]{CMS-PAS-EXO-17-026}%
  \BibitemOpen
  \href {https://cds.cern.ch/record/2637847} {\emph {\bibinfo {title}
  {{Searches for dijet resonances in pp collisions at
  $\sqrt{s}=13~\mathrm{TeV}$ using the 2016 and 2017 datasets}}}},\ \bibinfo
  {type} {Tech. Rep.}\ \bibinfo {number} {CMS-PAS-EXO-17-026}\ (\bibinfo
  {institution} {CERN},\ \bibinfo {address} {Geneva},\ \bibinfo {year}
  {2018})\BibitemShut {NoStop}%
\bibitem [{\citenamefont {Aaboud}\ \emph {et~al.}(2018)\citenamefont {Aaboud}
  \emph {et~al.}}]{Aaboud:2018fzt}%
  \BibitemOpen
  \bibfield  {author} {\bibinfo {author} {\bibfnamefont {M.}~\bibnamefont
  {Aaboud}} \emph {et~al.} (\bibinfo {collaboration} {ATLAS}),\ }\href
  {\doibase 10.1103/PhysRevLett.121.081801} {\bibfield  {journal} {\bibinfo
  {journal} {Phys. Rev. Lett.}\ }\textbf {\bibinfo {volume} {121}},\ \bibinfo
  {pages} {081801} (\bibinfo {year} {2018})},\ \Eprint
  {http://arxiv.org/abs/1804.03496} {arXiv:1804.03496 [hep-ex]} \BibitemShut
  {NoStop}%
\bibitem [{\citenamefont {Sirunyan}\ \emph
  {et~al.}(2018{\natexlab{b}})\citenamefont {Sirunyan} \emph
  {et~al.}}]{Sirunyan:2018pas}%
  \BibitemOpen
  \bibfield  {author} {\bibinfo {author} {\bibfnamefont {A.~M.}\ \bibnamefont
  {Sirunyan}} \emph {et~al.} (\bibinfo {collaboration} {CMS}),\ }\href
  {\doibase 10.1103/PhysRevLett.120.201801} {\bibfield  {journal} {\bibinfo
  {journal} {Phys. Rev. Lett.}\ }\textbf {\bibinfo {volume} {120}},\ \bibinfo
  {pages} {201801} (\bibinfo {year} {2018}{\natexlab{b}})},\ \Eprint
  {http://arxiv.org/abs/1802.06149} {arXiv:1802.06149 [hep-ex]} \BibitemShut
  {NoStop}%
\bibitem [{ATL(2018)}]{ATLAS-CONF-2018-015}%
  \BibitemOpen
  \href {https://cds.cern.ch/record/2621126} {\emph {\bibinfo {title} {{Search
  for dijet resonances in events with an isolated lepton using $\sqrt{s} =
  13$~TeV proton--proton collision data collected by the ATLAS detector}}}},\
  \bibinfo {type} {Tech. Rep.}\ \bibinfo {number} {ATLAS-CONF-2018-015}\
  (\bibinfo  {institution} {CERN},\ \bibinfo {address} {Geneva},\ \bibinfo
  {year} {2018})\BibitemShut {NoStop}%
\bibitem [{\citenamefont {Aaboud}\ \emph {et~al.}(2019)\citenamefont {Aaboud}
  \emph {et~al.}}]{Aaboud:2018zba}%
  \BibitemOpen
  \bibfield  {author} {\bibinfo {author} {\bibfnamefont {M.}~\bibnamefont
  {Aaboud}} \emph {et~al.} (\bibinfo {collaboration} {ATLAS}),\ }\href
  {\doibase 10.1016/j.physletb.2018.09.062} {\bibfield  {journal} {\bibinfo
  {journal} {Phys. Lett.}\ }\textbf {\bibinfo {volume} {B788}},\ \bibinfo
  {pages} {316} (\bibinfo {year} {2019})},\ \Eprint
  {http://arxiv.org/abs/1801.08769} {arXiv:1801.08769 [hep-ex]} \BibitemShut
  {NoStop}%
\bibitem [{\citenamefont {Baum}\ \emph {et~al.}(2019)\citenamefont {Baum},
  \citenamefont {Catena},\ and\ \citenamefont {Krauss}}]{Baum:2018lua}%
  \BibitemOpen
  \bibfield  {author} {\bibinfo {author} {\bibfnamefont {S.}~\bibnamefont
  {Baum}}, \bibinfo {author} {\bibfnamefont {R.}~\bibnamefont {Catena}}, \ and\
  \bibinfo {author} {\bibfnamefont {M.~B.}\ \bibnamefont {Krauss}},\ }\href
  {\doibase 10.1007/JHEP07(2019)015} {\bibfield  {journal} {\bibinfo  {journal}
  {JHEP}\ }\textbf {\bibinfo {volume} {07}},\ \bibinfo {pages} {015} (\bibinfo
  {year} {2019})},\ \Eprint {http://arxiv.org/abs/1812.01594} {arXiv:1812.01594
  [hep-ph]} \BibitemShut {NoStop}%
\bibitem [{\citenamefont {Abdallah}\ \emph {et~al.}(2014)\citenamefont
  {Abdallah} \emph {et~al.}}]{Abdallah:2014hon}%
  \BibitemOpen
  \bibfield  {author} {\bibinfo {author} {\bibfnamefont {J.}~\bibnamefont
  {Abdallah}} \emph {et~al.},\ }\href@noop {} {\  (\bibinfo {year} {2014})},\
  \Eprint {http://arxiv.org/abs/1409.2893} {arXiv:1409.2893 [hep-ph]}
  \BibitemShut {NoStop}%
\bibitem [{\citenamefont {Buchmueller}\ \emph {et~al.}(2014)\citenamefont
  {Buchmueller}, \citenamefont {Dolan},\ and\ \citenamefont
  {McCabe}}]{Buchmueller:2013dya}%
  \BibitemOpen
  \bibfield  {author} {\bibinfo {author} {\bibfnamefont {O.}~\bibnamefont
  {Buchmueller}}, \bibinfo {author} {\bibfnamefont {M.~J.}\ \bibnamefont
  {Dolan}}, \ and\ \bibinfo {author} {\bibfnamefont {C.}~\bibnamefont
  {McCabe}},\ }\href {\doibase 10.1007/JHEP01(2014)025} {\bibfield  {journal}
  {\bibinfo  {journal} {JHEP}\ }\textbf {\bibinfo {volume} {01}},\ \bibinfo
  {pages} {025} (\bibinfo {year} {2014})},\ \Eprint
  {http://arxiv.org/abs/1308.6799} {arXiv:1308.6799 [hep-ph]} \BibitemShut
  {NoStop}%
\bibitem [{\citenamefont {Dent}\ \emph {et~al.}(2015)\citenamefont {Dent},
  \citenamefont {Krauss}, \citenamefont {Newstead},\ and\ \citenamefont
  {Sabharwal}}]{Dent:2015zpa}%
  \BibitemOpen
  \bibfield  {author} {\bibinfo {author} {\bibfnamefont {J.~B.}\ \bibnamefont
  {Dent}}, \bibinfo {author} {\bibfnamefont {L.~M.}\ \bibnamefont {Krauss}},
  \bibinfo {author} {\bibfnamefont {J.~L.}\ \bibnamefont {Newstead}}, \ and\
  \bibinfo {author} {\bibfnamefont {S.}~\bibnamefont {Sabharwal}},\ }\href
  {\doibase 10.1103/PhysRevD.92.063515} {\bibfield  {journal} {\bibinfo
  {journal} {Phys. Rev.}\ }\textbf {\bibinfo {volume} {D92}},\ \bibinfo {pages}
  {063515} (\bibinfo {year} {2015})},\ \Eprint
  {http://arxiv.org/abs/1505.03117} {arXiv:1505.03117 [hep-ph]} \BibitemShut
  {NoStop}%
\bibitem [{\citenamefont {Frandsen}\ \emph {et~al.}(2012)\citenamefont
  {Frandsen}, \citenamefont {Kahlhoefer}, \citenamefont {Preston},
  \citenamefont {Sarkar},\ and\ \citenamefont
  {Schmidt-Hoberg}}]{Frandsen:2012rk}%
  \BibitemOpen
  \bibfield  {author} {\bibinfo {author} {\bibfnamefont {M.~T.}\ \bibnamefont
  {Frandsen}}, \bibinfo {author} {\bibfnamefont {F.}~\bibnamefont
  {Kahlhoefer}}, \bibinfo {author} {\bibfnamefont {A.}~\bibnamefont {Preston}},
  \bibinfo {author} {\bibfnamefont {S.}~\bibnamefont {Sarkar}}, \ and\ \bibinfo
  {author} {\bibfnamefont {K.}~\bibnamefont {Schmidt-Hoberg}},\ }\href
  {\doibase 10.1007/JHEP07(2012)123} {\bibfield  {journal} {\bibinfo  {journal}
  {JHEP}\ }\textbf {\bibinfo {volume} {07}},\ \bibinfo {pages} {123} (\bibinfo
  {year} {2012})},\ \Eprint {http://arxiv.org/abs/1204.3839} {arXiv:1204.3839
  [hep-ph]} \BibitemShut {NoStop}%
\bibitem [{\citenamefont {An}\ \emph {et~al.}(2012)\citenamefont {An},
  \citenamefont {Ji},\ and\ \citenamefont {Wang}}]{An:2012va}%
  \BibitemOpen
  \bibfield  {author} {\bibinfo {author} {\bibfnamefont {H.}~\bibnamefont
  {An}}, \bibinfo {author} {\bibfnamefont {X.}~\bibnamefont {Ji}}, \ and\
  \bibinfo {author} {\bibfnamefont {L.-T.}\ \bibnamefont {Wang}},\ }\href
  {\doibase 10.1007/JHEP07(2012)182} {\bibfield  {journal} {\bibinfo  {journal}
  {JHEP}\ }\textbf {\bibinfo {volume} {07}},\ \bibinfo {pages} {182} (\bibinfo
  {year} {2012})},\ \Eprint {http://arxiv.org/abs/1202.2894} {arXiv:1202.2894
  [hep-ph]} \BibitemShut {NoStop}%
\bibitem [{\citenamefont {An}\ \emph {et~al.}(2013)\citenamefont {An},
  \citenamefont {Huo},\ and\ \citenamefont {Wang}}]{An:2012ue}%
  \BibitemOpen
  \bibfield  {author} {\bibinfo {author} {\bibfnamefont {H.}~\bibnamefont
  {An}}, \bibinfo {author} {\bibfnamefont {R.}~\bibnamefont {Huo}}, \ and\
  \bibinfo {author} {\bibfnamefont {L.-T.}\ \bibnamefont {Wang}},\ }\href
  {\doibase 10.1016/j.dark.2013.03.002} {\bibfield  {journal} {\bibinfo
  {journal} {Phys. Dark Univ.}\ }\textbf {\bibinfo {volume} {2}},\ \bibinfo
  {pages} {50} (\bibinfo {year} {2013})},\ \Eprint
  {http://arxiv.org/abs/1212.2221} {arXiv:1212.2221 [hep-ph]} \BibitemShut
  {NoStop}%
\bibitem [{\citenamefont {Dobrescu}\ and\ \citenamefont
  {Yu}(2013)}]{Dobrescu:2013coa}%
  \BibitemOpen
  \bibfield  {author} {\bibinfo {author} {\bibfnamefont {B.~A.}\ \bibnamefont
  {Dobrescu}}\ and\ \bibinfo {author} {\bibfnamefont {F.}~\bibnamefont {Yu}},\
  }\href {\doibase 10.1103/PhysRevD.88.035021, 10.1103/PhysRevD.90.079901}
  {\bibfield  {journal} {\bibinfo  {journal} {Phys. Rev.}\ }\textbf {\bibinfo
  {volume} {D88}},\ \bibinfo {pages} {035021} (\bibinfo {year} {2013})},\
  \bibinfo {note} {[Erratum: Phys. Rev.D90,no.7,079901(2014)]},\ \Eprint
  {http://arxiv.org/abs/1306.2629} {arXiv:1306.2629 [hep-ph]} \BibitemShut
  {NoStop}%
\bibitem [{\citenamefont {Alves}\ \emph {et~al.}(2014)\citenamefont {Alves},
  \citenamefont {Profumo},\ and\ \citenamefont {Queiroz}}]{Alves:2013tqa}%
  \BibitemOpen
  \bibfield  {author} {\bibinfo {author} {\bibfnamefont {A.}~\bibnamefont
  {Alves}}, \bibinfo {author} {\bibfnamefont {S.}~\bibnamefont {Profumo}}, \
  and\ \bibinfo {author} {\bibfnamefont {F.~S.}\ \bibnamefont {Queiroz}},\
  }\href {\doibase 10.1007/JHEP04(2014)063} {\bibfield  {journal} {\bibinfo
  {journal} {JHEP}\ }\textbf {\bibinfo {volume} {04}},\ \bibinfo {pages} {063}
  (\bibinfo {year} {2014})},\ \Eprint {http://arxiv.org/abs/1312.5281}
  {arXiv:1312.5281 [hep-ph]} \BibitemShut {NoStop}%
\bibitem [{\citenamefont {Arcadi}\ \emph {et~al.}(2014)\citenamefont {Arcadi},
  \citenamefont {Mambrini}, \citenamefont {Tytgat},\ and\ \citenamefont
  {Zaldivar}}]{Arcadi:2013qia}%
  \BibitemOpen
  \bibfield  {author} {\bibinfo {author} {\bibfnamefont {G.}~\bibnamefont
  {Arcadi}}, \bibinfo {author} {\bibfnamefont {Y.}~\bibnamefont {Mambrini}},
  \bibinfo {author} {\bibfnamefont {M.~H.~G.}\ \bibnamefont {Tytgat}}, \ and\
  \bibinfo {author} {\bibfnamefont {B.}~\bibnamefont {Zaldivar}},\ }\href
  {\doibase 10.1007/JHEP03(2014)134} {\bibfield  {journal} {\bibinfo  {journal}
  {JHEP}\ }\textbf {\bibinfo {volume} {03}},\ \bibinfo {pages} {134} (\bibinfo
  {year} {2014})},\ \Eprint {http://arxiv.org/abs/1401.0221} {arXiv:1401.0221
  [hep-ph]} \BibitemShut {NoStop}%
\bibitem [{\citenamefont {Chala}\ \emph {et~al.}(2015)\citenamefont {Chala},
  \citenamefont {Kahlhoefer}, \citenamefont {McCullough}, \citenamefont
  {Nardini},\ and\ \citenamefont {Schmidt-Hoberg}}]{Chala:2015ama}%
  \BibitemOpen
  \bibfield  {author} {\bibinfo {author} {\bibfnamefont {M.}~\bibnamefont
  {Chala}}, \bibinfo {author} {\bibfnamefont {F.}~\bibnamefont {Kahlhoefer}},
  \bibinfo {author} {\bibfnamefont {M.}~\bibnamefont {McCullough}}, \bibinfo
  {author} {\bibfnamefont {G.}~\bibnamefont {Nardini}}, \ and\ \bibinfo
  {author} {\bibfnamefont {K.}~\bibnamefont {Schmidt-Hoberg}},\ }\href
  {\doibase 10.1007/JHEP07(2015)089} {\bibfield  {journal} {\bibinfo  {journal}
  {JHEP}\ }\textbf {\bibinfo {volume} {07}},\ \bibinfo {pages} {089} (\bibinfo
  {year} {2015})},\ \Eprint {http://arxiv.org/abs/1503.05916} {arXiv:1503.05916
  [hep-ph]} \BibitemShut {NoStop}%
\bibitem [{\citenamefont {Fairbairn}\ \emph {et~al.}(2016)\citenamefont
  {Fairbairn}, \citenamefont {Heal}, \citenamefont {Kahlhoefer},\ and\
  \citenamefont {Tunney}}]{Fairbairn:2016iuf}%
  \BibitemOpen
  \bibfield  {author} {\bibinfo {author} {\bibfnamefont {M.}~\bibnamefont
  {Fairbairn}}, \bibinfo {author} {\bibfnamefont {J.}~\bibnamefont {Heal}},
  \bibinfo {author} {\bibfnamefont {F.}~\bibnamefont {Kahlhoefer}}, \ and\
  \bibinfo {author} {\bibfnamefont {P.}~\bibnamefont {Tunney}},\ }\href
  {\doibase 10.1007/JHEP09(2016)018} {\bibfield  {journal} {\bibinfo  {journal}
  {JHEP}\ }\textbf {\bibinfo {volume} {09}},\ \bibinfo {pages} {018} (\bibinfo
  {year} {2016})},\ \Eprint {http://arxiv.org/abs/1605.07940} {arXiv:1605.07940
  [hep-ph]} \BibitemShut {NoStop}%
\bibitem [{\citenamefont {Cacciari}\ \emph {et~al.}(2008)\citenamefont
  {Cacciari}, \citenamefont {Salam},\ and\ \citenamefont
  {Soyez}}]{Cacciari:2008gp}%
  \BibitemOpen
  \bibfield  {author} {\bibinfo {author} {\bibfnamefont {M.}~\bibnamefont
  {Cacciari}}, \bibinfo {author} {\bibfnamefont {G.~P.}\ \bibnamefont {Salam}},
  \ and\ \bibinfo {author} {\bibfnamefont {G.}~\bibnamefont {Soyez}},\ }\href
  {\doibase 10.1088/1126-6708/2008/04/063} {\bibfield  {journal} {\bibinfo
  {journal} {JHEP}\ }\textbf {\bibinfo {volume} {04}},\ \bibinfo {pages} {063}
  (\bibinfo {year} {2008})},\ \Eprint {http://arxiv.org/abs/0802.1189}
  {arXiv:0802.1189 [hep-ph]} \BibitemShut {NoStop}%
\bibitem [{\citenamefont {Cacciari}\ and\ \citenamefont
  {Salam}(2006)}]{Cacciari:2005hq}%
  \BibitemOpen
  \bibfield  {author} {\bibinfo {author} {\bibfnamefont {M.}~\bibnamefont
  {Cacciari}}\ and\ \bibinfo {author} {\bibfnamefont {G.~P.}\ \bibnamefont
  {Salam}},\ }\href {\doibase 10.1016/j.physletb.2006.08.037} {\bibfield
  {journal} {\bibinfo  {journal} {Phys. Lett.}\ }\textbf {\bibinfo {volume}
  {B641}},\ \bibinfo {pages} {57} (\bibinfo {year} {2006})},\ \Eprint
  {http://arxiv.org/abs/hep-ph/0512210} {arXiv:hep-ph/0512210 [hep-ph]}
  \BibitemShut {NoStop}%
\bibitem [{\citenamefont {Alwall}\ \emph {et~al.}(2014)\citenamefont {Alwall},
  \citenamefont {Frederix}, \citenamefont {Frixione}, \citenamefont {Hirschi},
  \citenamefont {Maltoni}, \citenamefont {Mattelaer}, \citenamefont {Shao},
  \citenamefont {Stelzer}, \citenamefont {Torrielli},\ and\ \citenamefont
  {Zaro}}]{Alwall:2014hca}%
  \BibitemOpen
  \bibfield  {author} {\bibinfo {author} {\bibfnamefont {J.}~\bibnamefont
  {Alwall}}, \bibinfo {author} {\bibfnamefont {R.}~\bibnamefont {Frederix}},
  \bibinfo {author} {\bibfnamefont {S.}~\bibnamefont {Frixione}}, \bibinfo
  {author} {\bibfnamefont {V.}~\bibnamefont {Hirschi}}, \bibinfo {author}
  {\bibfnamefont {F.}~\bibnamefont {Maltoni}}, \bibinfo {author} {\bibfnamefont
  {O.}~\bibnamefont {Mattelaer}}, \bibinfo {author} {\bibfnamefont {H.~S.}\
  \bibnamefont {Shao}}, \bibinfo {author} {\bibfnamefont {T.}~\bibnamefont
  {Stelzer}}, \bibinfo {author} {\bibfnamefont {P.}~\bibnamefont {Torrielli}},
  \ and\ \bibinfo {author} {\bibfnamefont {M.}~\bibnamefont {Zaro}},\ }\href
  {\doibase 10.1007/JHEP07(2014)079} {\bibfield  {journal} {\bibinfo  {journal}
  {JHEP}\ }\textbf {\bibinfo {volume} {07}},\ \bibinfo {pages} {079} (\bibinfo
  {year} {2014})},\ \Eprint {http://arxiv.org/abs/1405.0301} {arXiv:1405.0301
  [hep-ph]} \BibitemShut {NoStop}%
\bibitem [{\citenamefont {Sjostrand}\ \emph {et~al.}(2006)\citenamefont
  {Sjostrand}, \citenamefont {Mrenna},\ and\ \citenamefont
  {Skands}}]{Sjostrand:2006za}%
  \BibitemOpen
  \bibfield  {author} {\bibinfo {author} {\bibfnamefont {T.}~\bibnamefont
  {Sjostrand}}, \bibinfo {author} {\bibfnamefont {S.}~\bibnamefont {Mrenna}}, \
  and\ \bibinfo {author} {\bibfnamefont {P.~Z.}\ \bibnamefont {Skands}},\
  }\href {\doibase 10.1088/1126-6708/2006/05/026} {\bibfield  {journal}
  {\bibinfo  {journal} {JHEP}\ }\textbf {\bibinfo {volume} {05}},\ \bibinfo
  {pages} {026} (\bibinfo {year} {2006})},\ \Eprint
  {http://arxiv.org/abs/hep-ph/0603175} {arXiv:hep-ph/0603175 [hep-ph]}
  \BibitemShut {NoStop}%
\bibitem [{\citenamefont {Sj{\"o}strand}\ \emph {et~al.}(2015)\citenamefont
  {Sj{\"o}strand}, \citenamefont {Ask}, \citenamefont {Christiansen},
  \citenamefont {Corke}, \citenamefont {Desai}, \citenamefont {Ilten},
  \citenamefont {Mrenna}, \citenamefont {Prestel}, \citenamefont {Rasmussen},\
  and\ \citenamefont {Skands}}]{Sjostrand:2014zea}%
  \BibitemOpen
  \bibfield  {author} {\bibinfo {author} {\bibfnamefont {T.}~\bibnamefont
  {Sj{\"o}strand}}, \bibinfo {author} {\bibfnamefont {S.}~\bibnamefont {Ask}},
  \bibinfo {author} {\bibfnamefont {J.~R.}\ \bibnamefont {Christiansen}},
  \bibinfo {author} {\bibfnamefont {R.}~\bibnamefont {Corke}}, \bibinfo
  {author} {\bibfnamefont {N.}~\bibnamefont {Desai}}, \bibinfo {author}
  {\bibfnamefont {P.}~\bibnamefont {Ilten}}, \bibinfo {author} {\bibfnamefont
  {S.}~\bibnamefont {Mrenna}}, \bibinfo {author} {\bibfnamefont
  {S.}~\bibnamefont {Prestel}}, \bibinfo {author} {\bibfnamefont {C.~O.}\
  \bibnamefont {Rasmussen}}, \ and\ \bibinfo {author} {\bibfnamefont {P.~Z.}\
  \bibnamefont {Skands}},\ }\href {\doibase 10.1016/j.cpc.2015.01.024}
  {\bibfield  {journal} {\bibinfo  {journal} {Comput. Phys. Commun.}\ }\textbf
  {\bibinfo {volume} {191}},\ \bibinfo {pages} {159} (\bibinfo {year}
  {2015})},\ \Eprint {http://arxiv.org/abs/1410.3012} {arXiv:1410.3012
  [hep-ph]} \BibitemShut {NoStop}%
\bibitem [{\citenamefont {de~Favereau}\ \emph {et~al.}(2014)\citenamefont
  {de~Favereau}, \citenamefont {Delaere}, \citenamefont {Demin}, \citenamefont
  {Giammanco}, \citenamefont {Lemaître}, \citenamefont {Mertens},\ and\
  \citenamefont {Selvaggi}}]{deFavereau:2013fsa}%
  \BibitemOpen
  \bibfield  {author} {\bibinfo {author} {\bibfnamefont {J.}~\bibnamefont
  {de~Favereau}}, \bibinfo {author} {\bibfnamefont {C.}~\bibnamefont
  {Delaere}}, \bibinfo {author} {\bibfnamefont {P.}~\bibnamefont {Demin}},
  \bibinfo {author} {\bibfnamefont {A.}~\bibnamefont {Giammanco}}, \bibinfo
  {author} {\bibfnamefont {V.}~\bibnamefont {Lemaître}}, \bibinfo {author}
  {\bibfnamefont {A.}~\bibnamefont {Mertens}}, \ and\ \bibinfo {author}
  {\bibfnamefont {M.}~\bibnamefont {Selvaggi}} (\bibinfo {collaboration}
  {DELPHES 3}),\ }\href {\doibase 10.1007/JHEP02(2014)057} {\bibfield
  {journal} {\bibinfo  {journal} {JHEP}\ }\textbf {\bibinfo {volume} {02}},\
  \bibinfo {pages} {057} (\bibinfo {year} {2014})},\ \Eprint
  {http://arxiv.org/abs/1307.6346} {arXiv:1307.6346 [hep-ex]} \BibitemShut
  {NoStop}%
\bibitem [{167(2018)}]{1676214}%
  \BibitemOpen
  \href {\doibase 10.17182/hepdata.80166} {\enquote {\bibinfo {title} {{Search
  for narrow and broad dijet resonances in proton-proton collisions at $
  \sqrt{s}=13 $ TeV and constraints on dark matter mediators and other new
  particles}},}\ } (\bibinfo {year} {2018}),\ \bibinfo {note} {hepData
  Repository (based on JHEP 08 (2018) 130), 2018)}\BibitemShut {NoStop}%
\bibitem [{\citenamefont {Kilian}\ \emph {et~al.}(2011)\citenamefont {Kilian},
  \citenamefont {Ohl},\ and\ \citenamefont {Reuter}}]{Kilian:2007gr}%
  \BibitemOpen
  \bibfield  {author} {\bibinfo {author} {\bibfnamefont {W.}~\bibnamefont
  {Kilian}}, \bibinfo {author} {\bibfnamefont {T.}~\bibnamefont {Ohl}}, \ and\
  \bibinfo {author} {\bibfnamefont {J.}~\bibnamefont {Reuter}},\ }\href
  {\doibase 10.1140/epjc/s10052-011-1742-y} {\bibfield  {journal} {\bibinfo
  {journal} {Eur. Phys. J.}\ }\textbf {\bibinfo {volume} {C71}},\ \bibinfo
  {pages} {1742} (\bibinfo {year} {2011})},\ \Eprint
  {http://arxiv.org/abs/0708.4233} {arXiv:0708.4233 [hep-ph]} \BibitemShut
  {NoStop}%
\bibitem [{\citenamefont {Moretti}\ \emph {et~al.}(2001)\citenamefont
  {Moretti}, \citenamefont {Ohl},\ and\ \citenamefont
  {Reuter}}]{Moretti:2001zz}%
  \BibitemOpen
  \bibfield  {author} {\bibinfo {author} {\bibfnamefont {M.}~\bibnamefont
  {Moretti}}, \bibinfo {author} {\bibfnamefont {T.}~\bibnamefont {Ohl}}, \ and\
  \bibinfo {author} {\bibfnamefont {J.}~\bibnamefont {Reuter}},\ }\href@noop {}
  {\  (\bibinfo {year} {2001})},\ \Eprint {http://arxiv.org/abs/hep-ph/0102195}
  {arXiv:hep-ph/0102195 [hep-ph]} \BibitemShut {NoStop}%
\bibitem [{\citenamefont {Baum}\ \emph {et~al.}(2018)\citenamefont {Baum},
  \citenamefont {Catena}, \citenamefont {Conrad}, \citenamefont {Freese},\ and\
  \citenamefont {Krauss}}]{Baum:2017kfa}%
  \BibitemOpen
  \bibfield  {author} {\bibinfo {author} {\bibfnamefont {S.}~\bibnamefont
  {Baum}}, \bibinfo {author} {\bibfnamefont {R.}~\bibnamefont {Catena}},
  \bibinfo {author} {\bibfnamefont {J.}~\bibnamefont {Conrad}}, \bibinfo
  {author} {\bibfnamefont {K.}~\bibnamefont {Freese}}, \ and\ \bibinfo {author}
  {\bibfnamefont {M.~B.}\ \bibnamefont {Krauss}},\ }\href {\doibase
  10.1103/PhysRevD.97.083002} {\bibfield  {journal} {\bibinfo  {journal} {Phys.
  Rev.}\ }\textbf {\bibinfo {volume} {D97}},\ \bibinfo {pages} {083002}
  (\bibinfo {year} {2018})},\ \Eprint {http://arxiv.org/abs/1709.06051}
  {arXiv:1709.06051 [hep-ph]} \BibitemShut {NoStop}%
\bibitem [{\citenamefont {Buckley}\ \emph {et~al.}(2015)\citenamefont
  {Buckley}, \citenamefont {Ferrando}, \citenamefont {Lloyd}, \citenamefont
  {Nordstr{\"o}m}, \citenamefont {Page}, \citenamefont {R{\"u}fenacht},
  \citenamefont {Sch{\"o}nherr},\ and\ \citenamefont {Watt}}]{Buckley:2014ana}%
  \BibitemOpen
  \bibfield  {author} {\bibinfo {author} {\bibfnamefont {A.}~\bibnamefont
  {Buckley}}, \bibinfo {author} {\bibfnamefont {J.}~\bibnamefont {Ferrando}},
  \bibinfo {author} {\bibfnamefont {S.}~\bibnamefont {Lloyd}}, \bibinfo
  {author} {\bibfnamefont {K.}~\bibnamefont {Nordstr{\"o}m}}, \bibinfo {author}
  {\bibfnamefont {B.}~\bibnamefont {Page}}, \bibinfo {author} {\bibfnamefont
  {M.}~\bibnamefont {R{\"u}fenacht}}, \bibinfo {author} {\bibfnamefont
  {M.}~\bibnamefont {Sch{\"o}nherr}}, \ and\ \bibinfo {author} {\bibfnamefont
  {G.}~\bibnamefont {Watt}},\ }\href {\doibase 10.1140/epjc/s10052-015-3318-8}
  {\bibfield  {journal} {\bibinfo  {journal} {Eur. Phys. J.}\ }\textbf
  {\bibinfo {volume} {C75}},\ \bibinfo {pages} {132} (\bibinfo {year}
  {2015})},\ \Eprint {http://arxiv.org/abs/1412.7420} {arXiv:1412.7420
  [hep-ph]} \BibitemShut {NoStop}%
\bibitem [{\citenamefont {Cacciari}\ \emph {et~al.}(2012)\citenamefont
  {Cacciari}, \citenamefont {Salam},\ and\ \citenamefont
  {Soyez}}]{Cacciari:2011ma}%
  \BibitemOpen
  \bibfield  {author} {\bibinfo {author} {\bibfnamefont {M.}~\bibnamefont
  {Cacciari}}, \bibinfo {author} {\bibfnamefont {G.~P.}\ \bibnamefont {Salam}},
  \ and\ \bibinfo {author} {\bibfnamefont {G.}~\bibnamefont {Soyez}},\ }\href
  {\doibase 10.1140/epjc/s10052-012-1896-2} {\bibfield  {journal} {\bibinfo
  {journal} {Eur. Phys. J.}\ }\textbf {\bibinfo {volume} {C72}},\ \bibinfo
  {pages} {1896} (\bibinfo {year} {2012})},\ \Eprint
  {http://arxiv.org/abs/1111.6097} {arXiv:1111.6097 [hep-ph]} \BibitemShut
  {NoStop}%
\bibitem [{\citenamefont {Brun}\ and\ \citenamefont
  {Rademakers}(1997)}]{Brun:1997pa}%
  \BibitemOpen
  \bibfield  {author} {\bibinfo {author} {\bibfnamefont {R.}~\bibnamefont
  {Brun}}\ and\ \bibinfo {author} {\bibfnamefont {F.}~\bibnamefont
  {Rademakers}},\ }\bibfield  {booktitle} {\emph {\bibinfo {booktitle} {{New
  computing techniques in physics research V. Proceedings, 5th International
  Workshop, AIHENP '96, Lausanne, Switzerland, September 2-6, 1996}}},\ }\href
  {\doibase 10.1016/S0168-9002(97)00048-X} {\bibfield  {journal} {\bibinfo
  {journal} {Nucl. Instrum. Meth.}\ }\textbf {\bibinfo {volume} {A389}},\
  \bibinfo {pages} {81} (\bibinfo {year} {1997})}\BibitemShut {NoStop}%
\bibitem [{\citenamefont {Cowan}\ \emph {et~al.}(2011)\citenamefont {Cowan},
  \citenamefont {Cranmer}, \citenamefont {Gross},\ and\ \citenamefont
  {Vitells}}]{Cowan:2010js}%
  \BibitemOpen
  \bibfield  {author} {\bibinfo {author} {\bibfnamefont {G.}~\bibnamefont
  {Cowan}}, \bibinfo {author} {\bibfnamefont {K.}~\bibnamefont {Cranmer}},
  \bibinfo {author} {\bibfnamefont {E.}~\bibnamefont {Gross}}, \ and\ \bibinfo
  {author} {\bibfnamefont {O.}~\bibnamefont {Vitells}},\ }\href {\doibase
  10.1140/epjc/s10052-011-1554-0, 10.1140/epjc/s10052-013-2501-z} {\bibfield
  {journal} {\bibinfo  {journal} {Eur. Phys. J.}\ }\textbf {\bibinfo {volume}
  {C71}},\ \bibinfo {pages} {1554} (\bibinfo {year} {2011})},\ \bibinfo {note}
  {[Erratum: Eur. Phys. J.C73,2501(2013)]},\ \Eprint
  {http://arxiv.org/abs/1007.1727} {arXiv:1007.1727 [physics.data-an]}
  \BibitemShut {NoStop}%
\bibitem [{\citenamefont {Catena}\ \emph {et~al.}(2018)\citenamefont {Catena},
  \citenamefont {Conrad},\ and\ \citenamefont {Krauss}}]{Catena:2017xqq}%
  \BibitemOpen
  \bibfield  {author} {\bibinfo {author} {\bibfnamefont {R.}~\bibnamefont
  {Catena}}, \bibinfo {author} {\bibfnamefont {J.}~\bibnamefont {Conrad}}, \
  and\ \bibinfo {author} {\bibfnamefont {M.~B.}\ \bibnamefont {Krauss}},\
  }\href {\doibase 10.1103/PhysRevD.97.103002} {\bibfield  {journal} {\bibinfo
  {journal} {Phys. Rev.}\ }\textbf {\bibinfo {volume} {D97}},\ \bibinfo {pages}
  {103002} (\bibinfo {year} {2018})},\ \Eprint
  {http://arxiv.org/abs/1712.07969} {arXiv:1712.07969 [hep-ph]} \BibitemShut
  {NoStop}%
\bibitem [{\citenamefont {Sirunyan}\ \emph
  {et~al.}(2018{\natexlab{c}})\citenamefont {Sirunyan} \emph
  {et~al.}}]{Sirunyan:2018wcm}%
  \BibitemOpen
  \bibfield  {author} {\bibinfo {author} {\bibfnamefont {A.~M.}\ \bibnamefont
  {Sirunyan}} \emph {et~al.} (\bibinfo {collaboration} {CMS}),\ }\href
  {\doibase 10.1140/epjc/s10052-018-6242-x} {\bibfield  {journal} {\bibinfo
  {journal} {Eur. Phys. J.}\ }\textbf {\bibinfo {volume} {C78}},\ \bibinfo
  {pages} {789} (\bibinfo {year} {2018}{\natexlab{c}})},\ \Eprint
  {http://arxiv.org/abs/1803.08030} {arXiv:1803.08030 [hep-ex]} \BibitemShut
  {NoStop}%
\bibitem [{\citenamefont {Sirunyan}\ \emph {et~al.}(2017)\citenamefont
  {Sirunyan} \emph {et~al.}}]{Sirunyan:2017ygf}%
  \BibitemOpen
  \bibfield  {author} {\bibinfo {author} {\bibfnamefont {A.~M.}\ \bibnamefont
  {Sirunyan}} \emph {et~al.} (\bibinfo {collaboration} {CMS}),\ }\href
  {\doibase 10.1007/JHEP07(2017)013} {\bibfield  {journal} {\bibinfo  {journal}
  {JHEP}\ }\textbf {\bibinfo {volume} {07}},\ \bibinfo {pages} {013} (\bibinfo
  {year} {2017})},\ \Eprint {http://arxiv.org/abs/1703.09986} {arXiv:1703.09986
  [hep-ex]} \BibitemShut {NoStop}%
\bibitem [{\citenamefont {Aad}\ \emph {et~al.}(2015)\citenamefont {Aad} \emph
  {et~al.}}]{Aad:2015eha}%
  \BibitemOpen
  \bibfield  {author} {\bibinfo {author} {\bibfnamefont {G.}~\bibnamefont
  {Aad}} \emph {et~al.} (\bibinfo {collaboration} {ATLAS}),\ }\href {\doibase
  10.1103/PhysRevLett.114.221802} {\bibfield  {journal} {\bibinfo  {journal}
  {Phys. Rev. Lett.}\ }\textbf {\bibinfo {volume} {114}},\ \bibinfo {pages}
  {221802} (\bibinfo {year} {2015})},\ \Eprint
  {http://arxiv.org/abs/1504.00357} {arXiv:1504.00357 [hep-ex]} \BibitemShut
  {NoStop}%
\bibitem [{\citenamefont {Khachatryan}\ \emph {et~al.}(2015)\citenamefont
  {Khachatryan} \emph {et~al.}}]{Khachatryan:2014cja}%
  \BibitemOpen
  \bibfield  {author} {\bibinfo {author} {\bibfnamefont {V.}~\bibnamefont
  {Khachatryan}} \emph {et~al.} (\bibinfo {collaboration} {CMS}),\ }\href
  {\doibase 10.1016/j.physletb.2015.04.042} {\bibfield  {journal} {\bibinfo
  {journal} {Phys. Lett.}\ }\textbf {\bibinfo {volume} {B746}},\ \bibinfo
  {pages} {79} (\bibinfo {year} {2015})},\ \Eprint
  {http://arxiv.org/abs/1411.2646} {arXiv:1411.2646 [hep-ex]} \BibitemShut
  {NoStop}%
\end{thebibliography}%
